%% file: paper_revised.tex
\documentclass[11pt,english]{article}
\usepackage{natbib}
\usepackage{graphicx} 
\usepackage{amsmath}
\usepackage{amssymb}
\usepackage{comment}
\usepackage{cancel}
\usepackage{babel}
\usepackage{amsthm}
\usepackage{mathrsfs}
\usepackage{etoolbox}
\usepackage{caption}
\usepackage{tabularx}
\usepackage{url}
\usepackage{booktabs}
\usepackage{pgf,tikz}
\usetikzlibrary{arrows,shapes.geometric,shapes.multipart,positioning,swigs}
\usepackage{adjustbox}
\usepackage{threeparttable}
\usepackage{thmtools}
\usepackage{enumerate}
\usepackage{subcaption}
\usepackage{hyperref}
\theoremstyle{remark}

\theoremstyle{plain}

\newtheorem{assumption}{Assumption}

\theoremstyle{definition}
\newtheorem{definition}{Definition}

\AtBeginEnvironment{example}{\vspace{5pt}}
\AtEndEnvironment{example}{\vspace{5pt}} 

\usepackage[T1]{fontenc}
\usepackage[utf8]{inputenc}

\usepackage{geometry}
\makeatletter

\usepackage{epsf}

\usepackage{setspace}
\newcommand{\ind}{\perp\!\!\!\perp} 
\newcommand{\IF}{\mathbb{IF}} 

\title{A Unifying Framework for Robust and Efficient Inference with Unstructured Data}
\author{Jacob Carlson and Melissa Dell\thanks{Corresponding author: Melissa Dell, Littauer Center, Cambridge, MA 02138. \texttt{melissadell@fas.harvard.edu}. We thank Isaiah Andrews, Xiaohong Chen, Pete Kyle, Ashesh Rambachan, Neil Shephard, Rahul Singh, Elie Tamer, and Davide Viviano, as well as seminar participants at The Bank of Japan, Berkeley, The Federal Reserve, Fordham, Harvard, Microsoft Research, Stanford, the University of Chicago, and the University of Toronto for their helpful comments and suggestions. We used Claude and Astra for proof checking and to provide suggestions on the exposition.}}
\date{}

\begin{document}

\maketitle

\abstract{AI can cheaply extract structured variables from unstructured data, with alternative implementation choices potentially yielding different measures of the same concept. This creates a credibility challenge, effectively moving specification search upstream. We confront this challenge with MAR-S (\textbf{M}issing \textbf{A}t \textbf{R}andom \textbf{S}tructured Data), a framework which adapts semiparametric missing data theory to deliver valid, efficient, robust inference with AI predictions. MAR-S anchors measurement in a transparent validation sample design, while AI scales measurement. We provide general recipes for MAR-S estimation and extend the framework to aggregated and transformed AI predictions, a common setting not addressed by existing frameworks.}

\section{Introduction}

Economists increasingly deploy predictions from artificial intelligence (AI) in empirical research. Deep neural networks, which are the foundation of modern AI, can cheaply extract low-dimensional structured variables from high-dimensional unstructured data, such as text, images, audio, and video \citep{goodfellow_deep_2016}. For example, researchers might extract the sentiment of a text, or use remote sensing to measure pollution or urbanization. 

Extracting structured variables from unstructured data traditionally required costly manual processing.
AI is now shifting this measurement bottleneck, from constructing \textit{any} plausible measure to determining which among \textit{many} plausible measures are credible. 
Even precisely defined economic constructs can be operationalized with AI in many different ways, and because neural networks are highly scaleable, creating many different implementations of the same concept can be cheap. 
Specifically, researchers must choose a model, a prompting strategy (when using generative AI), and system instructions. When training their own models, they must choose the architecture, training data distribution, decoding strategy, and various hyperparameters. Pre- and post-processing steps are also common.

Much as cheap compute made it possible for researchers to run not one but millions of regressions \citep{sala1997just}, AI moves specification search upstream, to choosing among many possible measurements of a construct.
The construct does not pin down AI implementation choices, nor can the researcher provide an interpretable basis for preferring one measure over another by directly inspecting the AI model. Deep neural networks contain interactions between millions to billions of learned parameters. Errors shift in complex ways with the distribution of inputs and implementation choices, and humans are poor at anticipating how these changes affect performance \citep{ben2010theory, vafa2024large}. 

The ``garden of forking paths'' \citep{leamer1978specification, gelman2013garden} created by AI measurement could be innocuous, if each AI-generated measure is simply the underlying construct plus classical measurement error. Theoretical and growing empirical evidence, however, suggest this is unlikely. AI predictions are generated by learned, highly nonlinear mappings whose errors can vary systematically with the model architecture, training data distribution, prompting strategy, and other implementation choices, violating classical measurement error. Empirically, changing the model or prompting strategy can materially alter economic estimates in magnitude, statistical significance, and sign, as \citet{LudwigMullainathanRambachan2026} document for applications with congressional bills and financial news headlines. Related evidence appears across a growing literature \citep{Sclar2024PromptFormatting, Atreja2025WhatsPrompt, Yin2026ExposureScores}. For instance, \citet{baumann2025large} replicate 21 published studies, showing that prompt paraphrasing can make virtually any hypothesis appear statistically significant. 

When using AI to measure economic variables, we argue that researchers should specify a precise, observable definition of their target measurement and validate AI predictions against this, adjusting their estimates for systematic errors. The former, \textit{construct definition}, translates a broader measurement objective into a criterion precise enough to be directly observed in a particular setting, whereas the latter,
\textit{observational validity}, takes the construct definition as fixed and examines how accurately it is implemented. 
For instance, suppose a researcher operationally defines economic policy uncertainty as responses to a set of survey questions or a rubric that specifies what it means for news articles to discuss it \citep{baker_measuring_2016}. AI can extend such definitions to a large sample, via synthetic survey respondents or text classification, and the researcher can assess observational validity against human survey responses or a sample of carefully validated news articles. Whether these measures capture latent economic policy uncertainty well enough to answer the research question is not directly observable and thus must be separately defended. When a researcher does not wish to specify an operational construct, discovery methods---which detect latent structure in data---offer advantages over prompting an AI with a concept that is too vague or implicit for the researcher to directly validate (see \cite{dell2026measurement} for a review).

This study focuses on \textit{observationally valid} inference with AI-generated measurements. It builds upon longstanding literatures on missing data, in particular the missing at random (MAR) mechanism of \citet{rubin_inference_1976}.\footnote{See also \citet{little_statistical_2019,robins_estimation_1994,robins_analysis_1995,robins_semiparametric_1995,bang_doubly_2005}.} Inference with AI predictions is cast as a missing data problem because the structured variables needed for empirical analysis (\textit{e.g.,} the topic of a text or amount of pollution in an image) are imputed using an AI model. 
The missing at random approach collects a conditionally random, modestly-sized \textit{validation sample} using a high-quality, interpretable procedure to implement the researcher's target measurement. For example, validation data might consist of expert annotations, readings from a scientific instrument such as a pollution monitor, survey responses, or administrative data. Validation data define the researcher's measurement objective. In some contexts, they record verifiable facts, whereas in others they have well-recognized limitations. Regardless, they represent the target measurement that the researcher would like an \textit{imputation technology} such as AI to cheaply scale. 
Validation data reveal how the imputed measurements differ from the researcher’s target. Semiparametric methods account for these differences in downstream estimation, while still exploiting information from the imputed sample to improve precision. 

To make credible inference with random validation samples accessible to applied economists, we develop MAR-S (\textit{\textbf{M}issing \textbf{A}t \textbf{R}andom \textbf{S}tructured Data}), which synthesizes and extends results from semiparametric missing data theory to a variety of settings where structured variables are predicted from unstructured data. 
There is a growing literature on inference with AI predictions using validation data \citep{angelopoulos_prediction-powered_2023,angelopoulos_ppi_2024,zrnic_cross-prediction-powered_2024,zrnic_active_2024,ji_predictions_2025,kluger_prediction-powered_2025,egami_using_2024,gligoric-etal-2025-unconfident}. However, there is limited guidance on inference with AI predictions for economists, as this work is dispersed across varied specialized settings, has largely developed outside economics, and does not address many common empirical economics settings. 
MAR-S articulates a common theoretical foundation for the growing set of black-box AI estimators and highlights that results from the broader missing data literature - which includes causal inference - can be applied to inference with AI-generated measures. MAR-S naturally accommodates efficiency analysis, which is beyond the scope of much of the existing literature; non-uniform or unknown validation probabilities; and big data settings in which only a vanishingly small share of observations is validated \citep{zhang_double_2023,kallus_role_2024}. 

Centrally, MAR-S addresses aggregated and transformed AI predictions, a pervasive setting in applied economics not covered by existing approaches. Existing methods assume that validation data are available for the variable used to estimate the target parameter. In many economic applications, however, researchers can validate only granular observations (\textit{e.g.,} information extracted from individual texts or images), while the parameter of interest depends on aggregated AI predictions that are often nonlinearly transformed. For example, researchers may validate individual newspaper articles but use an annual index constructed from millions of article-level predictions, that is logged, differenced, and interacted with other variables. Directly validating the aggregate would typically be prohibitively costly, and infeasible in a finite sample when the target is a population quantity. We develop methods that use granular validation data for parameters estimated from aggregated and transformed missing structured data.

More generally, MAR-S makes robust, efficient inference with AI predictions more accessible in settings routinely encountered by economists. Common estimands in applied economics fall outside the cases treated explicitly in the existing literature on inference with AI predictions. We provide a simple recipe for constructing MAR-S estimators whenever the target parameter is characterized by population moment conditions, by applying longstanding methods in the measurement error literature to high-dimensional unstructured data \citep{chen_semiparametric_2008}. We apply this recipe to a range of variables (\textit{e.g.}, outcome, conditioning variable, instrument, treatment) in common estimands. While deep neural networks are complicated, this recipe provides practical insights that have not been emphasized in the existing literature about what information they should use, what they should learn, and how they should behave asymptotically if efficiency is
important. We also provide an accompanying implementation package.\footnote{\url{https://github.com/jscarlson/mar-s}}

Rubin's missing at random framework closely parallels the potential outcomes framework \citep{rubin1974estimating, rubin1978bayesian}, as causal inference is likewise a missing data problem in which counterfactual outcomes are unobserved. Random validation samples play a statistical role analogous to random treatment assignment, addressing AI measurement challenges. 
With appropriately designed validation, AI predictions need not be accurate for inference to be valid.
Different AI implementations will yield similarly consistent estimates of the target parameter, with better AI predictions improving precision. This provides a principled basis for deciding whether investments in improving AI predictions are worthwhile. It also addresses the reproducibility concern that frontier AI models are frequently deprecated, as validated estimates using successor AI models should remain stable up to differences in precision.

What if validation data exist or can be collected, but not for a random sample? Identification then requires additional assumptions about how the AI model errs \citep{rambachan2024program}, such that the non-random validation sample can be informative for the target population; see \citet{dell2026measurement} for a review. 

Validation may be infeasible because no technology can access the target measurement, or a researcher might prefer to view all measures as noisy approximations of the target. This perspective suggests combining multiple AI measurements, using one as an instrument for another, or interpreting robustness across AI measures as evidence of validity. In the absence of validation, multiple measurement methods rely on strong assumptions about error structure, with point identification typically requiring the measurement errors to be conditionally independent \citep{Dawid1979ObserverErrorRates}. There is little reason to expect this to hold for AI generated measures. Different implementation choices can lead to different predictions. Yet frontier models also share architectures, train on similar corpora, and are tuned toward similar preferred responses, which can lead to ``response homogenization'' \citep{kirk2024understanding, jiang2026artificial}. AI benchmarks show that frontier models disproportionately err on the same items, with more capable models exhibiting more correlated errors \citep{Kim2025CorrelatedErrors}, and similar patterns arise in economic measurement \citep{LudwigMullainathanRambachan2026, Chen2026LLMPrompts}. Appendix Section \ref{ssec:robustness} documents that correlated errors also arise for the applications examined in this study---the economic policy uncertainty index and an index of political news coverage---with errors most strongly correlated among commercial frontier models. 

In short, agreement across AI-generated measures may reflect shared bias rather than identifying the construct.\footnote{\citet{Chen2026LLMPrompts} develop a method that bounds the degree of dependence, calibrating those bounds using external data in which the target construct is observed, in order to estimate an identified set.} 
This highlights the value, when feasible, of defining constructs that allow direct validation. If random validation data contain well-understood error, MAR-S can accommodate it if the errors are specified. By dramatically reducing the costs of scaling measurement, AI shifts the bottleneck towards having high quality measurements to scale in the first place. 

This paper is organized as follows. Section \ref{lit} situates MAR-S within the relevant literatures. Section \ref{sec:MAR-SFramework} introduces the MAR-S framework. Section \ref{sec:gmm} develops a general recipe for estimation with MAR-S, applies it to common estimands, and discusses the implications for efficient inference with unstructured data. Section \ref{sec:agg} extends MAR-S to aggregated and transformed AI predictions, and Section \ref{sec:dropext} addresses validation design in practice. Section \ref{Sec:conclusion} concludes.

\section{Literature} \label{lit}

MAR-S brings the classical semiparametric missing data framework to inference with AI-generated measures, building on Rubin's missing at random framework \citep{rubin_inference_1976} and the semiparametric literature on augmented inverse-probability weighting and efficient estimation \citep{robins_estimation_1994,robins_semiparametric_1995,scharfstein_adjusting_1999,little_statistical_2019,bang_doubly_2005}. These frameworks are particularly well suited to AI-generated measures because they allow nuisance functions---here, predictions of structured variables from unstructured data---to be estimated flexibly. At the same time, they impose relatively little structure on the underlying data generating process and provide principled benchmarks for evaluating efficiency \citep{pfanzagl_contributions_1982,bickel_efficient_1998,newey_asymptotic_1994,van_der_vaart_asymptotic_1998,ackerberg_asymptotic_2014,kennedy_semiparametric_2018,chernozhukov_locally_2022}. MAR-S also inherits the close connection between Rubin's missing data framework and the Rubin Causal Model \citep{rubin1974estimating,rubin1978bayesian}, facilitating the application of methods and intuition from causal inference to inference with AI predictions.

Bringing this semiparametric perspective to inference with AI predictions helps unify a rapidly growing literature on inference with ``black-box'' AI predictions using validation samples, which is summarized in Table \ref{tab:cp}.
The table illustrates that existing black-box AI estimators arise naturally as special cases of the more general semiparametric missing data perspective underlying MAR-S.  This clarifies how the varied estimators in this dispersed literature relate, as the distinctions between these frameworks can be viewed as choices within the more general MAR-S framework.\footnote{We focus here on formal statistical frameworks. More applied treatments include \citet{LudwigMullainathanRambachan2026}; \citet{modarressi_causal_2025}, who incorporate a similar perspective into an LLM-based framework for causal inference with text; and \citet{list_using_2024}, whose regression-adjustment results in experiments are also encompassed by the MAR-S perspective.} 
For example, the smooth, convex M-estimands required by many frameworks can be represented as more general Z-estimands; the known or constant validation probabilities assumed by many frameworks are special cases of the estimated and nonuniform probabilities that the more general framework admits; fixed predictions are a special case of allowing predictors to be learned; and low-dimensional predictors are accommodated within a framework that also permits high-dimensional features. 

\begin{table}[ht]
  \centering
  \footnotesize
  \begin{threeparttable}
  \begin{tabularx}{\textwidth}{%
      >{\centering\arraybackslash}r  
      >{\centering\arraybackslash}X
      >{\centering\arraybackslash}X
      >{\centering\arraybackslash}X
      >{\centering\arraybackslash}X
      >{\centering\arraybackslash}X
      >{\centering\arraybackslash}X
      >{\centering\arraybackslash}X
      >{\centering\arraybackslash}X
      >{\centering\arraybackslash}X
    }
    \toprule
     
      & \textbf{PPI(++)} 
      & \textbf{XPPI} 
      & \textbf{ASI}
      & \textbf{RePPI} 
      & \textbf{DSL} 
      & \textbf{CDI}
      & \textbf{PTD} 
      & \textbf{CEP-GMM} 
      & \textbf{MAR-S} \\
    \midrule
    Estimands & M & M  & M  & M & M & M & Z & Z & Z \\
    Assumptions & MAR+ & MAR+ & MAR+& MAR+ & MAR+ & MAR+ & MAR & MAR & MAR \\
     & PPI & PPI & PPI & PPI & PPI & PPI \\
    Allows unknown $\pi$? & No & No & No & No & No & No & No & Yes & Yes \\
    Allows nonuniform $\pi$? & No & No & Yes & No & Yes & Yes & Yes & Yes & Yes \\
    Semip. Efficiency? & No & No & No & Yes & No & No & No & Yes & Yes \\
    Finetuning? & No & Yes & No & Yes & Yes & No & No & Yes & Yes \\
    Sample Splitting? & No & Yes & No & Yes & Yes & No & No & No & Yes \\
    High-Dim. Features? & Yes & Yes & Yes & Yes & Yes & Yes & Yes & No & Yes \\
    Aggregation? & No & No & No & No & No & No & No & No & Yes \\
    \bottomrule
  \end{tabularx}

  \begin{tablenotes}
    \footnotesize
    \item[] Note: PPI(++) is ``Prediction-Powered Inference'' \citep{angelopoulos_prediction-powered_2023,angelopoulos_ppi_2024}; XPPI is ``Cross-Prediction-Powered Inference'' \citep{zrnic_cross-prediction-powered_2024}; ASI is ``Active Statistical Inference'' \citep{zrnic_active_2024}; RePPI is ``Recalibrated PPI'' \citep{ji_predictions_2025}; DSL is ``Design-based Supervised Learning'' \citep{egami_using_2023,egami_using_2024}; CDI is ``Confidence-Driven Inference'' \citep{gligoric-etal-2025-unconfident}; PTD is ``Predict-Then-Debias'' \citep{kluger_prediction-powered_2025}; and CEP-GMM is ``Conditional Expectation Projection GMM'' \citep{chen_semiparametric_2008}. ``Estimands'' records which classes of target parameters are covered. ``M'' denotes M-estimands, defined as minimizers of a population loss function, while ``Z'' denotes Z-estimands, defined as solutions to population estimating equations. ``Assumptions'' records the core assumptions used for identification: ``MAR'' denotes Rubin's missing at random conditions, while ``PPI'' additionally denotes the assumptions in the original PPI framework \citep{angelopoulos_prediction-powered_2023}, including a convex and sufficiently smooth loss function (differentiable and locally Lipschitz). ``Allows unknown $\pi$?'' indicates whether validation probabilities may be unknown and estimated; ``Allows nonuniform $\pi$?'' whether they may vary with observed characteristics; and ``Semiparametric efficiency?'' whether the framework supports semiparametric efficiency analysis and a path to efficient estimation. ``Finetuning?'' indicates whether the core framework accommodates researcher-led model training, and ``Sample splitting?'' whether it invokes sample splitting or cross-fitting. ``High-Dimensional Features?'' records whether imputations may be learned from high-dimensional features. Finally, ``Aggregation?'' indicates whether the framework accommodates settings in which validation data are observed at a more granular level than the unit at which the target parameter is defined.
  \end{tablenotes}
  \end{threeparttable}
  \caption{Comparison of Debiased Estimation Frameworks}
  \label{tab:cp}
\end{table}

The semiparametric perspective underlying MAR-S brings practical insights as well. MAR-S provides specific prescriptions about what information AI prediction should use, what it should learn, and how it should behave asymptotically if efficiency is important, tailoring the prediction task to the estimand to improve precision. 
This contrasts to the black-box AI literature, which typically takes the prediction target as given and emphasizes robustness to prediction quality.

The overarching goal of MAR-S is to make it more straightforward for economists to integrate validation samples into the vast array of contexts where they are increasingly deploying AI predictions. 
Many estimators that are common in economics are not explicitly treated in the black-box AI literature, which has made limited inroads in economics. 

To make these methods more accessible to economists, MAR-S applies general missing data results for moment-based estimands from \citet{chen_semiparametric_2008}, summarized in the CEP-GMM column of Table \ref{tab:cp}, to inference with high-dimensional, unstructured data. This framework, in contrast to others in the table, pre-dates modern AI and admits low-dimensional features. The application of its machinery to inference with AI-generated measures yields a systematic recipe for constructing efficient estimators with validated AI predictions, significantly simplifying derivation. This reduces the need to derive and justify separate procedures for each estimand, which as Table \ref{tab:cp} illustrates has been the norm. The distinction between low and high-dimensional features matters, underscoring that weak conditions for valid inference apply when {validation probabilities are known} and highlighting issues created by learning low-dimensional measures from high-dimensional data. These include neural network training and sample splitting, unknown neural network convergence behavior, and massive unlabeled samples. 

Centrally, MAR-S also addresses aggregated and transformed AI predictions, where validation occurs at a more granular level than the variable(s) entering the estimating equation. This setting is extremely common in applied economics and not accommodated by existing frameworks.

Beyond its relation to the black-box AI literature, MAR-S also connects to the literatures on surrogacy and double/debiased machine learning (DML). Its imputations can be viewed as ``imperfect surrogates'' \citep{kallus_role_2024,ji_predictions_2025}: proxies that, when combined with random validation labels, can improve efficiency even when standard surrogacy conditions fail \citep{athey_surrogate_2019}. This contrasts with settings in which validation data are unavailable in the target sample, where the relevance of surrogate values to the target estimand requires stronger theoretical justification \citep{rambachan2024program,nakamura_surrogate_2025}. MAR-S likewise shares DML's semiparametric foundations and robustness to flexible estimation of high-dimensional nuisance functions \citep{chernozhukov_doubledebiased_2018,chernozhukov_riesznet_2022,chernozhukov_automatic_2022,ichimura_influence_2022}. This shared foundation yields formal connections between the approaches: for example, under certain local product structure, MAR-S estimators can be represented through Neyman-orthogonal estimating equations \citep{chen_equivalence_2026}.

Several econometric frameworks for inference with AI predictions do not require a random validation sample. The tradeoff is that identification must instead rely on assumptions about the AI model or how its errors behave. \cite{rambachan2024program} study remotely sensed outcomes when validation data exist but for a different population, requiring assumptions that permit information about the AI imputed variable to transfer across samples. \cite{battaglia_inference_2024} develop corrections for imperfect AI and machine learning (ML) imputations without a validation sample, but focus on linear regression and rely on a variety of assumptions linking measurement error, sampling uncertainty, and the underlying data generating process.\footnote{Their baseline asymptotic framework assumes that measurement error and sampling uncertainty vanish at comparable rates; outside this case, implementation requires additional structure, such as a joint model of the measurement and outcome processes, with assumptions that depend on the application.} Finally, \cite{christensen_unstructured_2026} develop bias corrections for AI/ML predictions in BLP-type structural models, obtaining identification from restrictions on the structural model and the relationship between the proxies and latent characteristics.\footnote{Among other conditions, the approach requires sufficiently accurate estimation of a composite structural parameter relative to sampling uncertainty and assumes that the AI/ML proxies have the same dimensionality as the latent differentiation characteristics they proxy.} These approaches are valuable when random validation is infeasible. When random validation is possible, it shifts the identifying burden away from potentially difficult-to-assess assumptions about the accuracy, stability, or structural correctness of highly complex AI models, and onto a transparent sampling design.

\section{Missing At Random Structured Data} \label{sec:MAR-SFramework}

This section introduces the MAR-S framework. Section \ref{mmd} formalizes missing structured data, and Section \ref{assumptions} states the assumptions underlying MAR-S. The notation and assumptions closely parallel those used in semiparametric causal inference and missing data analysis more broadly \citep{tsiatis_semiparametric_2006}. Section \ref{dnn_bigdata} then discusses the novel challenges posed by deep neural networks and large datasets, and why the missing at random framework is well suited to address them.

\subsection{Modeling Missing Structured Data}\label{mmd}
To enable robust, efficient inference with \textit{unstructured data}, we recast the problem as inference on \textit{missing structured data}. Structured data \(M \in \mathcal{M}\) are low-dimensional variables that can enter estimating equations directly, whereas unstructured data \(U \in \mathcal{U}\)---such as raw images or text---are high-dimensional and generally unsuitable for direct use in econometric specifications.

Under MAR-S, structured data are observed through \textit{validation}. A non-scalable measurement technology---such as an expert annotator, survey team, or scientific instrument---records \(M\) for a given \(U\) according to a well-specified measurement procedure. Let \(A \in \{0,1\}\) indicate whether an observation is validated, with \textit{validation score}
\[
\pi(x):=P(A=1\mid X=x),
\]
where \(X \in \mathcal{X}\) denotes observed low-dimensional covariates that determine validation.

Since validation is costly, researchers leverage the full sample by learning an \textit{imputation function} \(\hat{\mu}\) (\textit{e.g.,} a neural network) to predict missing structured data from unstructured inputs. 

We use potential outcomes notation, which provides unified notation for missing structured data and other forms of missingness, including those arising in causal inference.
Let \(M^{a=1}\) denote the target structured variable that would be observed if a data instance were validated, and define \(M^*:=M^{a=1}\). When \(A=1\), \(M^*\) is observed; when \(A=0\), it is missing. For notational convenience, we follow Rubin in encoding the missing state as zero and write the observed structured data as
\begin{equation*}
    M = A M^*.
\end{equation*}
Because \(A\) is itself observed, this convention entails no loss of information: the pair \((A,AM^*)\) distinguishes an unobserved value from an observed value of zero. The observed data therefore take the canonical missing data form
\begin{equation*}
    \{(A_i,A_iM_i^*,\ldots)\}_{i=1}^n,
\end{equation*}
while the researcher is interested in functionals of the distribution of \(M^*\). Just as causal inference requires observing some units under treatment, valid inference on \(M^*\) requires some validated observations: \(P(A=1)>0\). 

We impose no general assumption on the relationship between \(M^*\) and \(U\). In some settings, \(M^*\) may be a deterministic function of \(U\), for example, when the content of the text fully reveals the target measure. This is referred to as $U$-measurability.\footnote{For example, \citet{nakamura_surrogate_2025} explicitly makes this assumption.} In other settings, \(M^*\) need not be determined by \(U\). For example, a satellite image does not fully reveal household consumption, although it may be highly predictive. 

\subsection{Assumptions}\label{assumptions}

We now outline the assumptions underlying the MAR-S framework. MAR-S is a superpopulation framework that assumes data are independently and identically distributed (iid), although we discuss extensions that allow for clustering and other error structures. 

\begin{assumption}[Consistency of validation]\label{cpo}
For target potential outcome $M^{\ast} \in \mathcal{M}$, structured data $M \in \mathcal{M} \times \{0\}$, and validation indicator $A \in \{0,1\}$, we have 
\[ M = A\, M^{\ast}.\]
\end{assumption}
This requires that validation reveal the researcher's target, $M^*$. 
The motivation for deploying AI is often to cheaply extend an existing target, whether it is text classifications from a well-defined rubric, ground pollution readings using remotely sensed images, or human survey responses using synthetic respondents. In such settings, the target measure should be accessible by definition, albeit in a sample that may be small.

More generally, MAR-S requires an assumption on either the absence or the form of validation error. Intuitively, arbitrary errors in imputing $M^*$---which MAR-S allows---cannot be corrected using validation data that contain unspecified error.  If researchers expect validation to measure $M^*$ with well-understood error, this could be accommodated by specifying the structure of that error.\footnote{See \cite{nakamura_surrogate_2025} for an example of a framework that permits labeling noise under additional assumptions.} 
We additionally maintain the usual non-interference condition: the validation for one observation does not depend on which other observations are validated. In practice, a stable definition and validation procedure should be applied consistently across observations.

The second assumption is the cornerstone of MAR-S, adapting Rubin's missing at random (MAR) condition to unstructured data.
Define $X$ as the set of observed covariates that condition validation probabilities and $V$ as auxiliary observed variables that do not condition validation but instead enter the target estimating equation (e.g., controls in a linear regression).

\begin{assumption}[Missing-at-random structured data]\label{aoo}
For target potential outcome $M^* \in \mathcal{M}$, validation indicator $A \in \{0,1\}$, observed covariates $X \in \mathcal{X}$, unstructured data $U \in \mathcal{U}$, and auxiliary observed variables $V \in \mathcal{V}$:
\begin{align*}
    [(U, M^*, V) \ind A] \mid X.
\end{align*}
\end{assumption}
Assumption \ref{aoo} is analogous to \textit{selection on observables} in causal inference \citep{hirano_efficient_2003} and could equivalently be described as \textit{validation on observables.} Conditional on the observed covariates $X$ that were used to determine which observations to validate, the observations selected for validation are representative of those not selected.\footnote{When there are no auxiliary variables related to the estimation equation, the assumption simplifies to just $[(U, M^*) \ind A] \mid X$.} %Define $\tilde X:=(X,U,V)$. This assumption implies that $[M^* \ind A]\mid \tilde X$ - 

When researchers construct their own validation samples, this assumption can be imposed by design---through random validation, or through stratified random validation with probabilities determined by observed $X$. This is particularly useful in AI applications: rather than making assumptions about a complex model's errors, researchers can secure identification through a validation design they directly control.

The third assumption requires that validation probabilities be known and bounded away from zero. 

\begin{assumption}[Known and bounded validation score function]\label{kas}
We define the ``validation score function'' to be $\pi(x):=P(A = 1 \mid X = x)$. We assume that $\pi(x)$ is fixed, known, and bounded away from zero, i.e., $\pi(x) \in [\delta,1]$ for $0<\delta<1$ for all $x \in \mathcal{X}$.
\end{assumption}

The requirement that the validation function is known holds by definition when researchers control the validation design. When a validation sample lacking documentation is inherited, this can be relaxed and $\pi(X)$ estimated, at the cost of additional rate conditions on estimation of the validation and imputation functions, discussed below.

The latter part of Assumption \ref{kas} is an overlap condition: every observation in the target population must have positive probability of being validated.
The naming convention \textit{validation score function} mimics the terminology of the \textit{propensity score function}, which serves an analogous role in causal inference. As in causal inference, overlap ensures that the data contain information about the missing structured variable throughout the target population. 

Common validation practices in social science violate overlap. For example, researchers sometimes validate only documents containing selected keywords but use all documents in estimation. This is particularly common with class-imbalanced data, when the class of interest is relatively rare (\textit{e.g.,} texts discussing a given topic). Documents outside the filter have zero probability of validation, so prediction errors for those documents cannot be learned from the validation data. A related problem arises when validation data come from one period, but imputed structured data are used over a broader or different period. 
A basic limitation of machine learning is that performance tends to deteriorate with shifts away from the training data distribution \citep{ben2010theory}, exacerbating this problem when the model is trained on a split of the validation data, or otherwise on data that are more similar to it. 
Section \ref{sec:dropext} discusses the design of validation schemes that satisfy overlap even in class-imbalanced settings. 
%For example, remote sensing models for crop classification can deteriorate substantially outside their training period because farming practices and weather change over time \citep{wang2020mapping}. %A basic limitation of machine learning is that performance tends to deteriorate with shifts away from the training data distribution \citep{ben2010theory}. 
%One solution is to redefine the target population to match the support of the validation design, though doing so changes the interpretation of the estimand. Alternatively, researchers can design validation schemes that oversample informative observations while retaining positive probability of validation throughout the target population (Section \ref{sec:dropext}). When some types of observations cannot be validated, additional non-innocuous assumptions about how imputation errors transfer across samples are required \citep{rambachan2024program}.

The high-dimensional nature of unstructured data distinguishes inference with AI predictions from many other missing data contexts. Importantly, validation in MAR-S is conditioned on low-dimensional $X$, not on high-dimensional $U$, which by Assumption \ref{aoo} is conditionally independent of validation status.
This restriction matters for the validation design. For example, if a researcher is less likely to validate a text after reading it and finding it difficult to assess, validation depends on \(U\). One might then try to justify this by conditioning on \(U\), but doing so could create serious overlap problems because $U$ is very high-dimensional \citep{damour_overlap_2021}.\footnote{If corrupted data prevent validation---for instance, because a text has been mangled by optical character recognition (OCR) or an image's quality makes it illegible---it is preferable to design a procedure for removing those observations from the corpus, rather than simply ignoring them for validation purposes.}
Importantly, however, validation probabilities can be conditioned on low-dimensional measures $X$ derived from $U$. For instance, distance of a text to a query in embedding space, uncertainty measures derived from neural classifiers, or verbalized confidence scores from an LLM could be derived for a corpus and used to stratify validation sampling.

As shown in Section \ref{sec:gmm}, when $\pi(X)$ is known, poor imputation of the missing data does not invalidate large sample inference; MAR-S accommodates any arbitrary fixed predictors, including poor ones. Assumptions about the imputation function are not needed for identification, which is the key advantage of validation. 

In contrast, we will see that efficiency does make requirements of the imputation function. Let the imputation function take the form
\[
\mu(\tilde x) := E[M\mid A=1,\tilde X=\tilde x],
\]
where $\tilde X:=(X,U,V)$. In practice, $\mu$ is unknown and must be estimated; we denote its estimator by $\hat\mu$. The following assumption is necessary for efficient inference.

\begin{assumption}[Universal consistency of the imputation function]\label{mse}
For conditional expectation function $\mu(\tilde x)$, estimated imputation function $\hat\mu(\tilde x)$, and features $\tilde X \in \tilde{\mathcal{X}}$,
\[
E\left[\int\left(\hat\mu(\tilde x)-\mu(\tilde x)\right)^2
dP_{\tilde X}(\tilde x)\right]=o(1).
\]
\end{assumption}

Assumption \ref{mse}, sometimes referred to as ``universal consistency'' \citep{wager_causal_2024}, requires the expected $L^2(P)$ error of the imputation function to vanish asymptotically. 
%Among regular, asymptotically linear estimators, the semiparametric efficiency bound gives the lowest attainable asymptotic variance \citep{van_der_vaart_asymptotic_1998}. 
% This lower bound corresponds to the variance of the efficient influence function (EIF) associated with the functional. 
% We will see that better approximating $\mu$, \textit{e.g.,} through fine-tuning on a split of the validation data, can move estimates toward this bound. %Poor predictions reduce precision but do not invalidate large sample inference.
  Recent theoretical work establishes universal consistency for certain classes of deep neural networks trained by gradient descent \citep{drews_universal_2024}. 
  
In contrast, if the validation score function is estimated (relaxing Assumption \ref{kas}), efficient, $\sqrt{n}$-consistent, and asymptotically normal inference instead requires sufficiently accurate estimation of both the imputation and validation score functions, summarized by a product-rate condition on their $L^2(P)$ errors \citep{chernozhukov_doubledebiased_2018,kennedy_semiparametric_2023,wager_causal_2024}, as further elaborated in Section \ref{sec:mars_implications}. %In observational causal inference, the (analogous) propensity score function usually has to be estimated, whereas in validation, it will often be under the direct control of the researcher, or at least known. 

\subsection{Deep Neural Networks, Big Data, and MAR-S} \label{dnn_bigdata}
The intuition behind MAR-S is the same as in the broader semiparametric missing data literature, but unstructured data introduce several features that are especially relevant in practice. First, \textit{the imputation function is typically a deep neural network}, often a transformer \citep{vaswani_attention_2017}, and its inputs include high-dimensional unstructured data. Classical nonparametric results imply that learning conditional expectations from high-dimensional covariates can be difficult without additional structure \citep{stone_optimal_1982}, motivating methods such as higher-order influence function (HOIF) estimators when nuisance functions must be estimated sufficiently well \citep{robins_higher_2008,robins_minimax_2017}. Under baseline MAR-S, Assumption \ref{kas} makes these convergence rate concerns irrelevant for validity: because the validation score function is known, valid large-sample inference does not require the imputation function to consistently estimate any particular conditional expectation. Slow or unknown convergence of a neural network may therefore reduce efficiency, but does not invalidate inference. This also eliminates the need for HOIF corrections aimed at removing bias from imperfect estimation of nuisance functions based on high-dimensional features. For efficiency, prediction quality matters. Recent theory establishes universal consistency or favorable convergence properties for some classes of neural networks \citep{drews_universal_2024,chen_causal_2024}, while empirical evidence documents strong predictive performance in high-dimensional settings \citep{klaassen_doublemldeep_2024}. 

Second, large-scale pretraining raises concerns about \textit{data contamination}. A neural network may perform unusually well on a benchmark because relevant examples or outcomes appeared in its pretraining data \citep{sainz2024data}; in economic applications, this can generate look-ahead bias when historical performance is used to infer how the model would perform using only information available at the time \citep{sarkar_lookahead_2024}. In MAR-S, however, the pretrained model is treated as a fixed prediction rule, while identification comes from the validation design.\footnote{See Section \ref{sec:mars_implications} for more on the connection to regression adjustment.} Valid inference therefore does not require that the pretrained model be uncontaminated, provided that the validation sample is representative of the population over which the target parameter is estimated. Contamination becomes problematic when validation performance from one population---for example, historical observations the model may have encountered during pretraining---is extrapolated to another, such as genuinely future observations. The validation sample is then no longer drawn from the target population, and MAR-S would need to be augmented with stability or transportability assumptions.

Third, \textit{the structured variable of interest may be rare relative to the size of the unstructured dataset}. For example, only a small fraction of social media posts or historical newspaper articles may discuss a particular topic. In such settings, simple random validation would devote most labels to uninformative negatives, yielding poor precision. Researchers can instead design the validation score function $\pi$ to oversample informative or rare observations while maintaining positive validation probability throughout the target population, drawing on ideas from importance sampling and active statistical inference \citep{mcbook,zrnic_active_2024}. We discuss such validation designs in Section \ref{sec:dropext}.

Finally, \textit{the fraction of observations that are validated may itself be very small}. 
This could motivate ``decaying overlap'' asymptotics in which the number of labeled observations diverges while their share of the full dataset converges to zero \citep{zhang_double_2023}.
MAR-S adopts a conventional asymptotic framework in which conditional validation probabilities are bounded away from zero, but the underlying semiparametric missing data structure makes clear that the MAR-S estimator need not change under decaying overlap. As shown in related settings, the resulting efficiency bound is increasingly determined by the size of the validation sample rather than the total sample \citep{kallus_role_2024,testa_semiparametric_2025}.\footnote{See, for example, Theorem 4.2 of \citet{kallus_role_2024}.} A central difficulty in the decaying overlap literature is estimating propensity scores \citep{zhang_double_2023,kallus_role_2024}. This is absent when the validation score function - analogous to a propensity score function - is known (Assumption \ref{kas}).\footnote{When the validation mechanism is unknown, researchers could instead use propensity score methods developed for the decaying overlap setting, such as those in \citet{zhang_double_2023} or \citet{kallus_role_2024}.} More generally, distinguishing the low-dimensional variables $X$ that govern validation from the potentially high-dimensional features $\tilde X:=(X,U,V)$ used for imputation makes conventional overlap substantially more plausible \citep{damour_overlap_2021}.

\section{Estimation with MAR-S}\label{sec:gmm}

We now show how to implement parameter estimation with MAR-S. Section \ref{ssec:recipe} describes a general recipe for constructing robust and efficient estimators for inference with missing structured data, adapting the missing data results of \citet{chen_semiparametric_2008} to unstructured data. The recipe has three steps: 1) define the target parameter $\theta$ and the imputation target; 2) construct an imputation function from observed information; and 3) derive the estimator for $\theta$. Section \ref{sec:common_mars} applies the recipe to common econometric estimands. The estimators take a familiar form, yet also make clear what information neural networks should use, what they should learn, and how they should behave asymptotically if efficiency is important. These implications are summarized in Section \ref{sec:mars_implications}.

\subsection{The General Recipe} \label{ssec:recipe}

\textbf{Step 1: Define the target parameter $\theta \in \Theta \subseteq \mathbb{R}^{d_\theta}$ and the imputation target.} We begin by defining our parameter of interest, $\theta$, through a ``full data'' population moment condition
\[
E_P[m(W;\vartheta)] = 0
\quad \text{iff} \quad
\vartheta=\theta,
\]
where \(W=(M^*,V)\) denotes the moment-relevant variables: fully observed random variables $V$ and missing data $M^*$. 
$P$ denotes the population distribution, and \(\vartheta\) is a candidate value. For each candidate \(\vartheta\), the moment function \(m(W;\vartheta)\) gives a \(d_m\)-dimensional vector of moment contributions, where \(d_m \geq d_\theta\).
We cannot directly compute the sample analogue of this moment condition since $M^*$ is not observed for unvalidated observations.

To make progress, we define an imputation target, $\mathcal{E}$, that depends only on observed information. First, we condition the expected value of the full data moment on the observed information $\tilde X:=(X,U,V)$, where $X$ contains covariates that determine the validation probabilities and $U$ is unstructured data: 
\[
\mathcal{E}(\tilde X;\vartheta)
:=
E_P[m(W;\vartheta)\mid\tilde X].
\]
This \textit{conditional moment} still contains $M^*$, through $W=(M^*, V)$, but under the missing at random assumption we can express it entirely in terms of observed data. Missing at random requires validation status, $A$, to be conditionally independent of the missing data given the observed information, $A\ind M^*\mid\tilde X$. The distribution of $M^*$ given $\tilde X$ is therefore the same for validated ($A=1$) and unvalidated ($A=0$) units, and we can use this to express the conditional moment $\mathcal{E}(\tilde X;\vartheta)$ only in terms of the observed data:
$$\mathcal{E}(\tilde X;\vartheta)=E_P[m(W;\vartheta)\mid\tilde X] = E_P[m((M,V);\vartheta)\mid\tilde X,A=1],$$
where $M=AM^*$. Importantly, $\mathcal{E}$, while a function of only observed information, identifies the same target parameter as the full data moment. By iterated expectations
$E[\mathcal{E}(\tilde X;\vartheta)]=E[m(W;\vartheta)],$
so $E[\mathcal{E}(\tilde X;\vartheta)]=0$ if and only if $\vartheta=\theta$. 

The $\mathcal{E}$ is identified, but it is not known. It is a population conditional expectation, whose value at each $\tilde x$ depends on the true conditional distribution of $M^*$ given $\tilde X=\tilde x$. Identification says this distribution is pinned down by the distribution of the observed data, but we must estimate it from a finite sample, which is the focus of Step 2.

\textbf{Step 2: Approximate the imputation function.}
We aim to approximate $\mathcal{E}$ with a function $\tilde x\mapsto\eta(\tilde x;\vartheta)$. 
Estimating a conditional moment function is an old problem. With discrete, low-dimensional covariates, estimating $\mathcal E$ often amounts to computing means within cells; with continuous low-dimensional covariates, kernel or series regression is common. These methods break down with high-dimensional $U$, given the curse of dimensionality. However, pretrained deep neural networks have proven effective in practice for this difficult task. 

While deep neural networks are powerful function approximators \citep{goodfellow_deep_2016}, their biases are complex and hard to characterize. 
This motivates a validation adjustment (Step 3), which preserves identification of $\theta$ for \textit{any fixed $\eta$}, including inaccurate ones. This provides the researcher with wide latitude in choosing $\eta$. The researcher may vary what observed information $\eta$ uses (\textit{e.g.,} unstructured data $U$ alone, or $U$ as well as observed structured variables in $\tilde X$) and what $\eta$ predicts (\textit{e.g.,} $M^*$, a function of $M^*$ that enters the estimating equation, or the full conditional moment $\mathcal{E}$). The $\eta$ may be an off-the-shelf model (\textit{e.g.,} a commercial LLM), a model trained on external data, a model fit on held-out validated observations, a combination of these, or some other black box, including a non-neural model.%If $\eta$ is tuned on the study's own data, sample splitting or cross-fitting allows it to be treated as fixed in estimation. 

These choices can yield much better or worse approximations of $\mathcal{E}$. This does not matter for identification, but it can matter enormously for precision. If the researcher aims for efficient inference, MAR-S gives specific prescriptions about what \textit{information} $\eta$ should use, what it should \textit{target}, and its \textit{asymptotic convergence}, as elaborated in Section \ref{sec:mars_implications}.

\textbf{Step 3: Derive the estimator for $\theta$.} Taking $\eta$ as given, we next construct an estimator of the target parameter $\theta$ that uses $\eta$ together with validation data. In the language of semiparametric inference, this is the second-step estimator. We consider identification, estimation and inference, and efficiency in turn. 

\textbf{Identification.} A model can predict well but still have non-zero imputation bias, $E[\eta(\tilde X;\vartheta)-m(W;\vartheta)]$. Hence, solving the moment condition with $\eta$ alone would give a biased estimate of $\theta$. 

Validation data let us estimate how the imputations differ from the researcher's target, and under the MAR-S assumptions this estimate is unbiased for the population. Let $\pi(X)$ be the conditional validation probability. We weight each validated unit's observed imputation error by $1/\pi(X)$ and subtract the weighted error from the imputed moment: 

\[
m_\eta(M,A,\tilde X;\vartheta)
:=
\underbrace{\eta(\tilde X;\vartheta)}_{\text{Imputed moment}}-
\underbrace{\frac{A}{\pi(X)}
[\eta(\tilde X;\vartheta)-m(W;\vartheta)]}_{\text{Re-weighted imputation error}},
\]
where $\pi(X)>0$ by Assumption \ref{kas}, and so the weights are finite. 
For unvalidated units ($A=0$), the correction vanishes and $m_\eta=\eta$.

If the full-data moment condition identifies $\theta$, so does the adjusted one, for any fixed $\eta$ for which the relevant expectations exist---even if $\eta$ is a poor approximation to $\mathcal{E}$.\footnote{One could instead use only the validated observations to estimate $\theta$, which is also valid but discards information in unvalidated units that can be used to improve precision.}
To see this, note that the missing at random assumption, together with the fact that validation depends only on $X$, implies $E[A\mid M^*,\tilde X]=\pi(X)$. Therefore by the law of iterated expectations:
\[
E\!\left[\frac{A}{\pi(X)}\big[\eta(\tilde X;\vartheta)-m(W;\vartheta)\big]\right]
=E\!\left[\frac{E[A\mid M^*,\tilde X]}{\pi(X)}\big[\eta(\tilde X;\vartheta)-m(W;\vartheta)\big]\right]
=E\big[\eta(\tilde X;\vartheta)-m(W;\vartheta)\big].
\]
It follows that
\[
E[m_\eta(M,A,\tilde X;\vartheta)]=E[\eta(\tilde X;\vartheta)]-E\big[\eta(\tilde X;\vartheta)-m(W;\vartheta)\big]=E[m(W;\vartheta)].
\]
This property is particularly useful with AI-based imputations, where the complexity and black-box nature of the imputation function complicates specifying how the model errs.

The next theorem formalizes this identification result; we focus on just-identified moment conditions for expositional simplicity. Efficient estimation in the over-identified setting is discussed in Appendix Section \ref{sec:overid}. As elaborated in Section \ref{sec:mars_implications}, this result requires a known validation score function, $\pi(X)$ (Assumption \ref{kas}).

\begin{restatable}{theorem}{Wdrgmm}\label{prop:wdrgmm}
Under Assumptions \ref{cpo}, \ref{aoo}, and \ref{kas}, with $d_m=d_\theta$, for any fixed function $\eta(\cdot;\vartheta)$,
\[
E_P[m_\eta(M,A,\tilde X;\vartheta)] = 0
\quad \text{iff} \quad
\vartheta=\theta.
\]
\end{restatable}

\textbf{Estimation and Inference.}
The adjusted moment $m_\eta$ is an ordinary function of observed data, as $\eta$ is fixed---potentially conditional on a tuning sample $\mathcal I'$ used to learn it---and hence standard GMM theory applies. For i.i.d. estimation sample $\{(M_i,A_i,\tilde X_i)\}_{i\in\mathcal I}$, the empirical moment is
\[
\bar m_\eta(\vartheta)
:=
|\mathcal{I}|^{-1}
\sum_{i\in\mathcal{I}}
m_\eta(M_i,A_i,\tilde X_i;\vartheta),
\]
In the just-identified case, the MAR-S GMM estimator solves
$\bar m_\eta(\hat\theta)=0$, equivalently
$\hat\theta=\arg\min_\vartheta \bar m_\eta(\vartheta)'\bar m_\eta(\vartheta)$.
Because $E[m_\eta(M,A,\tilde X;\vartheta)]=E[m(W;\vartheta)]$ for any fixed $\eta$, the
Jacobian $\mathcal J_\theta:=\partial E_P[m_\eta(M,A,\tilde X;\theta)]/\partial\theta$ does
not depend on $\eta$.

We now formally state the large sample properties of the MAR-S GMM estimator, under mild regularity conditions.

\begin{restatable}{theorem}{Thmeff}\label{thm:eff}
Consider the just-identified setting where $d_\theta=d_m$ and fix $\eta$. 
Assume the regularity conditions: (i) $\theta \in \text{Interior}(\Theta)$ and $\Theta$ is compact; (ii) $\vartheta \mapsto m_\eta(M,A,\tilde X;\vartheta)$ is continuous almost surely; (iii) $E[\sup_{\vartheta\in\Theta}\|m_\eta(M,A,\tilde X;\vartheta)\|]<\infty$; (iv) $\vartheta \mapsto m_\eta(M,A,\tilde X;\vartheta)$ is continuously differentiable in a neighborhood $\mathcal{N}$ of $\theta$ almost surely;  (v) $E_P[\|m_\eta(M,A,\tilde X;\theta)\|^2]<\infty$; (vi) $E_P[\sup_{\vartheta\in\mathcal N}\|\frac{\partial}{\partial\vartheta^\prime}m_\eta(M,A,\tilde X;\vartheta)\|]<\infty$; and (vii) $\mathcal J_{\theta}$ is nonsingular. Then, under the tuning-sample conditional law (holding fixed the data in the tuning sample \(\mathcal I'\)), as $|\mathcal{I}|\to\infty$
\[
\sqrt{|\mathcal{I}|}(\hat\theta-\theta)
\rightsquigarrow
N\left(
0,
\mathcal{J}_{\theta}^{-1}
E_P[
m_\eta(M,A,\tilde X;\theta)
m_\eta(M,A,\tilde X;\theta)^\prime
]
(\mathcal{J}_{\theta}^\prime)^{-1}
\right).
\]
\end{restatable}

This theorem states that as the estimation sample grows, $\hat\theta$ is consistent and
asymptotically normal at the usual rate, whatever fixed $\eta$ is used. The imputation
function affects only the middle term of the variance, $E[m_\eta m_\eta']$, so a better
$\eta$ improves precision. Inference proceeds with the usual
sandwich variance estimator, provided sample splitting is respected.\footnote{If the tuning sample also grows, the result can be extended to when the estimated imputation function converges to any well-defined limit and suitable regularity conditions hold uniformly.} 

\textbf{Efficiency.} Which $\eta$ gives the most precise estimator? The efficient
influence function answers this: its variance is the smallest asymptotic variance
attainable by regular estimators under MAR-S, so it is a natural benchmark against which any
choice of $\eta$ can be judged.

The following corollary of Theorem 2 in \citet{chen_semiparametric_2008} gives the efficient influence function for $\theta$ in the MAR-S setting under just-identified moment conditions; the over-identified case is in Appendix Section \ref{sec:overid}.

\begin{restatable}{corollary}{Corverifyjid}\label{cor:verify_jid}
Under Assumptions \ref{cpo}, \ref{aoo}, and \ref{kas}, if $d_m=d_\theta$ and $\mathcal{J}_\theta$ has full column rank equal to $d_\theta$, then the efficient influence function for $\theta$ is
\[
\varphi(M,A,\tilde X;\theta)
=
-\mathcal{J}_\theta^{-1}
\left(
\frac{A}{\pi(X)}
[m(W;\theta)-\mathcal{E}(\tilde X;\theta)]
+
\mathcal{E}(\tilde X;\theta)
\right).
\]
\end{restatable}

Recall that $\mathcal{E}$ is the conditional moment target that identifies $\theta$ using observable information.
Setting $\eta=\mathcal{E}$ in the adjusted moment gives $m_\eta(M,A,\tilde X;\theta)=\mathcal{E}(\tilde X;\theta)+\frac{A}{\pi(X)}[m(M,A,\tilde X;\theta)-\mathcal{E}(\tilde X;\theta)]$, which is exactly the bracketed term in $\varphi$. 
The estimator therefore attains the semiparametric lower bound asymptotically if the estimated $\eta$ converges to $\mathcal{E}$ in the sense of Assumption \ref{mse}. 
In other words, imputation should approximate the conditional moment from Step 1 as closely as possible when efficiency is desired. We now explore what this means in practice using a series of concrete examples. 

\subsection{MAR-S for Common Estimands} \label{sec:common_mars}
We apply the recipe from the previous section to common estimands, walking through several examples and then summarizing how MAR-S applies in a broader array of settings in Table \ref{tab:mars_estimators}. 

\subsubsection{Mean of missing structured data}
\textbf{Step 1: Define.} The $\theta$ is defined by the full-data moment condition
\[
E[m(W;\vartheta)]=0, \qquad m(W;\vartheta)=M^*-\vartheta .
\]
There are no other moment-relevant variables $V$, so the observed information is
$\tilde X=(X,U)$. The imputation target is the conditional moment:
\[\mathcal{E}(\tilde X;\vartheta)=\mu(\tilde X)-\vartheta \qquad \mu(\tilde X):=E_P[M^*\mid\tilde X].\]

\textbf{Step 2: Approximate.} We approximate $\mu$ with a function $\hat\mu$ (learned in the tuning sample or otherwise fixed), such that $\eta(\tilde X;\vartheta)=\hat\mu(\tilde X)-\vartheta$. Because the moment is linear in $M^*$, \textit{approximating $\mathcal{E}$ amounts to predicting $M^*$}. In this simple case, the common practice of imputing $M^*$ using $U$ targets the right object for efficiency.

\textbf{Step 3: Derive.} The adjusted moment is
\[
m_\eta(M,A,\tilde X;\vartheta)
=\widetilde M-\vartheta,
\qquad
\widetilde M_i
:=
\frac{A_i}{\pi(X_i)}\left[M_i-\hat\mu(\tilde X_i)\right]+\hat\mu(\tilde X_i).
\]
The $\widetilde M_i$ combines the imputation $\hat\mu(\tilde X_i)$
with the inverse probability-weighted difference between
the validation data and the imputation. It is not a prediction of the individual
value of $M^*$, but rather is constructed so that its average estimates the mean. Solving
$\bar m_\eta(\hat\theta)=0$ yields
\[
\hat\theta=\frac{1}{|\mathcal I|}\sum_{i\in\mathcal I}\widetilde M_i .
\]
By Theorems \ref{prop:wdrgmm} and \ref{thm:eff}, $\hat\theta$ is consistent and asymptotically normal based on any $\hat\mu$ that we can treat as fixed. 
The estimator is efficient to the extent that $\hat\mu$ approximates $\mu$, per Assumption \ref{mse}; for a formal statement, see Appendix Section \ref{sec:effstate}.

This estimator takes the familiar augmented inverse probability weighted (AIPW) form \citep{robins_estimation_1994} and can be further connected to two other estimators in the recent literature. To make the connection exact, consider the special case of MAR-S in which validation is unconditionally random, with validation probability $\pi(X_i)=\pi$ equal to the realized validation share,
$
\pi
=
\frac{|\mathcal{I}\cap\mathcal{J}|}{|\mathcal{I}|},
$
where $\mathcal{J}$ indexes validated observations and $\mathcal{I}$ indexes all observations used for estimation. We can alternatively write the MAR-S estimator as:
\begin{align*}
\hat\theta &=
\underbrace{
\frac{1}{|\mathcal{I}|}
\sum_{i\in\mathcal{I}}
\hat\mu(\tilde X_i)
}_{\textrm{A}}
+
\underbrace{
\frac{1}{|\mathcal{I}\cap\mathcal{J}|}
\sum_{i\in\mathcal{I}\cap\mathcal{J}}
(M_i-\hat\mu(\tilde X_i))
}_{\textrm{B}}
&& \text{(PPI)}
\\
&=
\underbrace{
\frac{1}{|\mathcal{I}\cap\mathcal{J}|}
\sum_{i\in\mathcal{I}\cap\mathcal{J}}
M_i
}_{\textrm{C}}
+
\underbrace{
\left(
\frac{1}{|\mathcal{I}|}
\sum_{i\in\mathcal{I}}
\hat\mu(\tilde X_i)
-
\frac{1}{|\mathcal{I}\cap\mathcal{J}|}
\sum_{i\in\mathcal{I}\cap\mathcal{J}}
\hat\mu(\tilde X_i)
\right)
}_{\textrm{D}}
&& \text{(FRA)}.
\end{align*}
PPI is the mean estimator in prediction powered inference \citep{angelopoulos_prediction-powered_2023}, and FRA corresponds to flexible regression adjustment \citep{list_using_2024}. 
Each representation emphasizes a different intuition. The AIPW form underscores that each observation contributes an imputed value, and validated observations additionally contribute an inverse probability weighted adjustment of the imputation error that accounts for the validation design. The PPI form expresses the same estimator directly as unweighted averages differenced across two distinct samples: term \textit{A} is the plug-in estimate obtained by imputing $M$ in the full estimation sample, while term \textit{B} uses the validation sample to adjust for imputation error. A large dataset can make \textit{A} precise but worse imputation will increase the variance of term \textit{B}, which may dominate. In the FRA form, term \textit{C} uses validation data alone, while term \textit{D} uses the imputation function as a nonparametric regression adjustment to improve precision.

\subsubsection{Linear regression with a missing outcome} 
\textbf{Step 1: Define.} Let $R$ denote a vector of regressors, observed for every observation, and let $\theta$ denote the population OLS coefficient vector. The $\theta$ is defined by the full data moment condition
\[
E[m(W;\vartheta)]=0, \qquad m(W;\vartheta)=R(M^*-R'\vartheta).
\]
Because $R$ enters the moment, we include it in the observed information, $\tilde X=(X,U,R)$. The imputation target is the conditional moment:
\[
\mathcal{E}(\tilde X;\vartheta)
=R\big(\mu(\tilde X)-R'\vartheta\big),
\qquad
\mu(\tilde X):=E[M^*\mid\tilde X].
\]

\textbf{Step 2: Approximate.} We approximate $\mu$ with a function $\hat\mu$ such that $\eta(\tilde X;\vartheta)=R(\hat\mu(\tilde X)-R'\vartheta)$. The moment is again linear in $M^*$, so approximating $\mathcal{E}$ again amounts to predicting $M^*$. The $\tilde X$ now includes the regressors $R$. Whether $R$ is used to predict $\hat \mu$ does not affect validity but matters for efficiency, an implication developed in Section \ref{sec:mars_implications}. 

\textbf{Step 3: Derive.} Because $R$ is always observed and is contained in $\tilde X$, it multiplies the adjustment but is not itself imputed or corrected. Therefore, the adjusted moment is the original moment with $M$ replaced by $\widetilde M$. 
\[
m_\eta(M,A,\tilde X;\vartheta)=R\big(\widetilde M-R'\vartheta\big),
\qquad
\widetilde M_i
:=
\frac{A_i}{\pi(X_i)}\left[M_i-\hat\mu(\tilde X_i)\right]+\hat\mu(\tilde X_i),
\]
Solving the sample moment condition
% \[
% \frac{1}{|\mathcal I|}\sum_{i\in\mathcal I}R_i\big(\widetilde M_i-R_i'\hat\theta\big)=0
% \]
yields
\[
\hat\theta
=
\Big(\sum_{i\in\mathcal I}R_iR_i'\Big)^{-1}\sum_{i\in\mathcal I}R_i\widetilde M_i .
\]
This is the familiar OLS form, with the missing outcome replaced by its validation-adjusted analogue $\widetilde M_i$. Again, by Theorems \ref{prop:wdrgmm} and \ref{thm:eff}, $\hat\theta$ is consistent and asymptotically normal for any fixed $\hat\mu$. It attains the efficiency bound of Corollary \ref{cor:verify_jid} to the extent that $\hat\mu$ approximates $\mu$ under Assumption \ref{mse}; again see Appendix Section \ref{sec:effstate} for a formal statement. 

\subsubsection{Linear regression with a missing regressor} 
\textbf{Step 1: Define.} Let $R$ denote the observed regressors and $Y$ the outcome, both observed for every observation, and let $R^*=\binom{M^*}{R}$ be the full regressor vector. Let $\theta$ denote the population OLS coefficient vector of $Y$ on $R^*$. The $\theta$ is defined by the full data moment condition
\[
E[m(W;\vartheta)]=0, \qquad m(W;\vartheta)=R^*\big(Y-R^{*\prime}\vartheta\big).
\]
Because $R$ and $Y$ enter the moment, they also enter the observed information,
$\tilde X=(X,U,R,Y)$. %The efficient imputation targets $\mu$ and $\mu_2$ are conditional expectations given \emph{all} observed information, and $Y$ is part of that information. An imputation function that ignores $Y$ discards it, making the estimator less efficient if it provides relevant information.

The terms containing $M^*$ are $M^*Y$, $M^*R'$, and
$(M^*)^2$. The first two are linear in $M^*$, as in the
previous examples, but $(M^*)^2$ is not.
The imputation target thus depends on two conditional expectations, $\mu$ and $\mu_2$:
\[
\mathcal{E}(\tilde X;\vartheta)
=
\binom{\mu(\tilde X)}{R}Y
-
\begin{pmatrix}
\mu_2(\tilde X) & \mu(\tilde X)R'\\
R\mu(\tilde X) & RR'
\end{pmatrix}\vartheta.
\]
\[
\mu(\tilde X):=E[M^*\mid\tilde X],
\qquad
\mu_2(\tilde X):=E[(M^*)^2\mid\tilde X].
\]

\textbf{Step 2: Approximate.} We approximate the two conditional moments with $\hat\mu$ and $\hat\mu_2$, both treated as fixed. Predicting only $M^*$  and squaring it will still lead to unbiased estimates if the MAR-S assumptions hold but is generally not the right target for efficiency, because $\mu_2(\tilde X)=\mu(\tilde X)^2+\mathrm{Var}(M^*\mid\tilde X)$. The exceptions are when $M^*$ is (nearly) determined by $\tilde X$, so that $\mathrm{Var}(M^*\mid\tilde X)\approx 0$, and when $M^*$ is binary, in which case the second moment equals the first: $\mu_2=\mu$. 

\textbf{Step 3: Derive.} Each missing quantity in the moment is replaced by its validation-adjusted analogue. The first is $\widetilde M_i$, defined as above. The second is
\[
\widetilde{M_i^2}
:=
\frac{A_i}{\pi(X_i)}\left[M_i^2-\hat\mu_2(\tilde X_i)\right]+\hat\mu_2(\tilde X_i),
\]
which is a single debiased object built from $M_i^2$, not the square of $\widetilde M_i$. The adjusted moment is
\[
m_\eta
=
\begin{pmatrix}\widetilde M\\R\end{pmatrix}Y
-
\begin{pmatrix}\widetilde{M^2}&\widetilde M R'\\R\widetilde M&RR'\end{pmatrix}\vartheta,
\]
and solving the sample moment condition gives
\[
\hat\theta
=
\left[\sum_{i\in\mathcal I}
\begin{pmatrix}\widetilde{M_i^2}&\widetilde M_iR_i'\\R_i\widetilde M_i&R_iR_i'\end{pmatrix}
\right]^{-1}
\sum_{i\in\mathcal I}\binom{\widetilde M_i}{R_i}Y_i .
\]
Again by Theorems \ref{prop:wdrgmm} and \ref{thm:eff}, $\hat\theta$ is consistent and asymptotically normal for any fixed
$\hat\mu,\hat\mu_2$. It
attains the efficiency bound of Corollary \ref{cor:verify_jid} to the extent
$\hat\mu$ and $\hat\mu_2$ approximate $\mu$ and $\mu_2$.

\subsubsection{The general pattern}
These three examples share a structure. Each moment is a sum of terms, where each term involves a function $g(M^*)$ of the missing variable (\textit{e.g.,} $M^*$, $(M^*)^2$), potentially multiplied by a known factor (such as $R$ or $Y$). Whenever a moment has this structure, it depends on the missing variable only through finitely many conditional means, $
\mu_g(\tilde X):=E_P[g(M^*)\mid\tilde X],
$
and $\mathcal E(\tilde X;\vartheta)$ is determined by them. %Because conditioning on $\tilde X$ treats the known factors as constants, they factor out of the conditional moment and the imputation error, and 
We replace each $g(M^*)$ with its validation-adjusted analogue
\[
\widetilde g
:=
\hat\mu_g(\tilde X)+\frac{A}{\pi(X)}\left[g(M)-\hat\mu_g(\tilde X)\right].
\]
Estimation then solves the adjusted empirical analog of the moment condition, which is based on the original moment with each missing $g(M_i^*)$ replaced by $\widetilde g_i$. When the moment is also linear in $\theta$, we can solve directly for $\hat\theta$, yielding a closed form expression.

\input{tab_estimators_revised}

Table \ref{tab:mars_estimators} applies the general recipe to a broader set of estimands. The difference in conditional means row shows how a common estimand can nest several research designs, as estimation in randomized control trials (RCTs), regression discontinuity design (RDD) under local randomization, and the building blocks of modern differences-in-differences (DiD) can all be expressed as differences in conditional means. In an RCT,
\[
\theta = E[M^*\mid D=1]-E[M^*\mid D=0],
\]
and under local randomization in an RDD with window $\mathcal B$ for running variable $S$,
\[
\theta = E[M^*\mid S\in\mathcal B,D=1]-E[M^*\mid S\in\mathcal B,D=0]
\]
\citep{cattaneo_regression_2022}. In difference-in-differences
under the assumptions of \citet{callaway_difference--differences_2021}, a group-time average treatment effect is the difference between the mean outcome change of a treated cohort and that of an appropriate comparison group, where $M^*$ is defined as the outcome change (so validation must cover the outcome at both dates). In each case, the variables defining the groups ($D$, $S$, or cohort membership) are observed for every unit, so the group indicator enters the adjusted moment as a known factor. 

\subsection{Efficient Inference with Unstructured Data}\label{sec:mars_implications}

While deep neural networks are complicated, semiparametric theory provides several practical insights about how they should be deployed when efficient estimation is important. These have not been a focus of the ``black-box AI'' literature, which largely treats imputation as a fixed black box, emphasizing robustness to prediction quality.

\paragraph{1. Information.}
Table \ref{tab:mars_estimators} underscores that the information efficient imputation should use is determined by the downstream estimand, not by the prediction task in isolation. In linear regression with a missing outcome, for example, efficiency generally requires approximating $E[M^*\mid X,U,R]$.
A model trained to predict $M^*$ using only unstructured data $U$ may be inefficient even if it performs well as a standalone prediction model, because the structured variables $R$ entering the downstream analysis can carry additional information about $M^*$. 

This pattern appears throughout Table \ref{tab:mars_estimators}. In the difference in conditional means, $\tilde X$ includes the group indicator $K$; in linear regression with a missing regressor, difference in conditional means with a missing conditioning variable, and linear IV with a missing instrument, $\tilde X$ includes the outcome $Y$ along with the observed regressors $R$. In the standard ``predict then regress'' approach, replacing $M^*$ with a prediction fit using $Y$ can mechanically build the outcome into the regressor, biasing coefficients towards a relationship that reflects the imputation model and not the population. That concern does not apply with MAR-S, because identification does not rely on the quality or construction of $\eta$. If the imputation is distorted by using $Y$, that is removed by the comparison with validation data. To the extent that $M^*$ is related to $Y$, the outcome carries information about $M^*$ that the unstructured data $U$ alone may not and hence can improve efficiency.

This departs from the common practice of imputing $M^*$ from $U$ alone. Learning jointly from unstructured and structured inputs is termed ``wide and deep learning'' % \citep{cheng_wide_2016}  
and has recently been studied in econometrics \citep{klaassen_doublemldeep_2024}.

There are important special cases in which these additional inputs are unnecessary. If the observed variable is conditionally independent of $M^*$, then conditioning on it does not improve efficiency. More strongly, if $M^*$ is $U$-measurable---that is, $M^*$ is a deterministic function of $U$---then $U$ alone contains all information needed to recover the researcher's target. This may be plausible for simple information extraction tasks, but not in the common scenario where the target depends on information external to $U$. For example, a satellite image may be highly predictive of household consumption but does not contain all information needed to recover it. 

\paragraph{2. Targets.}
The estimating equation determines not only which information the imputation model should use, but also what it should learn for efficient inference. For example, in linear regression with a missing regressor, the moment depends on $(M^*)^2$ as well as $M^*$, so semiparametric efficiency generally requires learning
$
\mu(\tilde X)=E[M^*\mid\tilde X],
$ and
$\mu_2(\tilde X)=E[(M^*)^2\mid\tilde X].
$

The examples in Table \ref{tab:mars_estimators} depended on $M^*$ only through conditional means. While common, this will not always be the case. For example, when $M^*$ enters a nonlinear estimating function together with $\theta$ (\textit{e.g.,} as in a Poisson regression), $\mathcal E(\tilde X;\vartheta)$ depends on $\vartheta$ through the full conditional distribution of $M^*$. Any $\eta$ still identifies $\theta$ if Assumptions \ref{cpo} through \ref{kas} hold. Efficient estimation, however, requires approximating the conditional moment $\mathcal E(\tilde X;\vartheta)$ directly.

This estimand-specific view of what the prediction model should learn
has not been emphasized, as existing approaches typically take the prediction target as given and emphasize robustness to prediction quality. This is natural when researchers rely on off-the-shelf or pre-existing predictions whose construction they do not control. The semiparametric perspective is most useful when researchers exert some control over imputation, showing how to tailor the predictor to the estimand.

\paragraph{3. Asymptotics.}
When the validation score function $\pi(X)$ is known---because the researcher constructs the validation data or uses an appropriately documented external dataset---efficiency requires the imputation function to be \emph{consistent} for its target (in the sense of Assumption \ref{mse}); unbiased inference does not. In practice, researchers may be able to achieve efficiency gains by tuning a customized model on a split of their validation data, improving imputation. 

If the validation score function must also be estimated (relaxing Assumption \ref{kas}), efficient, $\sqrt{n}$-consistent, and asymptotically normal inference requires sufficiently accurate estimation of both the target conditional moment and validation score function, as captured by a product-rate condition on their $L^2(P)$ estimation errors.\footnote{See, \textit{e.g.}, \citet{chernozhukov_doubledebiased_2018,kennedy_semiparametric_2023,wager_causal_2024}. For example, when the target is $E[M^*]$, convergence of both nuisance functions at order $n^{-1/4}$ is sufficient (Theorem 3.2 of \citet{wager_causal_2024}). Such rates are attainable for various classes of neural networks \citep{farrell_deep_2021}. %When $\tilde X$ is very high-dimensional and the validation score must be estimated, higher-order influence-function corrections may also be required to attain semiparametric efficiency \citep{liu_semiparametric_2023}.
}
This has also received little attention in the black-box AI literature, which typically assumes that validation probabilities are known. The connections to semiparametric missing data theory and observational causal inference---where the analogous propensity score function usually needs to be estimated---provide guidance in this scenario.\footnote{For parameters that can be written as means of missing structured data (see Appendix Section \ref{sec:marsmeans} for more on the generality of this notion), this guidance is explicit, and analogous to rate conditions on AIPW/DML estimation of average treatment effects.}

\subsection{An Empirical Example}\label{sec:epuapp}

We now turn to an empirical example. 
\citet{baker_measuring_2016} provide an observable definition of economic policy uncertainty (EPU), a latent underlying concept, deploying a keyword-based binary classifier to measure the share of articles in leading newspapers that discuss economic policy uncertainty.\footnote{For details on the keyword classifier training algorithm, see Section III.A of \citet{baker_measuring_2016}.} These shares are then scaled and normalized to construct the published index. Our focus is on the \textit{observational validity} of the defined EPU index. 

This application maps directly into MAR-S mean estimation. Unstructured data---newspaper texts---are used to estimate the expectation of a missing binary indicator $M_{it} \in \{0,1\}$ indicating the expression of economic policy uncertainty by article $i$ at time $t$, $U_{it} \in \mathcal{U} \subseteq \mathbb{R}^\ell$. 
%\citet{baker_measuring_2016} validated a simple random sample, so $\pi$ is constant. 
We can represent the EPU index's value at time $t$ as 
\[
\text{EPU}_t := c_t\theta_t, \quad \theta_t := E_{P_t}[M_{it}^*],
\]
where \(c_t\) collects the relevant scaling terms. Under the MAR-S assumptions, the robust, efficient estimator for $\theta_t$ is the MAR-S mean estimator.%, and the MAR-S EPU index is \(c_t\hat{\theta}_t\).

We chose this example because it is highly influential and because the unstructured data, a random validation sample, and all other observed variables are readily available, a relative rarity in text analysis applications in economics. Because the study predates transformer LLMs, it relies on a keyword-based imputation function, to which MAR-S applies equally well. 
To show applicability to the original approach as well as more modern methods, we construct annual MAR-S EPU indices for $\mathcal{T}=\{1985,1986,\dots,2010\}$ using two alternative imputation functions: their original keyword classifier and a deep neural classifier based on Answer.AI's ModernBERT \citep{warner_smarter_2024}, a lightweight model well suited to texts of this length.
To preserve the original definition of economic policy uncertainty, we use only the labels from the Baker, Bloom, and Davis audit sample for training and validation. Half of the validated observations are used to tune the imputation function, while the remaining half are used for validation. For the estimation sample in each period, we use randomly sampled unvalidated article abstracts from ProQuest at a ratio of nine unvalidated articles per validated article. Because validation is unconditionally random in the resulting sample, the validation score is $\pi=0.1$.%\footnote{As in the original paper, both validated and unvalidated articles are randomly sampled conditional on the text containing specified keywords related to the economy and uncertainty. Accordingly, ($\theta_t$) should be interpreted relative to the superpopulation of leading newspaper articles satisfying this keyword filter. The scaling factor $c_t$ effectively changes the denominator so that the published EPU index refers to the broader population of newspaper articles, under the assumption that articles discussing economic policy uncertainty are not excluded by the economy-and-uncertainty keyword filter.}

\begin{figure}[ht]
\centering
\caption{EPU and MAR-S EPU Indices}
\includegraphics[width=0.9\textwidth]{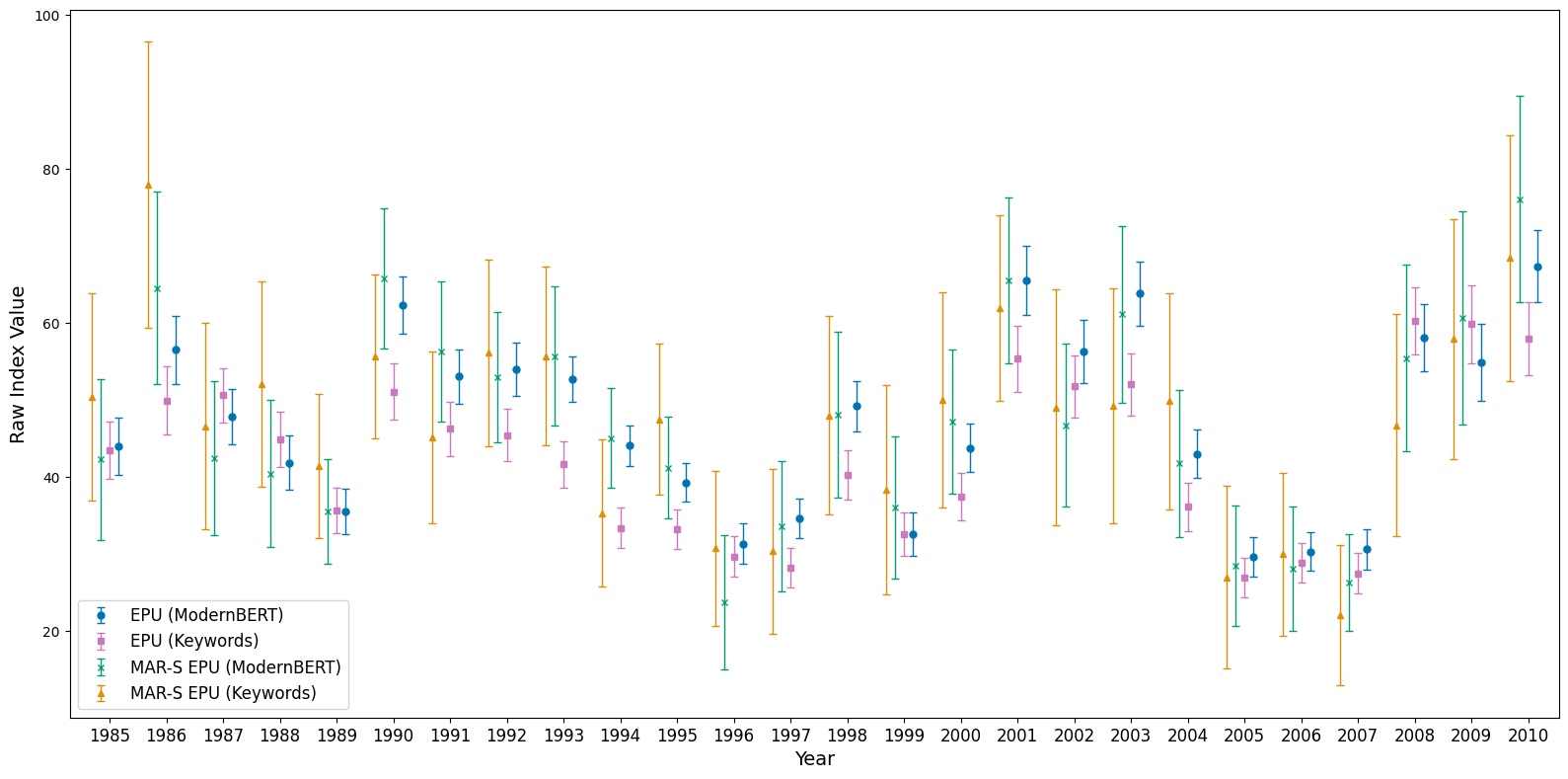}

\footnotesize
\raggedright 
This figure plots unadjusted EPU indices - computed using a keyword classifier or ModernBERT classifier - along with the MAR-S adjusted indices - again computed using a keyword classifier or ModernBERT classifier. 95\% confidence intervals are plotted for each index. 
\label{fig:epu}
\end{figure}

Figure \ref{fig:epu} plots both MAR-S EPU indices along with counterpart EPU indices that have not been adjusted using validation data: $\{c_t\tilde\theta_t\}_{t\in\mathcal{T}}$, where
$\tilde \theta_t = \frac{1}{|\mathcal{I}|}\sum_{i \in \mathcal{I}} \hat\mu(U_{it})$.\footnote{We compute 95\% confidence intervals using consistent, plug-in estimators of the asymptotic variances based on the known form of the efficient influence function.}
Confidence intervals widen substantially when they account for the uncertainty associated with adjusting for systematic measurement error. The unadjusted confidence intervals are generically valid for inference on \(E[\hat\mu(U_{it})]\), rather than the target \(E[M_{it}^*]\). They are valid for the latter only under additional restrictions on \(\hat\mu\), such as unbiasedness for \(\mu\). This assumption appears inconsistent with the downward bias, albeit modest, observed in many years of the index.

The MAR-S EPU index based on the deep neural classifier (81.5\% accuracy) has narrower confidence intervals than the index based on the keyword classifier (65.3\% accuracy). 
In principle, confidence intervals could be narrowed with additional training or validation data, a point further explored in Section \ref{sec:dropext}. 
To preserve legibility, we omit estimates that use only the validation data. These estimates show that across years, the neural MAR-S estimates have narrower confidence intervals than the validation sample-only estimates, illustrating how even imperfect AI predictions may improve precision. 

Appendix Section \ref{sec:gpr} provides a second empirical example, the geopolitical risk index \cite{caldara_measuring_2022}. The setup is analogous, with the raw imputed index showing substantial downward bias relative to the paper's validation sample.

\section{Aggregation under MAR-S}\label{sec:agg}

AI-generated measures are often aggregated and non-linearly transformed, as unstructured data are typically observed for individual texts or images, while other variables are often aggregated across time, space, or other dimensions. Existing frameworks require validation data for the imputed variable that enters the second stage estimating equation. In contrast, validation data are frequently available only for the underlying granular observations. In some cases, the missing variable of interest may itself be best represented as a population-level functional of granular missing structured data, e.g., a population-level mean, for which validation data are not observed in any finite sample. A central contribution of MAR-S is to extend valid inference to these aggregated and transformed measures.

\subsection{Random Finite Aggregation}\label{sec_rclust}

%When the aggregate is a known linear function of record-level missing data, the aggregate target can be reduced to record-level moments for which validation data are directly informative. 
Consider data ${W_{ij}}$ for $i\in[N_j]$ and $j\in[n]$, where $W_{ij}\overset{\mathrm{iid}}{\sim}P_1$ and $N_j\overset{\mathrm{iid}}{\sim}P_2$ for arbitrary distributions $P_1$ and $P_2$. The index $i$ denotes the granular, or ``record,'' level at which the unstructured data are observed (\textit{e.g.,} an individual text or image), while $j$ denotes the level of aggregation.  Let $W_{ij}=(M_{ij},A_{ij},X_{ij},U_{ij})$, where $M_{ij}$ is the missing structured variable, $A_{ij}$ is the validation indicator, $X_{ij}$ denotes observed covariates, and $U_{ij}$ denotes unstructured data. Suppose Assumptions \ref{cpo} and \ref{aoo} hold at the record level, and consider the aggregate-level target:
\[ \theta = E[M_j^*], \quad M_j^* = \alpha\left(\{M_{ij}^*\}_{i=1}^{N_j}\right),\]
For a general aggregation operator $\mathfrak{a}(\cdot)$, record-level validation data do not generally provide regular identification of $\theta$ \citep{khan_irregular_2010}. Evaluating $M_j^*$ may require observing the full collection ${M_{ij}^*}_{i=1}^{N_j}$, whereas $M^*$ is observed only for randomly validated records. When $N_j$ is large, the probability that every record within a group is validated becomes vanishingly small.

A simple special case arises when $\mathfrak{a}$ is a known linear function of the record-level structured data. Suppose
$$
M_j^*=\omega_j\sum_{i=1}^{N_j}M_{ij}^*,
\qquad \omega_j=h(N_j),
$$
with $N_j\ind M_{ij}^*$. Then, by the law of total expectation,
$$
\theta
=E[M_j^*]
=E[\omega_jN_j]E[M_{ij}^*]
\equiv \theta_1\theta_2,
\qquad
\theta_1:=E[\omega_jN_j],
\qquad
\theta_2:=E[M_{ij}^*].
$$
Thus, linearity reduces the aggregate-level target to an observed function of group size and a record-level mean for which record-level validation data are directly informative. For sums, $\omega_j=1$ and $\theta=E[N_j]E[M_{ij}^*]$; for averages, $\omega_j=1/N_j$ and $\theta=E[M_{ij}^*]$.

The efficient estimator is: 

$$
\hat\theta_2
=
\frac{1}{|\mathcal{I}|}
\sum_{ij\in\mathcal{I}}
\left[
\hat\mu(\tilde X_{ij})
+
\frac{A_{ij}}{\pi(X_{ij})}
\bigl(M_{ij}-\hat\mu(\tilde X_{ij})\bigr)
\right],
\qquad
\hat\mu(\tilde X_{ij})
:=
\hat E[M_{ij}\mid A_{ij}=1,\tilde X_{ij}].
$$
Because $\theta_1$ depends only on observed group sizes, it can be estimated by plug-in; combining $\hat\theta_1$ and $\hat\theta_2$ yields an efficient estimator of $\theta$.
The simplicity of this result comes from linearity: the aggregate target can be rewritten exactly in terms of moments identified at the record level. 

\subsection{Measurement Error Corrected Aggregation}\label{sec:mecols}

Several common features of empirical applications can make this simple approach inapplicable. Group size $N_j$ need not be iid or independent of the underlying structured data. Moreover, imputed structured data are often aggregated and then transformed nonlinearly (\textit{e.g.,} by taking logarithms). In the empirical example later in this section, we will see an example where predictions for individual texts are aggregated, differenced, logged, and interacted with firm characteristics, a series of transformations that is not uncommon. Although transformations can sometimes be approximated linearly, for example using a Taylor expansion (Appendix Section \ref{sec:app_aggregate}), this approach quickly becomes cumbersome and difficult to implement. More fundamentally, the above approach does not cover settings in which the missing variable of interest is itself defined as a population level functional, such as a population mean.

We develop a broadly applicable implementation of MAR-S that accommodates nonlinear transformations. In short, valid application of MAR-S to estimate the missing aggregate(s) (\textit{e.g.,} mean annual economic policy uncertainty estimated from newspaper article-level imputations) turns potentially systematic imputation error into classical measurement error arising from the random validation design. An estimated MAR-S aggregate may be noisy, but this reflects randomness in the validation design rather than potentially systematic imputation error. 
Therefore, using MAR-S aggregates (\textit{e.g.,} a MAR-S mean) in regression analysis becomes a familiar errors-in-variables problem \citep{deaton_panel_1985, fuller_measurement_1987} that we know how to address and extend to a variety of scenarios.

Consider the dataset $\{W_i\}_{i=1}^n$, where
$W_i \overset{\mathrm{iid}}{\sim} P$ and
$W_i=(\bar M_i^*,Y_i)$. 
$Y_i\in\mathcal{Y}\subseteq\mathbb{R}$ is an outcome of interest, and
$\bar M_i^*\in\bar{\mathcal{M}}\subseteq\mathbb{R}^K$ is a vector of
regressors. For expositional convenience, suppose that $Y_i$ and the
nonconstant components of $\bar M_i^*$ have been centered. 
%Assuming finite second moments and that $E[\bar M_i^*\bar M_i^{*\prime}]$ is nonsingular, 
Define $\beta$ as
the coefficients of the best linear predictor of $Y_i$ given
$\bar M_i^*$:
\[
    Y_i = \bar M_i^{*\prime}\beta + \varepsilon_i,
    \qquad
    E[\bar M_i^*\varepsilon_i] = 0.
\]

Not all components of $\bar M_i^*$ are observed, and so we will first need to estimate one or more of its components using a MAR-S first step. Let $\bar M_i$ denote the actually observed vector of regressors. Valid application of the MAR-S framework makes it reasonable to specify the measurement error model
\[
\bar M_i=\bar M_i^*+\upsilon_i,
\qquad
E[\upsilon_i\mid \bar M_i^*,\varepsilon_i]=0.
\]
MAR-S uses validation data (sampled independently of the regression estimation sample) to remove systematic discrepancies between the imputed data and $M^*$, leaving estimation error that is plausibly mean zero conditional on both the true regressors and the regression residual. %This does not eliminate measurement error: the remaining estimation noise still adds variation to the constructed regressors and, if ignored, distorts the regression coefficients, but a standard errors-in-variables framework can account for this. 

The variance of this remaining estimation error is characterized by $\Sigma_i:=\operatorname{Var}(\upsilon_i)$. For components that are observed directly,
$\bar M_{i,k}=\bar M_{i,k}^*$
and hence
$\upsilon_{i,k}=0$.
Consequently, $\operatorname{Var}(\upsilon_{i,k})=0$ and there is no covariance with the estimation error in any other component.
Based on large sample approximation for a MAR-S estimated component $k$, we assume that $\upsilon_{i,k}$ is mean zero and normally distributed with variance: 
\[
\sigma^2_{\upsilon,i,k}
:=
\operatorname{Var}(\upsilon_{i,k})
=
|\mathcal I_{i,k}|^{-1}\operatorname{Var}(\varphi)_{i,k},
\]
where $|\mathcal I_{i,k}|$ denotes the size of the estimation sample used for the MAR-S first step, and $\varphi_{i,k}$ is its influence function. 
The resulting variance depends both on sampling variation from constructing the regressor using a finite estimation sample and on uncertainty in the MAR-S adjustment, determined by the validation sample size and the accuracy of imputation.

For expositional simplicity, we impose two restrictions on the covariance structure of the MAR-S estimation errors. First, when multiple regressors are estimated using MAR-S first steps, we assume that the corresponding first steps are estimated from independent validation samples. Their estimation errors are therefore uncorrelated, so the off-diagonal elements of $\Sigma_i$ are zero. Second, we assume that the estimation error variances are homogeneous across observations, so that
$
\Sigma_i=\Sigma
$
for all $i$. 
Neither restriction is fundamental to the framework. Multiple MAR-S regressors may instead be estimated from the same validation sample---for example, when different features are measured from the same underlying texts---in which case their estimation errors may be correlated. Estimation error variances may also vary across observations. Appendix Section~\ref{sec:ime} develops the latter case, and Section~\ref{sec:bakermodel} develops further extensions needed for the empirical example below.
For expositional simplicity, we also treat $\Sigma$ as known, though appropriate consistency assumptions would allow the relevant estimation error variances to be estimated instead.

Although the estimation error in a MAR-S aggregate is conditionally mean zero, it still distorts the moment because it enters both the observed regressor and the residual:
\begin{align*}
E\left[
\bar M_i
\left(Y_i-\bar M_i^\prime\beta\right)
\right]
&=
E\left[
(\bar M_i^*+\upsilon_i)
(\varepsilon_i-\upsilon_i^\prime\beta)
\right] 
=
-E[\upsilon_i\upsilon_i^\prime]\beta
=
-\Sigma\beta,
\end{align*}
where the remaining terms vanish because
\(E[\bar M_i^*\varepsilon_i]=0\) and
\(E[\upsilon_i\mid\bar M_i^*,\varepsilon_i]=0\).
This requires the standard errors-in-variables correction.
Define
\[
g(\bar M_i,Y_i,\beta)
=
\bar M_i\left(Y_i-\bar M_i^\prime\beta\right)
+\Sigma\beta,
\]
so that \(E_P[g(\bar M_i,Y_i,\beta)]=0\). The corresponding just-identified
GMM estimator is
\[
\hat \beta
=
\left(
\frac{1}{n}\sum_{i=1}^n
(\bar M_i\bar M_i^\prime-\Sigma)
\right)^{-1}
\left(
\frac{1}{n}\sum_{i=1}^n \bar M_iY_i
\right)
=
(\bar M^\prime\bar M-n\Sigma)^{-1}\bar M^\prime Y.
\]
Under the standard regularity conditions for just-identified GMM
\citep{newey_chapter_1994}
\[
\sqrt{n}(\hat\beta-\beta)\rightsquigarrow
N\!\left(0,G^{-1}\Xi(G^{-1})^\prime\right),
\qquad
G:=E[\nabla_\beta g_i]
=-\!\left(E[\bar M_i\bar M_i^\prime]-\Sigma\right),
\qquad
\Xi:=\operatorname{Var}(g_i).
\]

To clarify how first-step estimation of the MAR-S aggregate affects precision, suppose additionally that the measurement error is jointly Gaussian (motivated by a multivariate CLT for MAR-S first-steps), $\upsilon_i\sim N(0,\Sigma)$, and independent of $(\bar M_i^*,\varepsilon_i)$. These assumptions are not required for the adjusted estimator or the GMM asymptotic distribution above; we use them only for illustrative purposes in order to obtain the following closed-form decomposition:
$$
\operatorname{AVar}(\hat\beta)
=
\underbrace{
\Omega^{-1}
E\!\left[\varepsilon_i^2\bar M_i^*\bar M_i^{*\prime}\right]
\Omega^{-1}
}_{\text{A}}
+
\underbrace{
\Omega^{-1}\Pi\Omega^{-1}
}_{\text{B}},
$$
where
$$
\Omega=E[\bar M_i^*\bar M_i^{*\prime}],
\qquad
\Pi
=
E[\varepsilon_i^2]\Sigma
+
\beta^\prime\Sigma\beta\,(\Omega+\Sigma)
+
\Sigma\beta\beta^\prime\Sigma.
$$
Term \textit{A} is the usual sandwich asymptotic variance of OLS if the true regressors $\bar M_i^*$ were observed. Term \textit{B} is the additional variance arising from estimation of the MAR-S regressors. Less accurate imputations affect the precision of the downstream regression through $\Sigma$. As the MAR-S aggregates become more precise, $\Sigma$ shrinks and so does this additional variance component; when $\Sigma=0$, term \textit{B} vanishes entirely.

All components of the asymptotic variance can be consistently estimated by their sample analogues. In particular,
$$
\hat V^{-1/2}\sqrt n(\hat\beta-\beta)\rightsquigarrow N(0,I),
$$
where
$$
\hat V
=
\hat G^{-1}\hat\Xi\hat G^{-\prime}
=
\left(\hat E[\bar M_i\bar M_i^\prime]-\Sigma\right)^{-1}
\widehat{\operatorname{Var}}\!\left(
\bar M_i(Y_i-\bar M_i^\prime\hat\beta)+\Sigma\hat\beta
\right)
\left(\hat E[\bar M_i\bar M_i^\prime]-\Sigma\right)^{-1}.
$$
Thus, standard Wald confidence intervals based on $\hat V$ are asymptotically valid.

This approach extends naturally to common econometric settings with richer dependence structures. Clustered data can be accommodated using the appropriate clustered central limit theorem, and the framework readily adapts to panel data, the original context in \citet{deaton_panel_1985}. The simplifying assumptions used above can also be relaxed: the outcome variable may itself be estimated in a MAR-S first step, the measurement error need not be normally distributed, and $\Sigma$ can be estimated rather than treated as known \citep{fuller_measurement_1987}.\footnote{For an alternative framing of this regression-centric method entirely in terms of nonparametrically identified functionals, see Appendix Section \ref{sec:fme}.} We now illustrate how the approach extends to more complicated panel regression settings, with full details provided in Appendix Section \ref{sec:bakermodel}.

\subsection{Empirical Example}
\citet{baker_measuring_2016} also use the EPU index, introduced in Section \ref{sec:epuapp}, in regression analysis.\footnote{Another application is real-time tracking of economic policy uncertainty. Because future texts cannot be included in the validation sample, an assumption about the stability of the imputation errors would be required to compute a real-time MAR-S EPU index.} To illustrate inference with aggregated missing data, we revisit a representative specification from their Table IV, column (5):
\[
Y_{it}
=
\beta_1 \Delta \log \mathrm{EPU}_t \times h_{it}
+
\beta_2 \Delta
\frac{\text{Federal purchases}_t}{\text{GDP}_t}
\times h_{it}
+
\alpha_i
+
\psi_t
+
u_{it},
\]
where \(i\) indexes firms and \(t\) indexes years; \(\alpha_i\) and \(\psi_t\) are firm and year fixed effects; \(h_{it}\) measures firm-year policy exposure; \(Y_{it}\) is firm-year employment growth; and \(\Delta (\text{Federal purchases}_t/\text{GDP}_t)\) is a control. The estimand of interest is \(\beta_1\).

We apply the framework for aggregated and transformed structured data to obtain valid inference. The MAR-S first step satisfies
$
\sqrt{|\mathcal{I}_t|}
\left(
\hat{\theta}_t-\theta_t
\right)
\rightsquigarrow
N\left(
0,
\operatorname{Var}(\varphi_t)
\right),
$
where \(\varphi_t\) is the efficient influence function. Since \(\mathrm{EPU}_t=c_t\theta_t\),
$$\sqrt{|\mathcal{I}_t|}
\left(\widehat{\text{EPU}}_t - \text{EPU}_t \right) \rightsquigarrow N(0, c_t^2\text{Var}(\varphi_t)).$$
Further using the delta method
$\sqrt{|\mathcal{I}_t|}
\left(\log\widehat{\text{EPU}}_t - \log \text{EPU}_t \right) \rightsquigarrow N(0, \theta_t^{-2}\text{Var}(\varphi_t))$.

Motivated by this result, we assume
\[
\log \widehat{\mathrm{EPU}}_t
=
\log \mathrm{EPU}_t
+
\tilde{\upsilon}_t,
\qquad
\tilde{\upsilon}_t
\overset{\mathrm{ind}}{\sim}
N(0,\sigma_t^2),
\qquad
\sigma_t^2
:=
|\mathcal{I}_t|^{-1}
\theta_t^{-2}
\operatorname{Var}(\varphi_t).
\]
The first regression component can then be written as
\[
\bar M_{it,1}
=
\bar M_{it,1}^*
+
\upsilon_{it,1},
\]
where
\[
\bar M_{it,1}
:=
\Delta \log \widehat{\mathrm{EPU}}_t \times h_{it},
\qquad
\bar M_{it,1}^*
:=
\Delta \log \mathrm{EPU}_t \times h_{it},
\qquad
\upsilon_{it,1}
:=
(\tilde{\upsilon}_t-\tilde{\upsilon}_{t-1})h_{it}.
\]
Thus, conditional on \(h_{it}\) - the exposure - the measurement error variance is
\[
\operatorname{Var}(\upsilon_{it,1}\mid h_{it})
=
(\sigma_t^2+\sigma_{t-1}^2)h_{it}^2,
\]
with dependence across adjacent years induced by the first differencing.

Because \(\upsilon_{it,1}\) and \(\bar M_{it,1}^*\) both depend on \(h_{it}\), this measurement error is not strongly classical. It is, however, weakly classical \citep{schennach_measurement_2022}, which is sufficient for identification of $\beta$.\footnote{Under the maintained independence assumptions,
$
E\left[
\upsilon_{it,1}
\mid
\bar M_{it}^*
\right]
=
0
\text{ a.s.},
$
and hence
$
E\left[
\upsilon_{it,1}\bar M_{it}^*
\right]
=
0.
$}

The MAR-S first-step motivated assumptions identify the measurement error variance matrix
$
\Sigma_{it}
:=
E\left[
\upsilon_{it}\upsilon_{it}^\prime
\right].
$
Accordingly, the measurement error-corrected least-squares estimator is
\[
\hat{\beta}
=
\left(
\sum_{i,t}
\left(
\bar M_{it}\bar M_{it}^\prime-\Sigma_{it}
\right)
\right)^{-1}
\left(
\sum_{i,t}
\bar M_{it}Y_{it}
\right),
\]
whose validity and inference properties we establish formally in Appendix Section \ref{sec:bakermodel}.

\begin{figure}[ht]
\centering
\caption{Estimates Using EPU and MAR-S EPU Indices}
\includegraphics[width=0.8\textwidth]{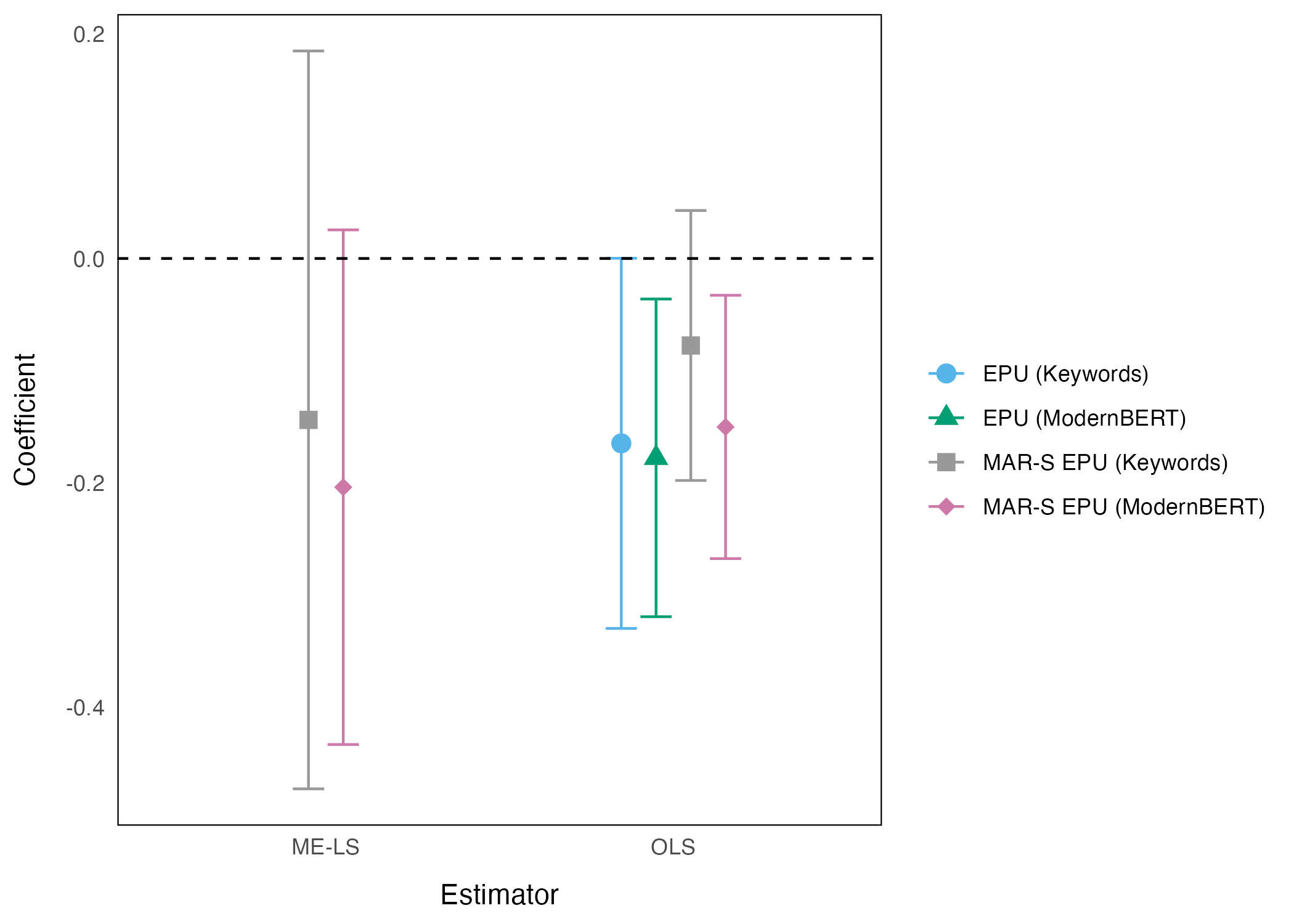}

\footnotesize
\raggedright 
This figure plots estimates of the impact of firm level exposure to economic policy uncertainty on employment growth. The ME-LS estimates use measurement error corrected least squares and the MAR-S adjusted EPU indices (constructed with either a keyword classifier or ModernBERT classifier). The OLS estimates use either the MAR-S EPU indices or the unadjusted EPU indices, again constructed with either a keyword or ModernBERT classifier.
\label{fig:coef}
\end{figure}

The left panel of Figure \ref{fig:coef} reports \(\hat{\beta}\) using the MAR-S EPU index and measurement error-corrected least squares (ME-LS). Confidence intervals account for firm-level clustering and dependence induced by common first-step measurement errors across firms and adjacent years (Appendix Section \ref{sec:bakermodel}). While the derivation above uses the efficient influence function, MAR-S aggregates remain valid when imputation is misspecified. The more accurate the imputation of the aggregate, the more precisely the target parameter is estimated when the remaining random measurement error is accounted for, and indeed we see that the MAR-S EPU index built from the deep neural classifier gives a narrower confidence interval than the keyword-based index. 

The right panel reports OLS estimates using the MAR-S EPU index---which ignores its random measurement error---and using the unadjusted EPU index---which ignores measurement error altogether.\footnote{The OLS estimate based on the unadjusted keyword EPU index differs somewhat from the corresponding estimate in \citet{baker_measuring_2016} for two reasons. First, our sample covers 1985--2010 rather than 1985--2012 because of ProQuest access restrictions with our university license. Second, we re-implement the \citet{baker_measuring_2016} keyword-classification procedure using their validation sample, so the selected keywords differ slightly from those used in the published EPU index.}
Plugging in the imputed index as if it were the true regressor yields confidence intervals that are too narrow, because they omit the uncertainty from imputation, and this problem becomes more severe with less accurate imputation.
Moreover, OLS estimates using the MAR-S EPU index exhibit the expected attenuation relative to the ME-LS estimates. The unadjusted EPU index contains both random measurement error, which induces attenuation, and modest systematic measurement error, visible in Figure \ref{fig:epu}. The resulting biases happen to roughly offset, but there is no reason to expect such offsetting generically.

\section{Validation in Practice} \label{sec:dropext}

Implementing inference with validation samples requires addressing important practical questions. In particular, the researcher needs to determine which observables $X$, if any, will condition validation, as well as how many observations to validate. This distinguishes MAR-S from observational causal inference settings, in which the propensity score---analogous to the validation score in MAR-S---is typically outside the researcher's control. This section provides some practical guidance, drawing from the related literature. 

\textbf{Which observations should be validated?}
While designing the validation score function can be very straightforward - \textit{e.g.,} in cases where validating a simple random sample is suitable - its design is less obvious when data are highly class-imbalanced. This occurs when the missing structured data effectively constitute a ``rare event,'' and can be common with massive scale data. For example, if the dataset is the universe of all social media posts on a major platform and the researcher wishes to impute which posts are about a given topic, the share of this content will likely be very small given the vast diversity of content. With rare event estimation, the coefficient of variation---the ratio of the variance of the indicator of the event over the probability of the event---can be large, making confidence intervals  uninformative. The literature suggests importance sampling to reduce the variance of estimates \citep{mcbook}. Such an approach is developed in machine learning work on ``batch active inference'' \citep{zrnic_active_2024}, which we now relate to the MAR-S framework as a drop-in extension.

In the setting of mean estimation, recall that the asymptotic variance of \(\hat\theta\) is
\begin{align*}
\sigma^2
&:= \operatorname{Var}\left(\mu(\tilde X)
+ \frac{A}{\pi(X)}(M-\mu(\tilde X))\right) \
= E\left[\frac{A}{\pi(X)^2}
(M-\mu(\tilde X))^2\right] + c,
\end{align*}
where, for fixed \(\mu\), \(c\) collects the terms that do not depend on the validation score function \(\pi(X)\). Under MAR-S,
$$
E\left[\frac{A}{\pi(X)^2}(M-\mu(\tilde X))^2\right]
=
E\left[
\frac{1}{\pi(X)}
E\left[(M^*-\mu(\tilde X))^2\mid X\right]
\right].
$$
Given an expected validation rate \(q\), the importance sampling validation rule solves
$$
\min_{\pi}\;
E\left[
\frac{E[(M^*-\mu(\tilde X))^2\mid X]}{\pi(X)}
\right]
\quad\text{subject to}\quad
E[\pi(X)]=q,\qquad
\underline{\pi}\leq\pi(X)\leq1.
$$
For an interior solution, the first-order condition implies a validation probability proportional to
$$
\pi_{IS}(X)
\propto
\left\{
E\left[(M^*-\mu(\tilde X))^2\mid X\right]
\right\}^{1/2}.
$$
This assigns higher validation probability to observations with larger conditional root mean squared prediction error.
This validation rule is infeasible because the conditional prediction error cannot be evaluated prior to validation. It nevertheless provides useful intuition: the validation score function should place greater weight on observations that are harder to impute. One could operationalize this using a feasible proxy for the uncertainty associated with the fixed imputation function: \textit{e.g.,} uncertainty measures derived from the predictive probabilities of neural network classifiers or verbalized confidence scores from LLMs \citep{yang_verbalized_2024}. Another approach that can work well in practice with class-imbalanced data is to stratify validation sampling on distance to a query in embedding space (\textit{e.g.,} an example query might be ``this article is about economic policy uncertainty''), as positive examples will tend to be over-represented near the query in embedding space (see \citet{dell_deep_2024} for more on embeddings). 

\textbf{How much of the estimation sample should be validated?}
To develop an intuition for this, consider the framework for power analysis with validation samples developed by \citet{broska_mixed_2025}. They define the effective sample size $n_0$ as the number of fully validated observations that would yield the same asymptotic precision as the combination of validation data and imputations. They derive $n_0$ for mean estimation

$$
n_0
:=
|\mathcal{I} \cap \mathcal{J}|
\times
\frac{|\mathcal{I}|}
{|\mathcal{I} \cap \mathcal{J}|
+
|\mathcal{I} \cap \mathcal{J}^c|
\left(1-\max\{\tilde{\rho}^2,0\}\right)},
$$
where \(\mathcal{J}\) is the set of validated indices, \(\mathcal{J}^c\) is the set of unvalidated indices, $\mathcal{I}$ is the estimation sample, and \(\tilde\rho\) is the asymptotic correlation between the validation sample-only and imputation-only estimators: 
$
\tilde\rho:=\operatorname{Cor}\!\left(M^*,\mu(\tilde X)\right)
$.\footnote{Note that many MAR-S estimands of interest can be expressed directly as means; see Section~\ref{sec:marsmeans}.}

When the imputations contain no useful information, \(n_0\) equals the number of validated observations; as their predictive content increases, so does \(n_0\), reaching the full sample size when imputation is perfect. $\tilde\rho$ is unknown ex ante since it depends on $M^*$, meaning $n_0$ cannot be calculated in practice, but researchers could posit a plausible value for $\tilde \rho$ based on prior information or could estimate it in a pilot sample.

We explore this empirically, using data assembled by \cite{dell_american_2023} that consist of 4,157 articles randomly sampled from historical U.S. newspapers and corresponding labels for whether the article discusses politics, denoted by \(M_i \in \{0,1\}\).\footnote{Politics was chosen because it is sufficiently common that simple random sampling yields many positive examples.} We randomly split the sample in half, using one set to tune a custom classifier and the other for estimation. Under Assumptions \ref{cpo}--\ref{mse}, the efficient estimator of the probability that an article discusses politics is the MAR-S mean.

\begin{figure}[!htbp]
\centering
\caption{Varying the Number of Validations}
\includegraphics[width=0.7\textwidth]{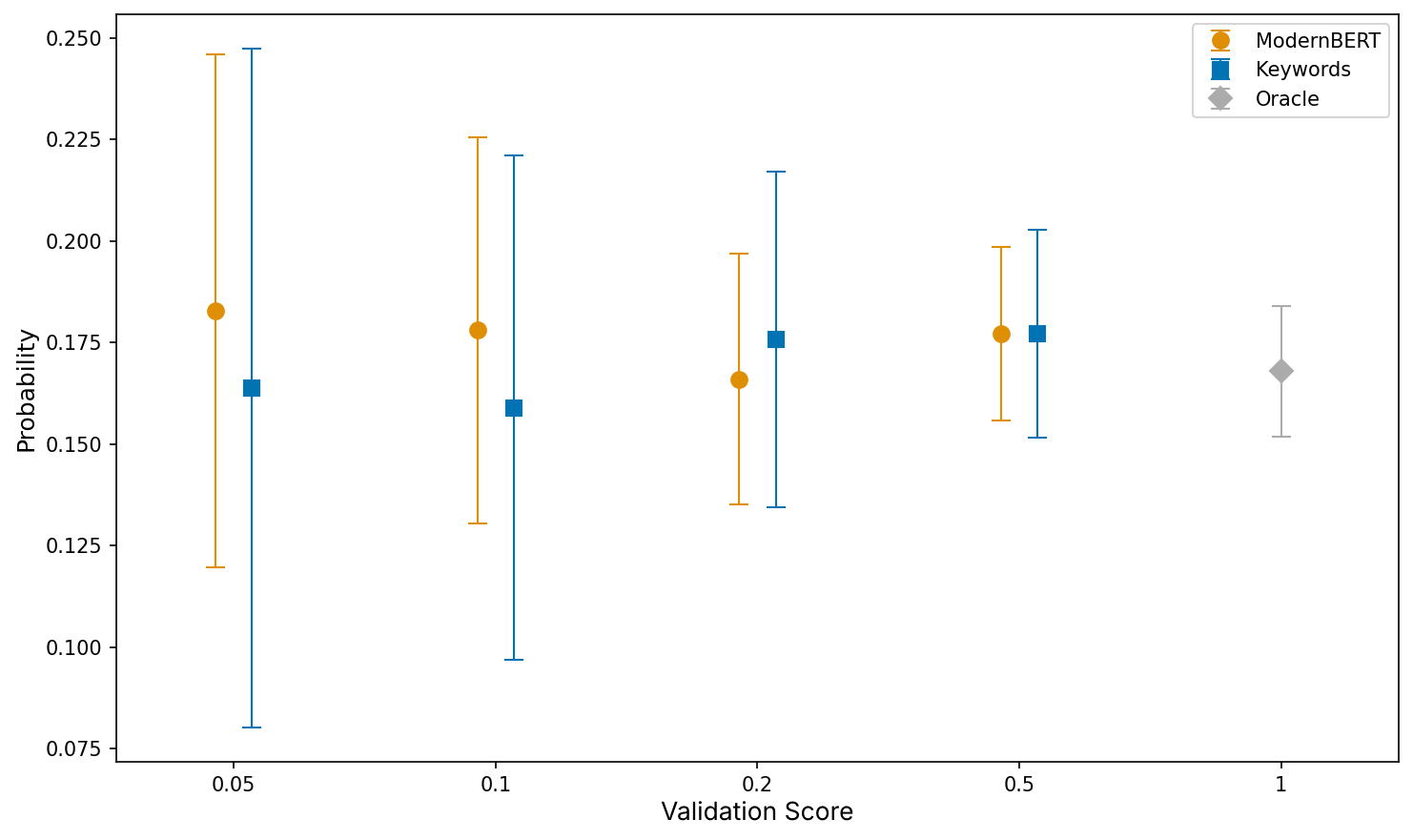}
\label{fig:pol_pi}

\footnotesize
\raggedright 
This figure estimates $P(\text{discussion of politics})$ in local newspaper articles, varying the size of the validation data. We use a validated share of $\left\{\frac{1}{20}, \frac{1}{10}, \frac{1}{5}, \frac{1}{2}\right\}$ of the estimation sample to compute the MAR-S adjusted means. We also compare to an ``oracle'' that uses the validation data for the full estimation sample. 
\end{figure}

Figure \ref{fig:pol_pi} reports MAR-S estimates using 5\%, 10\%, 20\%, and 50\% of the $n=2078$ estimation sample to validate the imputation of $M^*$, which we impute using either the keyword classifier in \citet{dell_american_2023} (around 77\% accuracy) or ModernBERT fine-tuned on the tuning split of the labeled data (around 92\% accuracy). We compare these to an oracle estimate computed from the labeled data for the full estimation sample.

Point estimates remain close to the oracle across validation sample sizes. Confidence intervals are wider for lower validation probabilities, particularly for the less accurate keyword classifier. As the validation sample grows, the intervals narrow toward the oracle's precision, and the precision gap between the keyword and ModernBERT classifiers diminishes.

We recommend beginning with a modest validation sample---for example, in the case of a binary text classification task, labeling a couple hundred texts---and prioritizing validation quality over quantity. If estimates are noisier than desired, the researcher can seek to improve the imputation model (\textit{e.g.,} by finding an external dataset to train on, by tuning on validated observations, by using a larger model, by changing the training objective, etc.), can expand the set of validated observations used in estimation, or both. The most cost-effective approach is likely to depend on the nature of the imputation task, the cost of collecting validation data, and the expertise of the researcher.

\section{Conclusion} \label{Sec:conclusion}
Deep learning is transforming economic research, by making it possible to utilize unstructured data at scale. However, it raises a major challenge: when data construction is cheap and many measures of the same concept can feasibly be built, what makes a measure credible, and how should researchers choose among them?

MAR-S addresses these challenges by framing the analysis of unstructured data as a missing structured data problem. The researcher defines an observable target and constructs a random validation sample through a transparent procedure. A cheaply scalable technology such as AI can then extend that measurement to a much larger sample. This approach enables the recovery of $\sqrt{n}$-consistent estimators with valid uncertainty quantification, even for poorly performing imputation functions, and offers a foundation for building estimators that are both efficient and robust. MAR-S makes inference with random validation samples accessible in a wide variety of empirical economics settings that have received little attention in the existing literature, including inference with aggregated AI predictions. 

Validation underscores the value of transparently defining the low-dimensional quantities that the researcher seeks to extract from unstructured data. Importantly, MAR-S reduces the scope for cheaply generating many alternative measures and then reporting only the measure(s) that yields the desired results. As the cost of extracting structured measures from unstructured data continues to fall dramatically, this discipline becomes increasingly important. Creating a validation sample can also expose ambiguities in the target definition that would otherwise be resolved implicitly—and opaquely—by an AI model, potentially producing measures that depart significantly from the intended construct. 

In the near future, could AI become so intelligent that validation is not necessary? AI, regardless of improvements in predictive accuracy, cannot ensure that the researcher has specified the task clearly enough for the AI to predict the desired target; careful validation can help ensure this. Moreover, AI is used for many tasks that go well beyond simple information extraction. The AI may not have access to the information required to perfectly predict $M^*$, no matter how well it utilizes the information available for the full sample. 

Random validation is not always feasible, and cannot tell us whether a given construct is well-suited to answering the research question, which is the domain of construct validity. Nevertheless, it provides a gold standard for observational validity, and it is our hope that MAR-S will make it more accessible across a wide range of empirical settings. 

\begingroup
\singlespacing
\bibliographystyle{aea}
\bibliography{paper}
\endgroup

\clearpage

\setcounter{table}{0}
\renewcommand{\thetable}{S-\arabic{table}} % Setting the table number output to letters 
\setcounter{figure}{0}
\renewcommand{\thefigure}{S-\arabic{figure}} % Setting the figure number output to letters 
\setcounter{section}{0}
\renewcommand{\thesection}{S-\arabic{section}}

\section{Appendix}

\subsection{MAR-S with Over-identification}\label{sec:overid}

If $d_m > d_\theta$, \textit{i.e.,} there exist over-identifying moment restrictions, then the efficient influence function for $\theta$ is given by the following corollary.

\begin{restatable}{corollary}{Corverifyoid}\label{cor:verify_oid}
Under Assumptions \ref{cpo}, \ref{aoo}, and \ref{kas}, if $d_m > d_\theta$ and  $\mathcal{J}_\theta$ has full column rank equal to $d_\theta$, then the efficient influence function for $\theta(P)$ is given by
\[
\varphi(M,A,\tilde X;\theta) = -(\mathcal{A}_\theta\mathcal{J}_\theta)^{-1}\mathcal{A}_\theta \mathcal{R}(W,A,\tilde X;\theta).
\]
where the $\mathcal{R}(W,A,\tilde X;\theta):=\frac{A}{\pi(X)}[m(W ; \theta)-\mathcal{E}(\tilde X; \theta )]+\mathcal{E}(\tilde X;\theta)$, the $\mathcal{A}_\theta := \mathcal{J}_\theta^\prime \Omega_\theta^{-1}$, and the $\Omega_\theta := E_P[\mathcal{R}(W,A,\tilde X;\theta)\mathcal{R}(W,A,\tilde X;\theta)^\prime].$
\end{restatable}

With the perspective of Theorem \ref{prop:wdrgmm} in mind, we can also see that, when many independently learned imputation functions are available, they can be combined as a single over-identifying set of moment restrictions, and the most efficient estimation of $\theta$ using all of them is given by an efficient GMM weighting matrix.

\subsection{More Derivations of MAR-S Estimators}\label{sec_addmars}

\subsubsection{Linear Regression with Missing Regressor}

We now consider a new setting, a linear regression with regressor based on missing structured data, and show that deriving an analogous MAR-S estimator is as simple as in the previously studied cases. Defining $R^* := (M^*, R^{\prime})^{\prime} \in \mathbb{R}^{d_r+1}$, and with $\theta = (\beta, \gamma^\prime)^\prime \in \Theta \subseteq \mathbb{R}^{d_r+1}$, the moment condition now of interest is
\[
m(W;\theta) =  R^* (Y- R^{*\prime} \theta).
\]
As such, the EIF for $\theta$ is derived by noting 
\[
\mathcal{E}(\tilde X ; \theta)=E[R^* (Y- R^{*\prime} \theta) \mid \tilde X] = E[R^*\mid \tilde X]Y- E[R^*R^{*\prime} \mid \tilde X]\theta,
\]
and thus
\begin{align*}
    &\varphi(W;\theta,P) = -\mathcal{J}_\theta^{-1}  \left(\frac{A}{\pi( X)}[m(W ; \theta)-\mathcal{E}(\tilde X; \theta )]+\mathcal{E}(\tilde X;\theta)\right) \\
    %= &E[R^* R^{*\prime}]^{-1} \left(\frac{A}{\pi(\tilde X)}[R^* (Y- R^{*\prime} \theta)-(E[R^*\mid \tilde X]Y- E[R^*R^{*\prime} \mid \tilde X]\theta)]+E[R^*\mid \tilde X]Y- E[R^*R^{*\prime} \mid \tilde X]\theta\right) \\
    = &E[R^* R^{*\prime}]^{-1} \left(\frac{A}{\pi( X)}\bigg[(R^*-E[R^*\mid \tilde X])Y- (R^*R^{*\prime} - E[R^*R^{*\prime} \mid \tilde X])\theta \bigg]+E[R^*\mid \tilde X]Y- E[R^*R^{*\prime} \mid \tilde X]\theta\right).
\end{align*}
Letting $\mu_2(\tilde X) := E[M^{*2}\mid \tilde X]= E[M^{2}\mid A=1, \tilde X]\in \mathbb{R}$, we may write
\[
E[R^*\mid \tilde X] = \begin{pmatrix}\mu(\tilde X) \\ R\end{pmatrix}, \quad E[R^* R^{*\prime}\mid \tilde X] = \begin{pmatrix} \mu_2(\tilde X)& \mu(\tilde X)R^\prime\\
R\mu(\tilde X) & RR^\prime\end{pmatrix}
\]
and so we can simplify the EIF as
\begin{align*}
    \varphi(W;\theta,P) &= E[R^* R^{*\prime}]^{-1} \left(\frac{A}{\pi( X)}\left[\begin{pmatrix}M-\mu(\tilde X) \\ 0\end{pmatrix}Y- \begin{pmatrix} M^{*2} - \mu_2(\tilde X)& (M-\mu(\tilde X))R^\prime\\
R(M-\mu(\tilde X)) & 0\end{pmatrix}\theta \right] \right.\\
&+ \left.\begin{pmatrix}\mu(\tilde X) \\ R\end{pmatrix}Y- \begin{pmatrix} \mu_2(\tilde X)& \mu(\tilde X)R^\prime\\
R\mu(\tilde X) & RR^\prime\end{pmatrix}\theta\right).
\end{align*}
As a result, we have that the estimating equation yields
\begin{align*}
|\mathcal{I}|^{-1} \sum_{i\in \mathcal{I}}
\Bigg(
&\frac{A_i}{\pi(X_i)}
\left[
\begin{pmatrix}
M_i-\hat\mu(\tilde X_i) \\
0
\end{pmatrix}Y_i
-
\begin{pmatrix}
M_i^{2} - \hat\mu_2(\tilde X_i) 
    & (M_i-\hat\mu(\tilde X_i))R_i^{\prime} \\
R_i(M_i-\hat\mu(\tilde X_i)) 
    & 0
\end{pmatrix}
\hat\theta
\right] \\
&+
\begin{pmatrix}
\hat\mu(\tilde X_i) \\
R_i
\end{pmatrix}Y_i
-
\begin{pmatrix}
\hat\mu_2(\tilde X_i) 
    & \hat\mu(\tilde X_i)R_i^{\prime} \\
R_i\hat\mu(\tilde X_i) 
    & R_iR_i^{\prime}
\end{pmatrix}
\hat\theta
\Bigg)
= 0 .
\end{align*}
Solving for the MAR-S estimator $\hat\theta$, we arrive at
\[
\begin{gathered}\hat{\theta}=\left[  \sum_{i\in \mathcal{I}}\left\{\frac{A_i}{\pi\left(X_i\right)}\left(\begin{array}{cc}M_i^{2}-\hat{\mu}_2(\tilde{X}_i) & \left(M_i-\hat{\mu}(\tilde{X}_i)\right) R_i^{\prime} \\ R_i\left(M_i-\hat{\mu}(\tilde{X}_i)\right) & 0\end{array}\right)+\left(\begin{array}{cc}\hat{\mu}_2(\tilde{X}_i) & \hat{\mu}(\tilde{X}_i) R_i^{\prime} \\ R_i \hat{\mu}(\tilde{X}_i) & R_i R_i^{\prime}\end{array}\right)\right\}\right]^{-1} \\ \cdot\left[ \sum_{i\in \mathcal{I}}\left\{\frac{A_i}{\pi\left(X_i\right)}\binom{M_i-\hat{\mu}(\tilde{X}_i)}{0}+\binom{\hat{\mu}(\tilde{X}_i)}{R_i}\right\} Y_i\right] .\end{gathered}
\]
%Again, this estimator contains many intuitive forms, which resemble the pseudo-outcomes seen before, and which we might call ``pseudo-regressors.'' 
This estimator naturally generalizes to more missing regressors (involving more nuisances).

\subsubsection{Linear IV with Missing Instrument}

Revisiting TSLS, we can assume the structural equation 
\[
Y=D \beta+C^{\prime} \gamma+\varepsilon, \quad R=\binom{D}{C}, \quad \theta=\binom{\beta}{\gamma}
\]
and instrument $Z^*=\binom{M^*}{C}$, where $D$ is the treatment and $\beta$ is the parameter of ultimate interest. Then the moment condition for $\theta$ is 
\[
m(W;\theta) = Z^* (Y - R^\prime \theta).
\]
Note then that 
\[
\mathcal{E}(\tilde X ; \theta)=\binom{\mu(\tilde X)}{C}Y - \binom{\mu(\tilde X)}{C}R^\prime\theta, \quad E_P[Z^* \mid \tilde X] = \binom{\mu(\tilde X)}{C},
\]
\[
\mathcal{J}_\theta = -E_P[Z^* R^\prime],
\]
and thus the EIF is given by
\begin{align*}
    &\varphi(W;\theta,P) = -\mathcal{J}_\theta^{-1}  \left(\frac{A}{\pi( X)}[m(W ; \theta)-\mathcal{E}(\tilde X; \theta )]+\mathcal{E}(\tilde X;\theta)\right) \\
    %= & E_P[Z^* R^\prime]^{-1} \left(\frac{A}{\pi( X)}[\binom{M^*}{C} (Y - R^\prime \theta)-(\binom{\mu(\tilde X)}{C}Y - \binom{\mu(\tilde X)}{C}R^\prime\theta)]+\binom{\mu(\tilde X)}{C}Y - \binom{\mu(\tilde X)}{C}R^\prime\theta\right) \\
    %= & E_P[Z^* R^\prime]^{-1} \left(\frac{A}{\pi( X)}[\binom{M}{C} Y - \binom{M}{C}R^\prime \theta-\binom{\mu(\tilde X)}{C}Y + \binom{\mu(\tilde X)}{C}R^\prime\theta]+\binom{\mu(\tilde X)}{C}Y - \binom{\mu(\tilde X)}{C}R^\prime\theta\right) \\
    = & E_P[Z^* R^\prime]^{-1} \left(\frac{A}{\pi( X)}\left[\binom{M-\mu(\tilde X)}{0} (Y-R^\prime \theta) \right]+\binom{\mu(\tilde X)}{C}(Y - R^\prime\theta)\right).
\end{align*}
The natural estimating equation is then
\[
|\mathcal{I}|^{-1} \sum_{i\in \mathcal{I}} \left(\frac{A_i}{\pi( X_i)}\left[\binom{M_i-\hat\mu(\tilde X_i)}{0} (Y_i-R_i^\prime \hat\theta) \right]+\binom{\hat\mu(\tilde X_i)}{C_i}(Y_i - R_i^\prime\hat \theta)\right) = 0
\]
which yields
\[
\hat\theta=
\left[
\sum_{i\in \mathcal{I}}
\binom{
\frac{A_i}{\pi(X_i)}(M_i-\hat\mu(\tilde X_i))+\hat\mu(\tilde X_i)
}{
C_i
}R_i^\prime
\right]^{-1}
\left[
\sum_{i\in \mathcal{I}}
\binom{
\frac{A_i}{\pi(X_i)}(M_i-\hat\mu(\tilde X_i))+\hat\mu(\tilde X_i)
}{
C_i
}Y_i
\right].
\]

%\subsubsection{Linear IV (Missing Outcome, Missing Treatment)}

%These will look identical to the use case of the linear regression with missing outcome, owing to identical structure in the FWL representation.

\subsubsection{Conditional Means (Missing Conditioning Variable)}

In general, we may represent the moment condition as, for binary conditioning indicator $M^* \in \{0, 1\}$ and outcome $Y \in \mathbb{R}$, and $\theta = (\beta, \gamma)$,
\[
m(W; \theta) = \binom{\gamma^{-1} M^* Y - \beta}{M^* - \gamma}.
\]
As such, we have that
\[
\gamma = E[M^*]=P(M^*=1), \quad \beta = E[\gamma^{-1}M^* Y] = E[Y\mid M^* = 1],
\]
where $\beta$ is the estimand of interest. 
Thus
\[
\mathcal{E}(\tilde X ; \theta)=E_P\left[\binom{\gamma^{-1} M^* Y - \beta}{M^* - \gamma}\mid \tilde X\right] = \binom{\gamma^{-1}  Y \mu(\tilde X)- \beta}{\mu(\tilde X) - \gamma},
\]
\[
\mathcal{J}_\theta = \left(\begin{array}{cc}-1 & -\gamma^{-2} E\left[M^* Y\right] \\ 0 & -1\end{array}\right),
\]
\begin{align*}
    &\varphi(W;\theta,P) = -\mathcal{J}_\theta^{-1}  \left(\frac{A}{\pi(X)}[m(W ; \theta)-\mathcal{E}(\tilde X; \theta )]+\mathcal{E}(\tilde X;\theta)\right) \\
    = &\left(\begin{array}{cc}1 & -\gamma^{-2} E\left[M^* Y\right] \\ 0 & 1\end{array}\right)  \left(\frac{A}{\pi( X)}\left[\binom{\gamma^{-1} M^* Y - \beta}{M^* - \gamma}-\binom{\gamma^{-1}  Y \mu(\tilde X)- \beta}{\mu(\tilde X) - \gamma}\right]+\binom{\gamma^{-1}  Y \mu(\tilde X)- \beta}{\mu(\tilde X) - \gamma}\right) \\
    = &\left(\begin{array}{cc}1 & -\gamma^{-2} E\left[M^* Y\right] \\ 0 & 1\end{array}\right)  
    \left(\frac{A}{\pi( X)}
    \left[\binom{\gamma^{-1} Y (M - \mu(\tilde X))}{M - \mu(\tilde X)}\right]+\binom{\gamma^{-1}  Y \mu(\tilde X)- \beta}{\mu(\tilde X) - \gamma}\right).
\end{align*}
Now we have that the estimating equations yield
\[
|\mathcal{I}|^{-1} \sum_{i\in \mathcal{I}} \left(\frac{A_i}{\pi( X_i)}
    \binom{\hat\gamma^{-1} Y_i (M_i - \hat\mu(\tilde X_i))}{M_i - \hat\mu(\tilde X_i)}+\binom{\hat\gamma^{-1}  Y_i \hat\mu(\tilde X_i)- \hat\beta}{\hat\mu(\tilde X_i) - \hat\gamma}\right) = 0
\]
\[
\hat\gamma=|\mathcal{I}|^{-1} \sum_{i\in \mathcal{I}}
\left[\frac{A_i}{\pi(X_i)}\bigl(M_i-\hat\mu(\tilde X_i)\bigr)
+
\hat\mu(\tilde X_i)
\right], \quad 
\hat\beta = 
|\mathcal{I}|^{-1} \sum_{i\in \mathcal{I}}
\frac{\left[
\frac{A_i}{\pi(X_i)}\bigl(M_i-\hat\mu(\tilde X_i)\bigr)
+
\hat\mu(\tilde X_i)
\right]}{\hat\gamma} Y_i.
\]
Notice that for efficient variance estimation, we are taking into account the estimation error in the event probability (through the Jacobian's impact).

\subsection{MAR-S Efficiency Results}\label{sec:effstate}

Consider estimator $\hat\theta_\text{mean}$ for estimating the population mean of missing structured data $\theta_\text{mean}$, per Table \ref{tab:mars_estimators}. Let $\varphi_\text{mean}$ be the corresponding EIF for $\theta_\text{mean}$.

\begin{restatable}{proposition}{Thmdesc}\label{thm:desc}
Under Assumptions \ref{cpo}, \ref{aoo}, \ref{kas}, and \ref{mse}, we have that, as $|\mathcal{I}|,|\mathcal{I}'|\to\infty$,
\[
\sqrt{|\mathcal{I}|}(\hat\theta_\text{mean} - \theta_\text{mean}) \rightsquigarrow N(0, \operatorname{Var}(\varphi_\text{mean}(M, A, \tilde X;\theta_\text{mean}))).
\]
\end{restatable}

Now consider estimator $\hat\theta_\text{reg}$ of linear regression coefficients $\theta_\text{reg}$ under missing structured outcomes per Table \ref{tab:mars_estimators}. Let $\varphi_\text{reg}$ be the corresponding EIF for $\theta_\text{reg}$. 

\begin{restatable}{proposition}{Thmlin}\label{thm:lin}
Under Assumptions \ref{cpo}, \ref{aoo}, \ref{kas}, and \ref{mse}, and given that $\|R\|$ is bounded almost surely, and that the smallest eigenvalue of $E[RR^\mathtt{T}]$ is bounded from below, as $|\mathcal{I}|,|\mathcal{I}'|\to\infty$,
\[
\sqrt{|\mathcal{I}|}(\hat\theta_\text{reg} - \theta_\text{reg}) \rightsquigarrow N(0, \operatorname{Var}(\varphi_\text{reg}(M, A, \tilde X;\theta_\text{reg}))).
\]
\end{restatable}

Now consider estimator $\hat\theta_\text{IV}$ of $\theta_\text{IV}$, the coefficients in a linear IV model with missing structured outcomes, per Table \ref{tab:mars_estimators}. Let $\varphi_\text{IV}$ be the corresponding EIF for $\theta_\text{IV}$.

\begin{restatable}{proposition}{Thmliniv}\label{thm:liniv}
Under Assumptions \ref{cpo}, \ref{aoo}, \ref{kas}, and \ref{mse}, and given that $\|Z\|$ is bounded almost surely, and that the smallest singular value of $E[ZR^\mathtt{T}]$ is bounded from below, as $|\mathcal{I}|,|\mathcal{I}'|\to\infty$,
\[
\sqrt{|\mathcal{I}|}(\hat\theta_\text{IV} - \theta_\text{IV}) \rightsquigarrow N(0, \operatorname{Var}(\varphi_\text{IV}(M, A, \tilde X;\theta_\text{IV}))).
\]
\end{restatable}

Finally, consider the estimator $\hat\theta_\text{diff}$ of  $\theta_\text{diff}$, the difference in conditional means of missing structured data, per Table \ref{tab:mars_estimators}. Let $\varphi_\text{diff}$ be the corresponding EIF for $\theta_\text{diff}$.

\begin{restatable}{proposition}{Thmdid}\label{thm:did}
Under Assumptions \ref{cpo}, \ref{aoo}, \ref{kas}, and \ref{mse}, so long as $p_{K=k}, p_{\tilde{K}=\tilde{k}}$ are bounded away from zero, we have that, as $|\mathcal{I}|,|\mathcal{I}'|\to\infty$,
\[
\sqrt{|\mathcal{I}|}(\hat\theta_\text{diff} - \theta_\text{diff}) \rightsquigarrow N(0, \text{Var}(\varphi_\text{diff}(M, A, \tilde X; \theta_\text{diff}))).
\]
\end{restatable}

\subsection{Graphical Models}

Per the discussion in Section \ref{mmd}, we have that $M^*$ can be thought of as a potential outcome in a causal inference framework, and, as such, we may also represent the core assumptions of the MAR-S framework as a (causal) graphical model. 

In particular, in Figure \ref{fig:swig}, we present examples of Single World Intervention Graphs (SWIGs) \citep{richardson_single_2013} compatible with MAR-S Assumption \ref{aoo}.\footnote{SWIGs are a proposal to concisely unify DAGs \citep{pearl_causal_1995} with the Rubin causal model.} The left-hand side SWIG in Figure \ref{fig:swig} could correspond to a setting where $X$ measures a household's membership to a politically significant group (e.g., membership in a particular ethnic group, determined at time $t=0$).\footnote{Consider, for example, \href{https://www.afrobarometer.org/surveys-and-methods/sampling}{Afrobarometer survey sampling methodology}.} $A$ is an indicator for being interviewed in an in-person household survey (at time $t=1$), $U$ is the roof (quality) of a household (e.g., measured using satellite data, at time $t=1$), and $M^*$ captures household income (at time $t=1$). The right-hand side SWIG could correspond to a setting where $U$ is a text, $X$ is a set of keywords that may or may not be present in that text, $A$ indicates validation, and $M^*$ is the sentiment represented in that text.

\begin{figure}[h!]
    \centering
    \begin{minipage}{0.45\textwidth}
    \centering
    \begin{tikzpicture}[
        line width=1pt,
        >=stealth,
        node distance=18mm and 18mm,
        ell/.style={draw, fill=white, inner sep=2pt, shape=ellipse}
    ]
    \tikzset{
      swig vsplit={
        gap=4pt
      }
    }
    \node[ell] (x) {$X$};
    \node[ell, right=of x] (u) {$U$};
    \node[shape=swig vsplit, below=of x] (a) {%
      \nodepart{left}{$A$}%
      \nodepart{right}{$a=1$}%
    };
    \node[ell, below=of u] (mstar) {$M^{*}$};
    \draw[->] (x) to[bend right=65] (a);
    %\draw[->] (mstar) -- (u);
    \draw[->] (a) -- (mstar);
    \draw[->] (x) -- (mstar);
    \draw[->] (x) -- (u); 
    \end{tikzpicture}
    \end{minipage}
    \begin{minipage}{0.45\textwidth}
    \centering
    \begin{tikzpicture}[
        line width=1pt,
        >=stealth,
        node distance=18mm and 18mm,
        ell/.style={draw, fill=white, inner sep=2pt, shape=ellipse}
    ]
    \tikzset{
      swig vsplit={
        gap=4pt
      }
    }
    \node[ell] (x) {$X$};
    \node[ell, right=of x] (u) {$U$};
    \node[shape=swig vsplit, below=of x] (a) {%
      \nodepart{left}{$A$}%
      \nodepart{right}{$a=1$}%
    };
    \node[ell, below=of u] (mstar) {$M^{*}$};
    \draw[->] (x) to[bend right=65] (a);
    \draw[->] (u) -- (mstar);
    \draw[->] (a) -- (mstar);
    \draw[->] (u) -- (x);
    \end{tikzpicture}
    \end{minipage}
    \caption{Examples of SWIGs compatible with the MAR-S framework.}
    \label{fig:swig}
\end{figure}

%\subsection{Comparison of Estimators}\label{sec:comparison}

\subsection{Independent Measurement Error Models}\label{sec:ime}

We return to the setting of Section \ref{sec:agg}, but now allow measurement errors to be independent and \textit{not} identically distributed:
\[
\bar{M}_{i,k} = \bar{M}_{i,k}^* + \upsilon_{i,k}, \quad \upsilon_{i,k} \overset{d}{\approx} N(0, \sigma^2_{\upsilon,i,k} ), \quad \sigma^2_{\upsilon,i,k} := |\mathcal{I}_{i,k}|^{-1}\text{Var}(\varphi)_{i,k},
\]
where $\text{Var}(\varphi)_{i,k}$ is the variance of the efficient influence function associated with $\bar{M}_{i,k}^*$.

We therefore will assume that
\[
\upsilon_i \overset{\text{ind}}{\sim} N(0, \Sigma_{i}), \quad \Sigma_{i} := \text{diag}(\sigma^2_{\upsilon,i,1} , \dots, \sigma^2_{\upsilon,i,K}),
\]
owing to the facts that (a) it will often be the case that each $\bar{M}_{i,k}$ will often be estimated on a separate sample, (b) for large $\mathcal{I}_{i,k}$ the normal approximation should be a reasonable one. We assume $\Sigma_i$ are known.

Let $\bar \Sigma := \frac{1}{n}\sum_{i=1}^n \Sigma_{i}$, and consider the estimator 
\[
\hat \beta := (\bar{M}^\prime\bar{M} - n\bar \Sigma)^{-1} \bar{M}^\prime Y.
\]
This estimator is consistent, as we can see that
\begin{align*}
    \hat \beta &:= \left(\bar{M}^\prime\bar{M} - n\bar \Sigma\right)^{-1} \bar{M}^\prime Y \\
     &= \left(\frac{1}{n} \bar{M}^\prime\bar{M} - \bar \Sigma\right)^{-1} \frac{1}{n} \bar{M}^\prime Y \\
     &= \left(\frac{1}{n} \sum_{i=1}^n \bar{M}_i \bar{M}_i^\prime - \bar \Sigma\right)^{-1} \frac{1}{n} \sum_{i=1}^n \bar{M}_iY_i \\
     &= \left(\frac{1}{n} \sum_{i=1}^n \bar{M}_i^* {\bar{M}_i^{*\prime}} + \frac{1}{n} \sum_{i=1}^n \upsilon_i {\bar{M}_i^{*\prime}}+\frac{1}{n} \sum_{i=1}^n \bar{M}_i^{*} {\upsilon_i}^\prime +\frac{1}{n} \sum_{i=1}^n \upsilon_i {\upsilon_i}^\prime  - \bar \Sigma\right)^{-1}  \left(\frac{1}{n}\sum_{i=1}^n \bar{M}^{*}_i Y_i + \frac{1}{n}\sum_{i=1}^n \upsilon_i Y_i\right)\\
     &\xrightarrow[p]{} \left(E[\bar{M}_i^*{\bar{M}_i^{*\prime}}]\right)^{-1}E[\bar{M}_i^* Y_i] = \beta.
\end{align*}
where the last line follows by the WLLN, the CMT, and Kolmogorov's variance criterion for averages (c.f., corollary 5.22, \cite{kallenberg_foundations_2021}), which yields
\[
\frac{1}{n} \sum_{i=1}^n \upsilon_i {\upsilon_i}^\prime   - \bar \Sigma = \frac{1}{n} \sum_{i=1}^n (\upsilon_i {\upsilon_i}^\prime - \Sigma_{i})\xrightarrow[p]{} 0
\]
assuming that, mildly, $\lim_{n\to\infty}\frac{1}{n^2} \sum_{i=1}^n \text{Var}(\upsilon_i {\upsilon_i}^\prime - \Sigma_{i}) < \infty$.

In order to understand the asymptotic variance of this estimator, note that
\begin{align*}
    \hat\beta &= \left(\frac{1}{n} \sum_{i=1}^n \bar{M}_i \bar{M}_i^\prime -  \bar\Sigma\right)^{-1} \frac{1}{n} \sum_{i=1}^n \bar{M}_iY_i \\
    &= \left(\frac{1}{n} \sum_{i=1}^n (\bar{M}_i \bar{M}_i^\prime - E[\bar{M}_i \bar{M}_i^\prime]) + E[\bar{M}_i^* \bar{M}_i^{*\prime}] \right)^{-1} \left(\frac{1}{n} \sum_{i=1}^n (\bar{M}_i Y_i - E[\bar{M}_i Y_i]) + E[\bar{M}_i^* Y_i]\right)  \\
    &= \left(\Delta_1 + \Omega \right)^{-1} (\Delta_2 + S) ,
\end{align*}
where $\Delta_1 := \frac{1}{n} \sum_{i=1}^n (\bar{M}_i \bar{M}_i^\prime - E[\bar{M}_i \bar{M}_i^\prime])$, $\Delta_2 := \frac{1}{n} \sum_{i=1}^n (\bar{M}_i Y_i - E[\bar{M}_i Y_i])$, $\Omega := E[\bar{M}_i^* \bar{M}_i^{*\prime}] $, and $S := E[\bar{M}_i^* Y]$. Thus, in this notation, $\beta = \Omega^{-1} S$. Noting that both $\Delta_1 = O_p(n^{-1/2})$ and $\Delta_2 = O_p(n^{-1/2})$ by the Lindeberg-Feller CLT, by a Taylor expansion argument, we have that 
\begin{align*}
    \hat\beta &=\left(\Delta_1 + \Omega \right)^{-1} (\Delta_2 + S) \\
    &=(\Omega^{-1} - \Omega^{-1}\Delta_1 \Omega^{-1} + O_p(n^{-1}) )(\Delta_2 + S) \\
    &=\Omega^{-1}\Delta_2  + \Omega^{-1}S - \Omega^{-1}\Delta_1 \Omega^{-1}S + O_p(n^{-1}) \\
    &=\beta + \Omega^{-1}\Delta_2 - \Omega^{-1}\Delta_1 \beta + O_p(n^{-1}) 
\end{align*}
and thus
\[
\hat\beta - \beta = \Omega^{-1}(\Delta_2 - \Delta_1 \beta) + O_p(n^{-1}).
\]
Substituting back in and scaling, we have that
\[
\sqrt{n}\left(\hat\beta - \beta\right) = \Omega^{-1}\frac{1}{\sqrt{n}} \sum_{i=1}^n \left(Z_i - E[Z_i]\right) + o_p(1),
\]
where $Z_i := \bar{M}_i (Y_i - \bar{M}_i^\prime\beta) \overset{\text{ind}}{\sim}$. Then by the Lindeberg-Feller CLT,
\[
\sqrt{n}\left(\hat\beta - \beta\right) \rightsquigarrow N(0, \Omega^{-1}\Psi \Omega^{-1}).
\]
where $\Psi := \lim_{n\to\infty} \frac{1}{n}\sum_i \text{Var}(Z_i)$, assuming this limit exists.

Let $\hat \Omega := \frac{1}{n} \sum_i (\bar{M}_i \bar{M}_i^\prime - \Sigma_i)$ and let $\hat \Psi := \frac{1}{n} \sum_{i=1}^n\left(\hat{Z}_i \hat{Z}_i^{\mathrm{T}}-\Sigma_i \hat{\beta} \hat{\beta}^{\mathrm{T}} \Sigma_i\right)$, for $\hat Z_i = \bar{M}_i (Y_i - \bar{M}_i^\prime \hat\beta)$.\footnote{The estimator $\hat \Psi := \frac{1}{n} \sum_{i=1}^n\left(\hat{Z}_i \hat{Z}_i^{\mathrm{T}}-\Sigma_i \hat{\beta} \hat{\beta}^{\mathrm{T}} \Sigma_i\right)$ is asymptotically equivalent to the estimator $\check \Psi := \frac{1}{n} \sum_{i=1}^n\left(\hat{Z}_i + \Sigma_i \hat\beta\right) \left(\hat{Z}_i + \Sigma_i \hat\beta\right)^\prime$, mirroring the form of the iid model's variance estimator.} Observe that, again by Kolmogorov's variance criterion for averages and the CMT, under mild conditions, we will have that
\[
\hat \Omega \xrightarrow[p]{} \Omega, \quad \hat \Psi \xrightarrow[p]{} \Psi.
\]
Then,
\[
\hat V^{-1/2}\sqrt{n}\left(\hat\beta - \beta\right) \rightsquigarrow N(0, I),
\]
where $\hat V := \hat \Omega^{-1}\hat \Psi \hat \Omega^{-1}$.

\subsection{Measurement Error Corrected Least Squares}\label{sec:bakermodel}

We consider the following regression from \cite{baker_measuring_2016}, the baseline specification in Table IV, column (5):
\[
Y_{it} = \beta_1 \Delta \log \mathrm{EPU}_t \times h_{it} + \beta_2 \Delta \frac{\text {Federal purchases}_t}{\text {GDP}_t} \times h_{it} + \alpha_i + \psi_t + u_{it},
\]
where $i$ indexes firms and $t$ indexes years; $\alpha_i$ and $\psi_t$ are firm and year fixed effects, respectively; $h_{it}$ is a firm-year policy exposure intensity variable, $Y_{it}$ is a firm-year employment growth variable, $\Delta \frac{\text {Federal purchases}_t}{\text {GDP}_t}$ is a relevant control variable; and $u_{it}$ is an error term. The parameter of interest is $\beta_1$.

We assume a short panel setup, with fixed $T$ and growing $N$ (i.e., fixed $T$, large $N$ asymptotics), where there may be arbitrary dependence across the same firm in time.

\paragraph{Re-labeling.} To harmonize notation with Section \ref{sec:mecols}, we denote within-firm transformed regressors as
\[\begin{aligned}
\bar{M}_{it,1}^*
&=
\Delta \log \mathrm{EPU}_t \times h_{it}
-\frac{1}{T}\sum_{s=1}^T
\Delta \log \mathrm{EPU}_s \times h_{is},\\
\bar{M}_{it,2}^*
&=
\Delta \frac{\text {Federal purchases}_t}{\text {GDP}_t} \times h_{it}
-\frac{1}{T}\sum_{s=1}^T
\Delta \frac{\text {Federal purchases}_s}{\text {GDP}_s} \times h_{is},
\end{aligned}\]
and so on, collecting all remaining regressors, including $T-1$ within-transformed year indicators and excluding firm indicators, in the fixed $K \times 1$ vector $\bar{M}_{it}^*$.

Similarly assuming $Y_{it},u_{it}$ are within-firm transformed, we can denote the moment condition
\[
g(Y_{it}, \bar{M}_{it}^*; \beta) = \bar{M}_{it}^*(Y_{it} - \bar{M}_{it}^{*\prime} \beta), \quad E[g(Y_{it}, \bar{M}_{it}^*; \beta)] = 0
\]
for $\beta = (\beta_1, \beta_2, \dots)^\prime$.

\paragraph{ME model.} Analysis in the main text, followed by the within-firm transformation, shows that
\[
\bar{M}_{it,1} = \bar{M}_{it,1}^* + \upsilon_{it,1}
\]
where
\[
\upsilon_{it,1}
:=
\check\upsilon_t \times h_{it}
-\frac{1}{T}\sum_{s=1}^T \check\upsilon_s \times h_{is},
\quad
\check\upsilon_t \sim N(0, \sigma_t^2 + \sigma_{t-1}^2),
\]
and $\operatorname{Cov}\left(\check\upsilon_t, \check\upsilon_{t+1}\right)=-\sigma_t^2$ and $\operatorname{Cov}\left(\check\upsilon_t, \check\upsilon_s\right)=0$ for $|t-s|>1$.

\paragraph{Matrix notation.} We let the $K\times 1$ vector of measurement errors simply be $\upsilon_{it} = (\upsilon_{it,1}, 0, \dots, 0)^\prime$. Then we may also write, collecting all $\check\upsilon_t$ as the jointly Gaussian random vector $\check\upsilon$,
\[
\upsilon_{it}=B_{it}\check\upsilon, \quad\bar{M}_{it}=\bar{M}_{it}^*+B_{it}\check\upsilon
\]
where $B_{it}$ is a $K \times T$ matrix equal to $\iota_1 b_{it}^{\prime}$, with $b_{it}
:=
h_{it}e_t
-\frac{1}{T}\sum_{s=1}^T h_{is}e_s$, using as notation $e_t$ the $t$-th standard basis vector in $\mathbb R^T$ and $\iota_1$ a $K\times 1$ vector with a one as its first element and zeros elsewhere.

\paragraph{Feasible GMM with ME.} Let $\mathcal{H}$ denote the information (sigma algebra) generated by the exposures. Since $B_{i t}$ is $\mathcal{H}$-measurable, and the mean-zero first-step error is assumed independent of $\mathcal{H}$,
\[
\Sigma_{i t}:=\operatorname{Var}\left(\upsilon_{i t} \mid \mathcal{H}\right)=E\left[\upsilon_{i t} \upsilon_{i t}^{\prime} \mid \mathcal{H}\right]=B_{i t} \Sigma_{\check{\upsilon}} B_{i t}^{\prime} .
\]

Weakly classical measurement error in the interacted MAR-S-first step regressor yields that $E[\upsilon_{it,1}\bar{M}_{it}^*]=0$. Also assuming $E[\upsilon_{it}u_{it}]=0$, we can consider the feasible moment condition, based on insights from Section \ref{sec:mecols},
\[
\tilde g(Y_{it}, \bar{M}_{it}; \beta) = \bar{M}_{it}(Y_{it} - \bar{M}_{it}^{\prime} \beta) + \Sigma_{it}\beta, \quad E[\tilde g(Y_{it}, \bar{M}_{it}; \beta)] = 0.
\]

\paragraph{Parameter Estimator.} This suggests the plug-in estimator
\[
\hat \beta = \left(\sum_{i,t}(\bar{M}_{it}\bar{M}_{it}^\prime - \Sigma_{it})\right)^{-1} \left(\sum_{i,t}\bar{M}_{it}Y_{it}\right) := \hat Q^{-1}\sum_{i,t}\bar{M}_{it}Y_{it}
\]
which satisfies
\[
\bar g(\hat\beta) := \sum_{i,t}\tilde g(Y_{it}, \bar{M}_{it}; \hat\beta) = 0.
\]

\paragraph{Inference.} Note that we may exactly write, since $b \mapsto \bar g(b)$ is affine,
\[
0 = \bar{g}(\beta) - \hat Q (\hat\beta - \beta).
\]
As such, we have that
\[
\sqrt{N} (\hat\beta - \beta) = (N\hat Q^{-1}) \frac{1}{\sqrt{N}}\bar g(\beta).
\]

We may then inspect the decomposition
$$
\bar{g}(\beta)=\sum_{i=1}^N \Phi_i+H_N \check{\upsilon}+R_N,
$$
where
$$
\Phi_i=\sum_{t=1}^T \bar{M}_{it}^* u_{i t},\quad
H_N=\sum_{i=1}^N \sum_{t=1}^T\left[B_{i t} u_{i t}-\bar{M}_{it}^* \beta^{\prime} B_{i t}\right], \quad R_N=-\sum_{i=1}^N \sum_{t=1}^T B_{i t}\left(\check{\upsilon} \check{\upsilon}^{\prime}-\Sigma_{\check{\upsilon}}\right) B_{i t}^{\prime} \beta .
$$

Note that the first term of the decomposition adheres to a standard firm-clustered CLT,
\[
\frac{1}{\sqrt{N}}\sum_{i=1}^N \Phi_i \rightsquigarrow N(0, S_{\mathrm{firm}}).
\]

For the second term, we assume that $N\Sigma_{\check{\upsilon}}$ converges to a finite matrix as $N\to\infty$, where dependence on $N$ is suppressed in the notation for $\Sigma_{\check{\upsilon}}$. This condition is justified by a proportional joint asymptotic thought experiment with $|\mathcal I_t|/N\to\text{constant}_t\in(0,\infty)$ for each MAR-S first-step. This assumption allows MAR-S first-step uncertainty to contribute at the same order as firm-level sampling uncertainty, while also permitting it to be negligible when the first-step sample sizes grow faster than $N$ in the relevant thought experiment. The condition only rules out first-step measurement-error (co)variance that decays more slowly than $N^{-1}$. This is comparable to modeling choices made in various other two-step estimation settings (see, e.g., \cite{newey_chapter_1994} Section 6.1, Theorem 6.1; footnote 2 of \cite{cocci_standard_2025}; various results from \cite{hahn_estimation_2020,hahn_central_2024}). By a firm-clustered WLLN, $\frac{H_N}{N} \xrightarrow{P} H$ for some $H$, and thus
$$
H_N \check{\upsilon}=N\left(\frac{H_N}{N}\right) \check{\upsilon}=N H \check{\upsilon}+o_P(N\|\check{\upsilon}\|).
$$
Then, by the Gaussian assumption and covariance matrix rate condition, $\check{\upsilon}=O_P(N^{-1/2})$, and so $H_N \check{\upsilon}=O_P(\sqrt{N})$.

For the third term, by the same fact, we have that $\check{\upsilon} \check{\upsilon}^{\prime}-\Sigma_{\check{\upsilon}}=O_P(N^{-1})$, and, provided $N^{-1}\sum_{i,t}\|B_{it}\|^2=O_P(1)$, we have that $R_N=O_P(1)$.

Therefore, we may write
\[
\frac{1}{\sqrt{N}} \bar{g}(\beta)=\frac{1}{\sqrt{N}} \sum_i \Phi_i+\left(\frac{H_N}{N}\right) \sqrt{N} \check{\upsilon}+o_P(1) .
\]
And so, by asymptotic independence between the first-step Gaussian error and the firm-level score, and provided $S_{\mathrm{firm}}+NH\Sigma_{\check{\upsilon}}H^\prime$ has a positive definite limit, using Slutsky
\[
\left[
S_{\mathrm{firm}}+NH\Sigma_{\check{\upsilon}}H^\prime
\right]^{-1/2}
\frac{1}{\sqrt{N}}\bar g(\beta)
\rightsquigarrow N(0,I_K).
\]
The preceding CLT and finite covariance limit imply that
$\bar g(\beta)/\sqrt{N}=O_P(1)$. Thus, using $\hat Q/N\xrightarrow{P}Q$
for a nonsingular $Q$, we obtain
\[
\sqrt{N}(\hat\beta-\beta)
=
\left(\frac{\hat Q}{N}\right)^{-1}
\frac{\bar g(\beta)}{\sqrt{N}}
=
Q^{-1}\frac{\bar g(\beta)}{\sqrt{N}}+o_P(1).
\]
Consequently,
\[
\left\{
Q^{-1}
\left[
S_{\mathrm{firm}}
+NH\Sigma_{\check{\upsilon}}H^\prime
\right]
(Q^{-1})^\prime
\right\}^{-1/2}
\sqrt{N}(\hat\beta-\beta)
\rightsquigarrow N(0,I_K).
\]
The implied standard error is therefore of the familiar two-step form of generated-regressor OLS \citep{newey_chapter_1994}.

\paragraph{Variance Estimator.} To define the plug-in variance estimator, let
\[
\hat u_{it}:=Y_{it}-\bar M_{it}^{\prime}\hat\beta,
\qquad
\hat s_i:=\sum_{t=1}^T\tilde g(Y_{it},\bar M_{it};\hat\beta),
\]
and define the unnormalized firm-clustered covariance sum and the estimated loading by
\[
\widehat S_{\mathrm{firm}}
:=
\sum_{i=1}^N\hat s_i\hat s_i^{\prime},
\qquad
\widehat H_N
:=
\sum_{i=1}^N\sum_{t=1}^T
\left[
B_{it}\hat u_{it}
-\bar M_{it}\hat\beta^{\prime}B_{it}
\right].
\]
Under the maintained moment and LLN conditions,
\[
\frac{\widehat S_{\mathrm{firm}}}{N}
\xrightarrow{P}S_{\mathrm{firm}},
\qquad
\frac{\widehat H_N}{N}
\xrightarrow{P}H,
\]
justifying the plug-in
\[
\widehat{\operatorname{Var}}(\hat\beta)
=
\hat Q^{-1}
\left[
\widehat S_{\mathrm{firm}}
+\widehat H_N\Sigma_{\check\upsilon}\widehat H_N^{\prime}
\right]
(\hat Q^{-1})^{\prime}.
\]

Indeed, under these conditions,
\[
N\widehat{\operatorname{Var}}(\hat\beta)
-
Q^{-1}
\left[
S_{\mathrm{firm}}+NH\Sigma_{\check{\upsilon}}H^\prime
\right]
(Q^{-1})^\prime
\xrightarrow{P}0,
\]
and hence
\[
\widehat{\operatorname{Var}}(\hat\beta)^{-1/2}
(\hat\beta-\beta)
\rightsquigarrow N(0,I_K),
\]
where the inverse square roots are the symmetric positive definite inverse square roots.

\subsection{Nonlinear Transformations of Aggregated Missing Data} \label{sec:app_aggregate}

When working with a large unstructured database, it is often the case that the functional of interest concerns the distribution of (a transformation of) missing structured data at an aggregate level, and the researcher does not observe any validation data at the aggregate level.

The aggregation of interest may be a nonlinear transformation of a linear combination of record-level structured data. We introduce the following assumption to represent this case.

\begin{definition}[Nonlinearly transformed linearly aggregated structured data]\label{nlagg}
For record-level structured data $M^*_{ij} \overset{\text{iid}}{\sim}$ and record count $N_j\overset{\text{iid}}{\sim}$, we define nonlinearly transformed linearly aggregated structured data as
\begin{align*}
    \tilde M_{j}^* := f(M_j^*), \quad M_j^* := \omega_j \sum_{i=1}^{N_j} M_{ij}^*.
\end{align*}
We further assume that $N_j \ind M_{ij}^*$, $\omega_j = h(N_j)$ for some measurable function $h$, and $f$ is thrice differentiable.
\end{definition}

A relevant example considers a researcher interested in inference on the logarithm of averaged structured data: 
\[ \theta = E[\tilde M_j^*] = E\left[\log\left(\frac{1}{N_j} \sum_{i=1}^{N_j} M_{ij}^*\right)\right].\]
In this setting, without validation data at the aggregate level $j$, we still do not recover regular point identification of the functional of interest. However, given sufficient smoothness of the transforming function, we can identify what may be a good approximation of $\theta$. Specifically, a Taylor expansion around $E[M_j^*]$ shows that
\begin{align*}
    \theta &= E\left[f\left(M_j^*\right)\right] \\
    &= f\left(E\left[M_j^*\right] \right)+\frac{1}{2}f^{\prime \prime}\left(E\left[M_j^*\right]\right) \text{Var}\left(M_j^* \right) + E[R(M_{j}^*)] \\
    &= f\left(\theta_1 \theta_2 \right)+\frac{1}{2}f^{\prime \prime}\left(\theta_1 \theta_2\right) \theta_3 + E[R(M_{j}^*)] \\
    &= \underline{\theta} + E[R(M_{j}^*)],
\end{align*}
where 
\[\underline{\theta} := f\left(\theta_1 \theta_2 \right)+\frac{1}{2}f^{\prime \prime}\left(\theta_1 \theta_2\right) \theta_3,\]
$ E[R(M_{j}^*)]$ is a remainder term, and
\[
\theta_3 := \text{Var}(\omega_j N_j) E[M_{ij}^*]^2 + E[\omega_j^2 N_j]\text{Var}(M_{ij}^*),
\]
which is regularly identified with validation data at record level $i$.\footnote{Notice, for example, when the aggregation of interest is an average, $\theta_3$ simplifies as
\[
\theta_3 = E[N_j^{-1}]\text{Var}(M_{ij}^*).
\]}

When is $\underline{\theta}$ close to $\theta$? In other words, when is this second order Taylor expansion a good approximation, or when is $E[R(M_{j}^*)]$ sufficiently small? Note that
\[
E[R(M_{j}^*)] = E\left[\frac{1}{6}f^{'''}\left(\xi\left(M_j^*\right)\right)\left(M_j^*-E[M_j^*]\right)^3\right]
\]
for $\xi(M_j^*)$ which lies between $M_j^*$ and $E[M_j^*]$ with probability 1, by the mean value representation of the Taylor remainder. We anticipate this term to be small, for example, when $\left|f^{'''}\left(\xi\left(M_j^*\right)\right)\right| < C$ with probability 1 (such as when $M_j^*$ is bounded with probability 1), and that $M_{j}^*$ has a law similar to a normal random variable (which we may indeed think is the case for averages, such as $\omega_j = 1/N_j$ and $N_j$ large), i.e., a small (absolute) third central moment:
\begin{align*}
    |E[R(M_{j}^*)]| &= |E\left[\frac{1}{6}f^{'''}\left(\xi\left(M_j^*\right)\right)\left(M_j^*-E[M_j^*]\right)^3\right]| \\
    &\leq \frac{C}{6}E\left[ \left|M_j^*-E[M_j^*]\right|^3\right].
\end{align*}

\subsection{Functionals for Linear Models with MAR-S Measurement Error}\label{sec:fme}

Consider the (univariate, for simplicity) linear regression coefficient functional:
\[
\theta = \theta_\text{den}^{-1} \theta_\text{num} = E[{M_j^*}^{2}]^{-1} E[Y_j M_j^*].
\]

We have that $A_j = 0$ w.p.1, i.e., we do not observe the researcher's target for missing structured data $M_j^*$, disabling identification in the traditional MAR-S framework. However, we do know that
\[
M_j^* := \frac{1}{N_j}\sum_{i=1}^{N_j} M_{ij}^* + \zeta_j
\]
where $\zeta_j$ is classical measurement error, and $N_j \sim_{iid}$, $N_j \ind (M_{ij}^*, Y_j)$. 
Then 
\begin{align*}
    \theta_\text{num} &= E\left[Y_j M_j^*\right] \\
    &= E\left[Y_j \left( \frac{1}{N_j}\sum_{i=1}^{N_j} M_{ij}^* + \zeta_j \right)\right] \\
    &= E\left[Y_j  \frac{1}{N_j}\sum_{i=1}^{N_j} M_{ij}^* \right] \\
    &= E\left[Y_j  M_{ij}^* \right]
\end{align*}
which is identified in the data using the MAR-S assumptions at the record level. Consider also that
\begin{align*}
    \theta_\text{den} &= E[{M_j^*}^{2}] \\
    &= E\left[\left(\frac{1}{N_j}\sum_{i=1}^{N_j} M_{ij}^* + \zeta_j\right)^2\right] \\
    &= E\left[\left(\frac{1}{N_j}\sum_{i=1}^{N_j} M_{ij}^*\right)^2\right] + E\left[\zeta_j^2\right] \\
    &= E\left[N_j^{-1}\right]\text{Var}\left(M_{i j}^*\right) +E\left[M_{i j}^*\right]^2+ E\left[\zeta_j^2\right].
\end{align*}
With assumed knowledge of the measurement error variance $E[\zeta_j^2]$, this functional is similarly identified. 

\subsection{MAR-S Mean Functionals}\label{sec:marsmeans}

An important and common class of functionals of interest accords to the following definition. This definition provides an alternative pathway to semiparametric efficient estimation under MAR-S, which some practitioners may find useful.

\begin{definition}\label{def:mmf}
A ``MAR-S mean functional'' is any functional that can be written as
\[
\theta(P) = E_P[ \tilde M^*]
\]
where $\tilde M^* = g(M^*, V)$ for a known deterministic function $g$ and known random variable $V$ (which is not itself a function of $\pi( X)$) and with $[V \ind A] \mid (X, U, M^*)$. The function $g$ is homogeneous of degree one in its first argument.
\end{definition}

Many interesting functionals of missing structured data can be written as MAR-S mean functionals.\footnote{Note that the definition of a MAR-S mean functional implies that $[(U,M^*,V) \ind A] \mid X$ by the contraction property of conditional independence.}
Focusing on this case is therefore intrinsically meaningful, but also permits further connections to related literatures that focus their attention on means, such as in causal inference \citep{aronow_nonparametric_2025} or semi-supervised learning \citep{zhang_double_2023}, meaning that insights in these literature can be directly applied to the present setting.

For MAR-S mean functionals, we have the following identification result.

\begin{restatable}{proposition}{Propid}\label{prop:id}
Under Assumptions \ref{cpo} and \ref{aoo}, a MAR-S mean functional $ E_P[ \tilde M^*]$ can be point identified as
\[
E_P[ \tilde M^*]=E_P[ \tilde \mu (\tilde X)],
\]
where $\tilde \mu (\tilde X) := E[\tilde M \mid A=1, \tilde X]$, for $\tilde M := g(M, V)$ and $\tilde X := (X, U, V)$.
\end{restatable}

We now state the following lemma, which greatly simplifies the derivations of efficient influence functions for the applications considered based on MAR-S mean functionals.\footnote{Similar identification results are possible by simply letting $\tilde X = X$ or $\tilde X = (X,V)$. The importance of incorporating unstructured data $U$ in $\tilde X$ is, naturally, to produce the most accurate possible imputations of the missing structured data, for which $X$ or $V$ may contain very little signal. The intuition mimics that of (nonparametric) regression adjustment. For formal results on the importance of imputation based on unstructured data in the settings we consider, see Appendix Section \ref{sec:fs}.}

\begin{restatable}{Lemma}{Lemmmf}\label{lem:mmf}
The efficient influence function of a point identified MAR-S mean functional $E_P[ \tilde \mu (\tilde X)]$ under a nonparametric statistical model $\mathcal{P} \ni P$ is the same as the efficient influence function of $E_P[ \tilde \mu (\tilde X)]$ under the semiparametric statistical model $\mathcal{P}_\pi \ni P$ induced by Assumption \ref{kas}.
\end{restatable}

Under the MAR-S framework, the statistical model is semiparametric when the validation score function $\pi$ is known. Deriving efficient influence functions under semiparametric models is typically more challenging than under fully nonparametric models, for which there is only one influence function, which is the efficient influence function. Lemma \ref{lem:mmf} shows that the EIF for a MAR-S mean functional $\theta$ under a nonparametric statistical model is also the correct EIF for $\theta$ under the semiparametric statistical model, simplifying derivation. 
Intuitively, perturbing the distribution given by the validation score does not change the value of the functional: if we had labeled the data in a different (but valid) way, the (in-population) value of the parameter being estimated would remain unchanged. This agrees with several related results in \cite{chen_semiparametric_2008}, e.g., the asymptotic variance lower bound of parameters identified using ``verify-in-sample'' auxiliary datasets is unaffected by knowledge of the ``propensity score'' (which is $1-\pi(x)$ in the MAR-S framework).

With Lemma \ref{lem:mmf}, we can compute the efficient influence function for a generic MAR-S mean functional, which we introduce in the next proposition.

\begin{restatable}{proposition}{Propmmfeif}\label{prop:mmfeif}
%\begin{proposition}\label{prop:mmfeif}
The efficient influence function for MAR-S mean functional $E_P\left[\tilde \mu(\tilde X)\right]$, under the semiparametric statistical model $\mathcal{P}_\pi \ni P$ induced by Assumption \ref{kas}, and under Assumption \ref{aoo}, is
\[ \varphi = \tilde \mu(\tilde X) + \frac{A}{\pi(X)}(\tilde M - \tilde \mu(\tilde X)) - \theta, \]
where $\pi$ is the validation score function.
\end{restatable}

\subsection{Feature Selection for Imputation}\label{sec:fs}

The MAR-S framework provides insights on the features that researchers should use to impute structured data, in order to achieve more efficient estimators. 

Point identification under Proposition \ref{prop:id} holds under many possible choices of $\tilde X$, so long as $\tilde X$ contains $X, V$ and any added components to $\tilde X$ are still independent of validation given $X$. The next two propositions formally help us unpack what a good choice of $\tilde X$ looks like, and why it should include $U$.

We know that the semiparametric variance lower bound for $\theta$ will be given by $\text{Var}(\varphi)$ for any given statistical model based on iid data $(\tilde M, A, \tilde X)$. In the following proposition, we highlight the dependence of $\text{Var}(\varphi)$ on $\tilde X$ itself, which echoes similar statements made by studies of (nonparametric) regression adjustment in experiments (e.g., \cite{cytrynbaum_optimal_2021,list_using_2024}).

\begin{restatable}{proposition}{Proprsq}\label{prop:r2}
Under Assumption \ref{aoo}, the semiparametric variance lower bound for functional $E_P\left[\tilde \mu(\tilde X_i)\right]$ is given by
 \[
\text{Var}(\varphi) \equiv \text{Var}(\varphi)(\tilde X) = \text{Var}(\tilde M^*) + E[(\pi(X)^{-1}-1)\text{Var}(\tilde M^* \mid \tilde X)].
 \]
Under a constant validation score $\pi$,
 \[
 \text{Var}(\varphi)(\tilde X) = \text{Var}(\tilde M^*)\left(1+ \left(\frac{1}{\pi}-1 \right) \left(1 - R^2(\tilde X) \right) \right),
 \]
where $R^2(\tilde X) := 1 - \frac{E[\operatorname{Var}(\tilde M^* \mid \tilde X)]}{\operatorname{Var}(\tilde M^*)} $, the nonparametric $R^2$.
\end{restatable}

The second expression in Proposition \ref{prop:r2} says that choices of $\tilde X$ that have higher nonparametric $R^2$ deflate the semiparametric variance lower bound, with the lowest possible bound achieved for by a $\tilde X$ that explains all variation in $\tilde M^*$, which recovers $\text{Var}(\varphi) = \text{Var}(\tilde M^*)$, the asymptotic variance we would expect if no data were missing. If $\pi = 1$, which is to say we validate all data, then we again recover $\text{Var}(\varphi) = \text{Var}(\tilde M^*)$, the asymptotic variance we would expect if no data were missing.

\subsection{Proofs}

\Corverifyjid*
\begin{proof}
We consider the case where functional/parameter of interest $\theta := \theta(P) \in \Theta \subseteq \mathbb{R}^{d_\theta}$ is defined via the moment condition(s)
\[
E_{P}[m(W ; \vartheta)]=0 \quad\text{ iff }\quad \vartheta = \theta
\]
where $W=(M^*, V)$ for some collection of non-missing random variables $V$ and $m(\cdot ; \vartheta)$ is a set of functions with dimension $d_m \geq d_\theta$. 

Define $\tilde X = (X,U,V)$. We may always write $m((M^*,V);\theta)=g((M^*,\tilde X);\theta) $ for some $g$ that simply ignores $(U,X)$, and so letting $\tilde W = (M^*, \tilde X)$, we have that
\[
E_{P}[g(\tilde W ; \vartheta)]=0 \quad\text{ iff }\quad \vartheta = \theta.
\]
Because $\tilde W_i \overset{\text{iid}}{\sim} P$, and using Assumption \ref{cpo}, we have recovered the exact setup of the ``verify-in-sample'' case of \cite{chen_semiparametric_2008}. 

Noting additionally the stated assumption that $\mathcal{J}_\theta$ has full column rank equal to $d_\theta$ and Assumption \ref{kas}, we recover all content of Assumption 1 of \cite{chen_semiparametric_2008} (for the verify-in-sample case); under Assumption \ref{aoo} 
%and reusing the conditional independence properties of $V$ outlined in Definition \ref{def:mmf}, 
we have that
\[
[M^* \ind A] \mid \tilde X
\]
thereby recovering Assumption 2 of \cite{chen_semiparametric_2008}.

Further define the conditional moment and the Jacobian
\[
\widetilde{\mathcal{E}}_P(\tilde X ; \theta):=E_P[g(\tilde W ; \theta) \mid \tilde X], \quad \widetilde{\mathcal{J}}_\theta(P)=\frac{\partial}{\partial \theta} E_P[g(\tilde W ; \theta)].
\]
Then Theorem 2 by way of Theorem 1 of \cite{chen_semiparametric_2008} immediately delivers that the efficient influence function when $\theta(P)$ is just identified is given by
\[
-\widetilde{\mathcal{J}}_\theta(P)^{-1} \left(\frac{A}{\pi(\tilde X)}[g(\tilde W ; \theta)-\widetilde{\mathcal{E}}_P(\tilde X; \theta )]+\widetilde{\mathcal{E}}_P(\tilde X;\theta)\right).
\]

We recall then that $m((M^*,V);\theta)=g((M^*,\tilde X);\theta) $, and additionally leverage the fact that $\pi(\tilde X)=\pi(X)$ almost surely under the MAR-S framework assumptions (as $(U,M^*,V) \ind A \mid X$), yielding the stated EIF.
\end{proof}

\Corverifyoid*
\begin{proof}
The structure of this proof follows identically to the proof of Corollary \ref{cor:verify_jid}, but instead uses the EIF for over-identified moments provided in the proof of Theorem 1 of \cite{chen_semiparametric_2008} (noting there is a typo in the EIF given above equation (29), which can be seen from a dimension conformity check, but does not substantively affect downstream results).
\end{proof}

\Wdrgmm*
\begin{proof}
Consider that for arbitrary choice of $\eta(\cdot;\vartheta)$ we have that
\begin{align*}
    E [m_{\eta}(W,A,\tilde X;\theta, P)] &= E\left[  \left(\frac{A}{\pi(X)}[m(W ; \theta)-\eta(\tilde X;\theta)]+\eta(\tilde X;\theta)\right)\right] \\
    &=  E\left[ \frac{A}{\pi(X)}m(W ; \theta)-\frac{A}{\pi(X)}\eta(\tilde X;\theta)+\eta(\tilde X;\theta)\right] \\
    &= E\left[ \frac{A}{\pi(X)}m(W ; \theta)+\left(1-\frac{A}{\pi(X)}\right)\eta(\tilde X;\theta)\right] \\
     &= E\left[ \frac{A}{\pi(X)}m(W ; \theta)\right]+E\left[\left(1-\frac{A}{\pi(X)}\right)\eta(\tilde X;\theta)\right].
\end{align*}
The second term in the parenthesis resolves to
\begin{align*}
    E\left[\left(1-\frac{A}{\pi(X)}\right)\eta(\tilde X;\theta)\right] &= E\left[E\left[\left(1-\frac{A}{\pi(X)}\right)\eta(\tilde X;\theta)\mid \tilde X\right]\right] \\
    &= E\left[\eta(\tilde X;\theta)E\left[\left(1-\frac{A}{\pi(X)}\right)\mid \tilde X\right]\right] \\
    &= E\left[\eta(\tilde X;\theta)\left(1-\frac{E[A\mid \tilde X]}{\pi(X)}\right)\mid \tilde X\right]  \\
    &= E\left[\eta(\tilde X;\theta)\cdot 0\mid \tilde X\right]  \tag{$\pi(\tilde X) = \pi(X)$ a.s.} \\
    &= 0,
\end{align*}
regardless of choice of $\eta$. Thus, if the first term in the parenthesis is equal to zero, then the stated conclusion has been proven. This is true, as we can see that
\begin{align*}
    E\left[ \frac{A}{\pi(X)}m(W ; \theta)\right] &= E\left[ E\left[\frac{A}{\pi(X)}m((M^*, V) ; \theta)\mid \tilde X\right]\right] \\
    &= E\left[\pi(X)^{-1} E\left[A\mid \tilde X\right]E\left[m((M^*, V) ; \theta)\mid \tilde X\right]\right] \\
    &= E\left[E\left[m((M^*, V) ; \theta)\mid \tilde X\right]\right] \\
    &= E\left[m((M^*, V) ; \theta)\right] \\
    &= 0.
\end{align*}
\end{proof}

\Thmeff*
\begin{proof}
We start by validating the consistency of $\hat\theta$. First note that, implicitly, for the just identified setting, the choice of weighting matrix will not end up mattering asymptotically, so implicitly we choose weighting matrix $\hat B := I_{d_m}$. 

To show consistency, we need to demonstrate or assume the assumptions of Theorem 2.6 of \cite{newey_chapter_1994}. Our data is iid by setup, and we may simply note $\hat B := I_{d_m} \xrightarrow[]{P} I_{d_m} := B$, and that $I_{d_m}$ is positive semi-definite. 

To establish the validity of the moment condition, note that if $\eta$ is fixed, then
by Theorem \ref{prop:wdrgmm} it is immediate that the moment condition holds only at $\theta$. If the nuisance is a function of the tuning split $\{(M_i, A_i, \tilde X_i)\}_{i\in \mathcal{I}'}$, which we may denote $\hat\eta$, then under the tuning set conditional law
\begin{align*}
    E[m_{\hat\eta}(W,A,\tilde X;\theta)\mid \hat\eta = \eta] 
    &=E[m_{\eta}(W,A,\tilde X;\theta)\mid \hat\eta = \eta] \\
    &=E[m_{\eta}(W,A,\tilde X;\theta)] \tag{Independent samples} \\
    &= 0 \tag{Theorem \ref{prop:wdrgmm}}
\end{align*}
and so the moment condition still holds. The by applying conditions (i), (ii), and (iii), we satisfy all assumptions of \cite{newey_chapter_1994} Theorem 2.6 are still satisfied, immediately yielding consistency: $\hat\theta \xrightarrow[]{P} \theta$.

With consistency established, we now consider asymptotic normality. To do so, we need to satisfy the assumptions of Theorem 3.2 of \cite{newey_chapter_1994}. Note that $\hat\theta$ already takes the form described in this theorem, and that the identity weighting matrix still satisfies the required regularity conditions, and that we have already shown consistency.

By condition (v), the moment function has a finite second moment at $\theta$. Using conditional i.i.d.\ sampling and $E_P[m_\eta(M,A,\tilde X;\theta)]=0$, the
multivariate Lindeberg-Levy CLT gives
\[
\sqrt{|\mathcal I|}\,\bar m_\eta(\theta)
\rightsquigarrow
N\!\left(
    0,
    E_P\!\left[
        m_\eta(M,A,\tilde X;\theta)
        m_\eta(M,A,\tilde X;\theta)^\prime
    \right]
\right).
\]

For $\vartheta\in\mathcal N$, define
\[
J_\eta(\vartheta)
:=
E_P\left[
    \frac{\partial}{\partial\vartheta^\prime}
    m_\eta(M,A,\tilde X;\vartheta)
\right].
\]
By conditions (iv) and (vi), dominated convergence implies that $J_\eta$ is continuous on $\mathcal N$. Moreover, since $\mathcal N$ is compact, conditions (iv) and (vi) allow us to apply Lemma 2.4 of \cite{newey_chapter_1994}, yielding
\[
\sup_{\vartheta\in\mathcal N}
\left\|
    \frac{\partial}{\partial\vartheta^\prime}
    \bar m_\eta(\vartheta)
    -
     J_\eta(\vartheta)
\right\|
\xrightarrow{P}0.
\]
Conditions (iii), (iv), and (vi) also justify differentiation under the expectation at $\theta$, so that
\[
 J_\eta(\theta)
=
\left.
\frac{\partial}{\partial\vartheta^\prime}
E_P[m_\eta(M,A,\tilde X;\vartheta)]
\right|_{\vartheta=\theta}
=
\mathcal J_{\eta,\theta}.
\]

By condition (vii) and $B=I_{d_m}$,
\[
J_\eta(\theta)^\prime B J_\eta(\theta)
=
\mathcal J_{\eta,\theta}^\prime\mathcal J_{\eta,\theta}
\]
is nonsingular. Condition (i) ensures that $\theta$ is an
interior point of $\Theta$, and condition (iv) ensures that $\bar m_\eta$ is continuously differentiable in a neighborhood
of $\theta$ almost surely. Together with consistency and the central limit theorem, uniform convergence, and continuity established above, these verify all the conditions of
Theorem 3.2 of \cite{newey_chapter_1994}. Thus,
\[
\sqrt{|\mathcal{I}|}(\hat\theta - \theta)\rightsquigarrow N(0,(\mathcal{J}_{\eta,\theta}^\prime B\mathcal{J}_{\eta,\theta})^{-1} \mathcal{J}_{\eta,\theta}^\prime BE[m_\eta(W,A,\tilde X;\theta)m_\eta(W,A,\tilde X;\theta)^\prime]B \mathcal{J}_{\eta,\theta} (\mathcal{J}_{\eta,\theta}^\prime B\mathcal{J}_{\eta,\theta})^{-1}).
\]
However, because $d_m = d_\theta$, we have that
\[
\sqrt{|\mathcal{I}|}(\hat\theta - \theta)\rightsquigarrow N(0,{J}_{\eta,\theta}^{-1}E[m_\eta(W,A,\tilde X;\theta)m_\eta(W,A,\tilde X;\theta)^\prime] (\mathcal{J}_{\eta,\theta}^\prime)^{-1})
\]
which is the stated result.
\end{proof}

\Thmdesc*
\begin{proof}
To prove a CLT for $\hat\theta$ and show that it attains asymptotic efficiency, we will (1) introduce an ``oracle'' version of the MAR-S estimator where $\mu$ is known and show that this oracle estimator attains the semiparametric efficiency lower bound and then (2) show that the feasible MAR-S estimator is asymptotically equivalent to the oracle estimator under our assumptions.

We now introduce the oracle estimator
\[ \tilde \theta := \frac{1}{|\mathcal{I}|}\sum_{i \in \mathcal{I}} \left[ \mu(\tilde X_i) + \frac{A_i}{\pi(X_i)}(M_i - \mu(\tilde X_i)) \right].\]
Notice that we can re-write the oracle estimator as
\[ \tilde \theta = \theta + \frac{1}{|\mathcal{I}|}\sum_{i \in \mathcal{I}} \varphi(M_i, A_i, \tilde X_i;\theta).\]
Rearranging and scaling, we have
\[ \sqrt{|\mathcal{I}|}\left(\tilde \theta - \theta\right) =  \frac{1}{\sqrt{|\mathcal{I}|}}\sum_{i \in \mathcal{I}} \varphi(M_i, A_i, \tilde X_i;\theta)\]
By application of Lindeberg-L\'evy CLT for iid data, we have that
\[
\sqrt{|\mathcal{I}|}\left(\tilde \theta - \theta\right) \rightsquigarrow N(0, \text{Var}(\varphi(M_i, A_i, \tilde X_i;\theta))),
\]
and as such we can see that $\tilde \theta$ is semiparametric efficient. 

We now want to show that, under our assumptions,
\[
\sqrt{|\mathcal{I}|}\left(\hat \theta - \tilde \theta\right)  \xrightarrow[]{P} 0 
\]
where, as introduced earlier,
\[ \hat \theta = \frac{1}{|\mathcal{I}|}\sum_{i \in \mathcal{I}} \left[ \hat\mu(\tilde X_i) + \frac{A_i}{\pi(X_i)}(M_i - \hat\mu(\tilde X_i)) \right]\]
and $\hat \mu$ is a random function of $\{(M_i, A_i, \tilde X_i)\}_{i \in \mathcal{I}'}$, the tuning sample. In particular, we will leverage Chebyshev's inequality, for which we need to show the mean of this scaled difference is zero, and its variance goes to zero in large samples.

To start, we can write
\begin{align*}
    \hat \theta - \tilde \theta &= \frac{1}{\left|\mathcal{I}\right|}\sum_{i\in \mathcal{I}} \left[\hat\mu(\tilde X_i) + \frac{A_i}{\pi(X_i)}\left(M_i - \hat\mu(\tilde X_i)\right) - \mu(\tilde X_i) - \frac{A_i}{\pi(X_i)}\left(M_i - \mu(\tilde X_i)\right)\right] \\
    &= \frac{1}{\left|\mathcal{I}\right|}\sum_{i\in \mathcal{I}} \left[\left(\hat\mu(\tilde X_i) - \mu(\tilde X_i)\right)\left(1-\frac{A_i}{\pi(X_i)}\right)\right].
\end{align*}
Consider the expectation of this difference,
\begin{align*}
    E\left[\hat \theta - \tilde \theta\right] 
    &= E\left[\frac{1}{\left|\mathcal{I}\right|}\sum_{i\in \mathcal{I}} \left[\left(\hat\mu(\tilde X_i) - \mu(\tilde X_i)\right)\left(1-\frac{A_i}{\pi(X_i)}\right)\right]\right] \\
    &= \frac{1}{\left|\mathcal{I}\right|}\sum_{i\in \mathcal{I}} E\left[\left(\hat\mu(\tilde X_i) - \mu(\tilde X_i)\right)\left(1-\frac{A_i}{\pi(X_i)}\right)\right].
\end{align*}
Further considering an arbitrary summand (and abusing notation slightly, where conditioning on $\mathcal{I}'$ indicates conditioning on the tuning dataset), we have that
\begin{align*}
    E\left[\left(\hat\mu(\tilde X_i) - \mu(\tilde X_i)\right)\left(1-\frac{A_i}{\pi(X_i)}\right)\right] 
    &= E\left[E\left[\left(\hat\mu(\tilde X_i) - \mu(\tilde X_i)\right)\left(1-\frac{A_i}{\pi(X_i)}\right)\mid \tilde X_i, \mathcal{I}' \right] \right]\\
    &= E\left[\left(\hat\mu(\tilde X_i) - \mu(\tilde X_i)\right)E\left[\left(1-\frac{A_i}{\pi(X_i)}\right)\mid \tilde X_i, \mathcal{I}'\right] \right] \\
    &= E\left[\left(\hat\mu(\tilde X_i) - \mu(\tilde X_i)\right)\left(1-\frac{E\left[A_i\mid \tilde X_i, \mathcal{I}'\right]}{\pi(X_i)}\right)\right]  \\
    &= E\left[\left(\hat\mu(\tilde X_i) - \mu(\tilde X_i)\right)\left(1-\frac{E\left[A_i\mid X_i\right]}{\pi(X_i)}\right)\right]  \\
    &= 0,
\end{align*}
and as such it is immediate that $E\left[\hat \theta - \tilde \theta\right]=0$, and thus that $E\left[\sqrt{|\mathcal{I}|}\left(\hat \theta - \tilde \theta\right)\right]=0$. Note that this holds in part because we already know
\[
%[A_i \ind (U_i, V_i) ] \mid X_i, \quad \{W_{i'}\}_{i'\in \mathcal{I}'}  \ind (X_i,U_i, V_i, A_i)
[A_i \ind U_i ] \mid X_i, \quad \{(M_i, A_i, \tilde X_i)\}_{i \in \mathcal{I}'}  \ind (X_i,U_i, A_i)
\]
so by the weak union and contraction properties of conditional independence
\[
[A_i \ind (U_i,\{(M_i, A_i, \tilde X_i)\}_{i \in \mathcal{I}'}) ] \mid X_i.
\]

Now, computing the variance of this difference, we have
\begin{align*}
    &\text{Var}\left[\hat \theta - \tilde \theta\right] \\
    &=\text{Var}\left[\frac{1}{\left|\mathcal{I}\right|}\sum_{i\in \mathcal{I}} \left[\left(\hat\mu(\tilde X_i) - \mu(\tilde X_i)\right)\left(1-\frac{A_i}{\pi(X_i)}\right)\right]\right] \\
    &= \frac{1}{\left|\mathcal{I}\right|^2}\text{Var}\left[\sum_{i\in \mathcal{I}} \left[\left(\hat\mu(\tilde X_i) - \mu(\tilde X_i)\right)\left(1-\frac{A_i}{\pi(X_i)}\right)\right]\right] \\
    &= \frac{1}{\left|\mathcal{I}\right|^2}E\left[\text{Var}\left[\left(\sum_{i\in \mathcal{I}} \left[\left(\hat\mu(\tilde X_i) - \mu(\tilde X_i)\right)\left(1-\frac{A_i}{\pi(X_i)}\right)\right]\right)\mid \{\tilde X_i\}, \mathcal{I}'\right]\right] \\
    &= \frac{1}{\left|\mathcal{I}\right|}E\left[\text{Var}\left[\left(\left(\hat\mu(\tilde X_i) - \mu(\tilde X_i)\right)\left(1-\frac{A_i}{\pi(X_i)}\right)\right)\mid \tilde X_i, \mathcal{I}'\right]\right] \\
    &= \frac{1}{\left|\mathcal{I}\right|}E\left[E\left[\left(\left(\hat\mu(\tilde X_i) - \mu(\tilde X_i)\right)\left(1-\frac{A_i}{\pi(X_i)}\right)\right)^2\mid \tilde X_i, \mathcal{I}'\right]\right] \\
    &= \frac{1}{\left|\mathcal{I}\right|}E\left[\left(\hat\mu(\tilde X_i) - \mu(\tilde X_i)\right)^2 \frac{1-\pi(X_i)}{\pi(X_i)}\right] \\
    &\leq \frac{1}{ \left|\mathcal{I}\right|}\frac{1-\delta}{\delta}E\left[\left(\hat\mu(\tilde X_i) - \mu(\tilde X_i)\right)^2\right] \tag{Assumption \ref{kas}} \\
    &= \frac{1}{ \left|\mathcal{I}\right|}\frac{1-\delta}{\delta}E\left[E\left[\left(\hat\mu(\tilde X_i) - \mu(\tilde X_i)\right)^2 \mid \mathcal{I}'\right]\right] \\
    &= \frac{1}{ \left|\mathcal{I}\right|}\frac{1-\delta}{\delta}E\left[\int\left(\hat\mu(\tilde x) - \mu(\tilde x)\right)^2 dP_{\tilde X}(\tilde x)\right] \tag{Data splitting} \\
    &= o(\left|\mathcal{I}\right|^{-1})\tag{Assumption \ref{mse}}
.\end{align*}
As such, it is immediate that
\[
\text{Var}\left[\sqrt{|\mathcal{I}|}\left(\hat \theta - \tilde \theta\right)\right] = |\mathcal{I}|\text{Var}\left[\left(\hat \theta - \tilde \theta\right)\right] = o(1). 
\]
Then, by Chebyshev's inequality, we have indeed shown that
\[
\sqrt{|\mathcal{I}|}\left(\hat \theta - \tilde \theta\right)  \xrightarrow[]{P} 0.
\]

Accordingly, we have that
\[
\sqrt{|\mathcal{I}|}\left(\hat \theta - \theta \right) = \sqrt{|\mathcal{I}|}\left(\tilde \theta - \theta \right) + o_p(1) \rightsquigarrow N(0, \text{Var}(\varphi(M_i, A_i, \tilde X_i;\theta)))
\]
by Slutsky's theorem.
\end{proof}

\Thmlin*
\begin{proof}
The task of this proof simply resolves to showing that $\hat \theta$ is asymptotically equivalent to $\tilde \theta$ using an argument based on Chebyshev's inequality (as the oracle estimator $\tilde \theta$ is immediately seen to be semiparametric efficient). Write
\[
\Delta_i
:=
R_i\{\hat\mu(\tilde X_i)-\mu(\tilde X_i)\}
\left(1-\frac{A_i}{\pi(X_i)}\right).
\]
Then
\[
\hat\theta-\tilde\theta
=
\hat Q^{-1}\frac{1}{|\mathcal I|}\sum_{i\in\mathcal I}\Delta_i,
\qquad
\hat Q:=\frac1{|\mathcal I|}\sum_{i\in\mathcal I}R_iR_i^\prime .
\]
Notice that, for any fixed $v$,
\begin{align*}
    E\left[\sqrt{|\mathcal I|}v^\prime(\hat \theta - \tilde \theta)\right] 
    &= E\left[
    v^\prime\hat Q^{-1}
    \frac{1}{\sqrt{|\mathcal I|}}\sum_{i\in \mathcal{I}}
    R_i(\hat\mu(\tilde X_i)-\mu(\tilde X_i))
    \left(1-\frac{A_i}{\pi(X_i)}\right)
    \right] \\
    &= E\left[
    v^\prime\hat Q^{-1}
    \frac{1}{\sqrt{|\mathcal I|}}\sum_{i\in \mathcal{I}}
    E\left\{
    R_i(\hat\mu(\tilde X_i)-\mu(\tilde X_i))
    \left(1-\frac{A_i}{\pi(X_i)}\right)
    \mid \tilde X_i,\mathcal I'
    \right\}
    \right].
\end{align*}
Moreover,
\begin{align*}
     E\left[ R_i\left(\hat\mu(\tilde X_i) - \mu(\tilde X_i)\right)\left(1-\frac{A_i}{\pi(X_i)}\right)\mid \tilde X_i, \mathcal{I}' \right] 
    &=  R_i\left(\hat\mu(\tilde X_i) - \mu(\tilde X_i)\right)E\left[\left(1-\frac{A_i}{\pi(X_i)}\right)\mid \tilde X_i, \mathcal{I}' \right]  \\
    &=  R_i\left(\hat\mu(\tilde X_i) - \mu(\tilde X_i)\right)\left(1-\frac{E\left[A_i\mid \tilde X_i, \mathcal{I}'\right]}{\pi(X_i)}\right)  \\
    &= R_i\left(\hat\mu(\tilde X_i) - \mu(\tilde X_i)\right)\left(1-\frac{E\left[A_i\mid \tilde X_i\right]}{\pi(X_i)}\right) \\
    &= R_i\left(\hat\mu(\tilde X_i) - \mu(\tilde X_i)\right)\left(1-\frac{E\left[A_i\mid  X_i\right]}{\pi(X_i)}\right)  \\
    &= 0,
\end{align*}
and so,
\[
E\left[\sqrt{|\mathcal I|}v^\prime(\hat \theta - \tilde \theta)\right] = 0.
\]

Finally, observe that, conditional on $\{\tilde X_i:i\in\mathcal I\}$ and $\mathcal I'$, the summands $\Delta_i$ are conditionally independent and conditionally mean-zero. Hence
\begin{align*}
&\operatorname{Var}\left[
\sqrt{|\mathcal I|}v^\prime(\hat\theta-\tilde\theta)
\mid \{\tilde X_i\}_{i\in\mathcal I},\mathcal I'
\right] \\
&=
v^\prime\hat Q^{-1}
\left\{
\frac1{|\mathcal I|}\sum_{i\in\mathcal I}
R_iR_i^\prime
\{\hat\mu(\tilde X_i)-\mu(\tilde X_i)\}^2
\operatorname{Var}\left[
1-\frac{A_i}{\pi(X_i)}
\mid \tilde X_i,\mathcal I'
\right]
\right\}
\hat Q^{-1}v \\
&=
v^\prime\hat Q^{-1}
\left\{
\frac1{|\mathcal I|}\sum_{i\in\mathcal I}
R_iR_i^\prime
\{\hat\mu(\tilde X_i)-\mu(\tilde X_i)\}^2
E\left[
\left(1-\frac{A_i}{\pi(X_i)}\right)^2
\mid \tilde X_i,\mathcal I'
\right]
\right\}
\hat Q^{-1}v \\
&=
v^\prime\hat Q^{-1}
\left\{
\frac1{|\mathcal I|}\sum_{i\in\mathcal I}
R_iR_i^\prime
\{\hat\mu(\tilde X_i)-\mu(\tilde X_i)\}^2
\frac{1-\pi(X_i)}{\pi(X_i)}
\right\}
\hat Q^{-1}v \\
&\leq
\frac{1-\delta}{\delta}
v^\prime\hat Q^{-1}
\left\{
\frac1{|\mathcal I|}\sum_{i\in\mathcal I}
R_iR_i^\prime
\{\hat\mu(\tilde X_i)-\mu(\tilde X_i)\}^2
\right\}
\hat Q^{-1}v
\tag{Assumption \ref{kas}}.
\end{align*}

Let
\[
\mathcal T:=\{\lambda_{\min}(\hat Q)\geq c_2\}.
\]
By the WLLN, $\hat Q \xrightarrow[]{P} E[R_i R_i^\prime]:=Q$, and by assumption $\lambda_{\min}(Q)\geq c_1$, and thus $P(\mathcal T^c)\to 0$. On $\mathcal T$, by operator norm bound, 
$\|\hat Q^{-1}v\|^2\leq c_2^{-2}\|v\|^2$. Therefore, using Cauchy-Schwarz and noting that $0 \leq \mathbf 1_{\mathcal T} \leq 1$,
\begin{align*}
& E\left[
\left\{\sqrt{|\mathcal I|}v^\prime(\hat\theta-\tilde\theta)\right\}^2
\mathbf 1_{\mathcal T}
\right] \\
&=
E\left[
\operatorname{Var}\left\{
\sqrt{|\mathcal I|}v^\prime(\hat\theta-\tilde\theta)
\mid \{\tilde X_i\}_{i\in\mathcal I},\mathcal I'
\right\}
\mathbf 1_{\mathcal T}
\right] \\
&\leq
\frac{1-\delta}{\delta}
E\left[
v^\prime\hat Q^{-1}
\left\{
\frac1{|\mathcal I|}\sum_{i\in\mathcal I}
R_iR_i^\prime
\{\hat\mu(\tilde X_i)-\mu(\tilde X_i)\}^2
\right\}
\hat Q^{-1}v
\mathbf 1_{\mathcal T}
\right] \\
&\leq
\frac{1-\delta}{\delta}
c_2^{-2}\|v\|^2
E\left[
\frac1{|\mathcal I|}\sum_{i\in\mathcal I}
\|R_i\|^2
\{\hat\mu(\tilde X_i)-\mu(\tilde X_i)\}^2
\right].
\end{align*}
If $\|R_i\|$ is uniformly bounded, say $\|R_i\|\leq c_3$, then
\begin{align*}
& E\left[
\left\{\sqrt{|\mathcal I|}v^\prime(\hat\theta-\tilde\theta)\right\}^2
\mathbf 1_{\mathcal T}
\right] \\
&\leq
\frac{1-\delta}{\delta}
c_2^{-2}c_3^2\|v\|^2
E\left[
\frac1{|\mathcal I|}\sum_{i\in\mathcal I}
\{\hat\mu(\tilde X_i)-\mu(\tilde X_i)\}^2
\right].
\end{align*}
Using the independence of $\mathcal I$ and $\mathcal I'$,
\begin{align*}
E\left[
\frac1{|\mathcal I|}\sum_{i\in\mathcal I}
\{\hat\mu(\tilde X_i)-\mu(\tilde X_i)\}^2
\right]
&=
E\left[
\int
\left(\hat\mu(\tilde x)-\mu(\tilde x)\right)^2
dP_{\tilde X}(\tilde x)
\right] \\
&=o(1).
\end{align*}
Hence
\[
E\left[
\left\{\sqrt{|\mathcal I|}v^\prime(\hat\theta-\tilde\theta)\right\}^2
\mathbf 1_{\mathcal T}
\right]
=o(1).
\]
Therefore, for every $\varepsilon>0$,
\begin{align*}
P\left(
\left|
\sqrt{|\mathcal I|}v^\prime(\hat\theta-\tilde\theta)
\right|>\varepsilon
\right)
&\leq
P(\mathcal T^c)
+
P\left(
\left|
\sqrt{|\mathcal I|}v^\prime(\hat\theta-\tilde\theta)
\right|>\varepsilon,
\mathcal T
\right) \\
&\leq
P(\mathcal T^c)
+
\frac{
E\left[
\left\{\sqrt{|\mathcal I|}v^\prime(\hat\theta-\tilde\theta)\right\}^2
\mathbf 1_{\mathcal T}
\right]
}{\varepsilon^2} \\
&\to 0.
\end{align*}
Therefore,
\[
\sqrt{|\mathcal I|}v^\prime(\hat\theta-\tilde\theta)=o_P(1)
\]
for every fixed $v$. By the Cramér-Wold device, this implies
\[
\sqrt{|\mathcal I|}(\hat\theta-\tilde\theta)=o_P(1).
\]
Thus $\hat\theta$ and $\tilde\theta$ are asymptotically equivalent. Since $\tilde\theta$ is semiparametric efficient, the same asymptotic distribution and efficiency bound carry over to $\hat\theta$.

\end{proof}

\Thmliniv*
\begin{proof}
The proof of this proposition is identical to that of Proposition \ref{thm:lin}, owing to their identical structure, and is immediate as such (the natural conditions on $E[ZR^\prime]$ replace those on $E[RR^\prime]$, and conditions on $Z$ replace those on $R$).
\end{proof}

\Thmdid*
\begin{proof}
In order to prove the efficiency of $\hat\theta$, we may proceed by: (1) proving the efficiency of $(\hat\theta_{K=k},\hat\theta_{\tilde K=\tilde k})$; and (2) applying the delta method to get a limiting distribution for $\hat\theta$.

Defining the oracle estimator
\[
\tilde\theta_{K=k} = |\mathcal{I}|^{-1} \sum_{i\in \mathcal{I}} \frac{\mathbf{1}_{K=k,i}}{p_{K=k}} \left(\frac{A_i}{\pi( X_i)}[M_i- \mu_{K=k}(\tilde X_i)]+\mu_{K=k}(\tilde X_i)\right).
\]
it is immediate by the Lindeberg-Levy CLT that $\tilde\theta_{K=k}$ is a semiparametric efficient estimator. Then, once again, we must simply prove that $\sqrt{|\mathcal{I}|}(\hat\theta - \tilde\theta) = o_P(1)$. To start, note that
\begin{align*}
    \hat \theta - \tilde \theta 
    &= \frac{1}{\left|\mathcal{I}\right|}\sum_{i\in \mathcal{I}} \frac{\mathbf{1}_{K=k,i}}{p_{K=k}}\left[\left(\hat\mu_{K=k}(\tilde X_i) - \mu_{K=k}(\tilde X_i)\right)\left(1-\frac{A_i}{\pi(X_i)}\right)\right].
\end{align*}
Consider the expectation of this difference,
\begin{align*}
    E\left[\hat \theta - \tilde \theta\right] 
    &= \frac{1}{\left|\mathcal{I}\right|}\sum_{i\in \mathcal{I}} E\left[\frac{\mathbf{1}_{K=k,i}}{p_{K=k}}\left(\hat\mu_{K=k}(\tilde X_i) - \mu_{K=k}(\tilde X_i)\right)\left(1-\frac{A_i}{\pi(X_i)}\right)\right].
\end{align*}
Further considering an arbitrary summand (and abusing notation slightly, where conditioning on $\mathcal{I}'$ indicates conditioning on the tuning dataset), we have that
\begin{align*}
    &E\left[\frac{\mathbf{1}_{K=k,i}}{p_{K=k}}\left(\hat\mu_{K=k}(\tilde X_i) - \mu_{K=k}(\tilde X_i)\right)\left(1-\frac{A_i}{\pi(X_i)}\right)\right] \\
    =& E\left[E\left[\frac{\mathbf{1}_{K=k,i}}{p_{K=k}}\left(\hat\mu_{K=k}(\tilde X_i) - \mu_{K=k}(\tilde X_i)\right)\left(1-\frac{A_i}{\pi(X_i)}\right)\mid \tilde X_i, \mathcal{I}' \right] \right]\\
    =& E\left[\frac{\mathbf{1}_{K=k,i}}{p_{K=k}}\left(\hat\mu_{K=k}(\tilde X_i) - \mu_{K=k}(\tilde X_i)\right)E\left[\left(1-\frac{A_i}{\pi(X_i)}\right)\mid \tilde X_i, \mathcal{I}'\right] \right] \\
    =& E\left[\frac{\mathbf{1}_{K=k,i}}{p_{K=k}}\left(\hat\mu_{K=k}(\tilde X_i) - \mu_{K=k}(\tilde X_i)\right)\left(1-\frac{E\left[A_i\mid \tilde X_i, \mathcal{I}'\right]}{\pi(X_i)}\right)\right]  \\
    =& E\left[\frac{\mathbf{1}_{K=k,i}}{p_{K=k}}\left(\hat\mu_{K=k}(\tilde X_i) - \mu_{K=k}(\tilde X_i)\right)\left(1-\frac{E\left[A_i\mid X_i\right]}{\pi(X_i)}\right)\right]  \\
    =& 0,
\end{align*}
and as such it is immediate that $E\left[\hat \theta - \tilde \theta\right]=0$, and thus that $E\left[\sqrt{|\mathcal{I}|}\left(\hat \theta - \tilde \theta\right)\right]=0$. Note that this holds in part because we already know
\[
%[A_i \ind (U_i, V_i) ] \mid X_i, \quad \{W_{i'}\}_{i'\in \mathcal{I}'}  \ind (X_i,U_i, V_i, A_i)
[A_i \ind U_i ] \mid X_i, \quad \{(M_i, A_i, \tilde X_i)\}_{i \in \mathcal{I}'}  \ind (X_i,U_i, A_i)
\]
so by the weak union and contraction properties of conditional independence
\[
[A_i \ind (U_i,\{(M_i, A_i, \tilde X_i)\}_{i \in \mathcal{I}'}) ] \mid X_i.
\]

Now, computing the variance of this difference, we have
\begin{align*}
    &\text{Var}\left[\hat \theta - \tilde \theta\right] \\
    &=\text{Var}\left[\frac{1}{\left|\mathcal{I}\right|}\sum_{i\in \mathcal{I}} \frac{\mathbf{1}_{K=k,i}}{p_{K=k}}\left[\left(\hat\mu_{K=k}(\tilde X_i) - \mu_{K=k}(\tilde X_i)\right)\left(1-\frac{A_i}{\pi(X_i)}\right)\right]\right] \\
    &= \frac{1}{\left|\mathcal{I}\right|^2}\text{Var}\left[\sum_{i\in \mathcal{I}} \frac{\mathbf{1}_{K=k,i}}{p_{K=k}}\left[\left(\hat\mu_{K=k}(\tilde X_i) - \mu_{K=k}(\tilde X_i)\right)\left(1-\frac{A_i}{\pi(X_i)}\right)\right]\right] \\
    &= \frac{1}{\left|\mathcal{I}\right|^2}E\left[\text{Var}\left[\left(\sum_{i\in \mathcal{I}}\frac{\mathbf{1}_{K=k,i}}{p_{K=k}} \left[\left(\hat\mu_{K=k}(\tilde X_i) - \mu_{K=k}(\tilde X_i)\right)\left(1-\frac{A_i}{\pi(X_i)}\right)\right]\right)\mid \{\tilde X_i\}, \mathcal{I}'\right]\right] \\
    &= \frac{1}{\left|\mathcal{I}\right|}E\left[\text{Var}\left[\left(\frac{\mathbf{1}_{K=k,i}}{p_{K=k}}\left(\hat\mu_{K=k}(\tilde X_i) - \mu_{K=k}(\tilde X_i)\right)\left(1-\frac{A_i}{\pi(X_i)}\right)\right)\mid \tilde X_i, \mathcal{I}'\right]\right] \\
    &= \frac{1}{\left|\mathcal{I}\right|}E\left[E\left[\left(\frac{\mathbf{1}_{K=k,i}}{p_{K=k}}\left(\hat\mu_{K=k}(\tilde X_i) - \mu_{K=k}(\tilde X_i)\right)\left(1-\frac{A_i}{\pi(X_i)}\right)\right)^2\mid \tilde X_i, \mathcal{I}'\right]\right] \\
    &= \frac{1}{\left|\mathcal{I}\right|}E\left[\frac{\mathbf{1}_{K=k,i}}{p_{K=k}^2}\left(\hat\mu_{K=k}(\tilde X_i) - \mu_{K=k}(\tilde X_i)\right)^2 \frac{1-\pi(X_i)}{\pi(X_i)}\right] \\
    &\leq \frac{1}{ \left|\mathcal{I}\right|}\frac{1-\delta}{\delta}p_{K=k}^{-2}E\left[\left(\hat\mu_{K=k}(\tilde X_i) - \mu_{K=k}(\tilde X_i)\right)^2\right] \tag{Assumption \ref{kas}} \\
    &= \frac{1}{ \left|\mathcal{I}\right|}\frac{1-\delta}{\delta}p_{K=k}^{-2}E\left[E\left[\left(\hat\mu_{K=k}(\tilde X_i) - \mu_{K=k}(\tilde X_i)\right)^2 \mid \mathcal{I}'\right]\right] \\
    &= \frac{1}{ \left|\mathcal{I}\right|}\frac{1-\delta}{\delta}p_{K=k}^{-2}E\left[\int\left(\hat\mu_{K=k}(\tilde x) - \mu_{K=k}(\tilde x)\right)^2 dP_{\tilde X}(\tilde x)\right] \tag{Data splitting} \\
    &= o(\left|\mathcal{I}\right|^{-1})\tag{Assumption \ref{mse}}
.\end{align*}
As such, it is immediate that
\[
\text{Var}\left[\sqrt{|\mathcal{I}|}\left(\hat \theta - \tilde \theta\right)\right] = |\mathcal{I}|\text{Var}\left[\left(\hat \theta - \tilde \theta\right)\right] = o(1). 
\]
Then, by Chebyshev's inequality, we have indeed shown that
\[
\sqrt{|\mathcal{I}|}\left(\hat \theta - \tilde \theta\right)  \xrightarrow[]{P} 0.
\]

% Indeed, we may use an identical proof for the efficiency of $\hat\theta_G$ as for the estimator of the numerator functional in Proposition \ref{thm:lin}, but substituting ${C}_{i,j}^\perp$ for $G_{i2} P(G_{i2} = 1)^{-1}$ (which is bounded almost surely without assumption), $\mu$ for $\mu_G$, and using the appropriate definition of $\tilde X_i$. So too does the proof of the efficiency of $\hat\theta_C$ proceed identically, given that it has identical structure to $\hat\theta_G$, simply replacing conditioning event $\{G_{i2}=1\}$ with $\{C_i=1\}$ in the proof. 

The output of the application of these identically structured proofs yields that, for the vector-valued estimators and estimand
\[
\boldsymbol{\hat\theta} := \begin{pmatrix}
\hat \theta_{K=k} \\
\hat \theta_{\tilde{K}=\tilde{k}}
\end{pmatrix}, \quad \boldsymbol{\tilde \theta} := \begin{pmatrix}
\tilde \theta_{K=k} \\
\tilde \theta_{\tilde{K}=\tilde{k}}
\end{pmatrix}, \quad\boldsymbol{\theta} := \begin{pmatrix}
 \theta_{K=k} \\
 \theta_{\tilde{K}=\tilde{k}}
\end{pmatrix},
\]
we can write
\[
\sqrt{|\mathcal{I}|}
(\boldsymbol{\hat\theta}-\boldsymbol{\theta}) = \sqrt{|\mathcal{I}|}
(\boldsymbol{\tilde\theta}-\boldsymbol{\theta}) + o_P(1) \rightsquigarrow N \left( 0, E[\boldsymbol{\varphi} \boldsymbol{\varphi}^\prime]\right)
\]
where $\boldsymbol{\varphi} := \begin{pmatrix}
    \varphi_{K=k} \\ \varphi_{\tilde{K}=\tilde{k}}
\end{pmatrix}$.

Lastly, note that 
\[
\hat \theta = \phi( \boldsymbol{\hat\theta}), \quad  \theta = \phi( \boldsymbol{\theta}),
\]
for $\phi(a,b) = a-b$, so, by application of the delta method, we have that
\[
\sqrt{|\mathcal{I}|}(\hat\theta - \theta) \rightsquigarrow N(0, \text{Var}(\varphi_{K=k} - \varphi_{\tilde{K}=\tilde{k}})).
\]

%The proof of the efficiency of the estimator in Proposition \ref{thm:rdd} follows by corollary, owing to its identical structure, i.e., simply substituting $G_{i2}  P(G_{i2} = 1)^{-1}$ for $\mathbb{I}\{R_i \in \mathcal{B}, D_i=1\} P(R_i \in \mathcal{B}, D_i=1)^{-1}$ and $C_{i}  P(C_{i} = 1)^{-1}$ for $\mathbb{I}\{R_i \in \mathcal{B}, D_i=0\} P(R_i \in \mathcal{B}, D_i=0)^{-1}$; and $\mu_G$ for $\mu_1$ and $\mu_C$ for $\mu_0$ in the above proof.
\end{proof}

\Propid*
\begin{proof}
We define a MAR-S mean functional as any functional that can be written as 
\[\theta(P) = E_P[\tilde M^*]\]
where $\tilde M^* = g(M^*, V)$ and where $g$ is a known deterministic function and $V$ is a known random variable. When $g$ is homogeneous of degree 1 in its first argument, under Assumption \ref{cpo} we have that
\[ A \tilde M^* = A g(M^*,V) = g(AM^*, V) = g(M,V) := \tilde M. \]
Under Assumption \ref{aoo}, and because $[V \ind A] \mid (X,U,M^*)$, by the contraction property of conditional independence we have that
\[ [(U,M^*,V) \ind A] \mid X. \]
Using the weak union property of conditional independence, then, $[(M^*,U,V) \ind A] \mid (X,U,V) $, and finally by the decomposition property of conditional independence:
\[ [g(M^*, V) \ind A] \mid (X, U, V).\]
Labeling $\tilde X := (X,U,V)$, we can also write this as
\[ [\tilde M^* \ind A] \mid \tilde X.\]

As such, for a MAR-S mean functional point identification can always be achieved as
\begin{align*}
    \theta &= E[\tilde M^*] \\
    &= E[E[\tilde M^*\mid \tilde X]] \\
    &= E[E[\tilde M^*\mid A=1, \tilde X]] \\
   &= E[E[\tilde M\mid A=1, \tilde X]]  \\
   &= E[\tilde \mu(\tilde X)] 
\end{align*}
where $\tilde \mu(\tilde X) := E[\tilde M \mid A=1 , \tilde X]$. Explicitly, we can write
\[
\theta = \int_{\tilde{\mathcal{X}}} \left(\int_{\tilde{\mathcal{M}}} \tilde m \, p(\tilde m\mid 1, \tilde x)\, d\tilde m \right) p(\tilde x)\, d\tilde x,
\]
a form we will rely on in subsequent proofs.
\end{proof}

\Lemmmf*
\begin{proof}
Let us first assume that $\mathcal{P} \ni P$ is a fully nonparametric statistical model, which is to say Assumption \ref{kas} does not hold. Under $\mathcal{P}$, without any restrictions, $P$ generically factors as:
\[ P(\tilde M, A, \tilde X) = P(\tilde M \mid A, \tilde X) P(A \mid \tilde X) P(\tilde X).\]
The tangent space $\mathcal{T}$ (i.e., the set of all valid scores $h$ associated with $P$ and $\mathcal{P}$) for this model can be constructed as the orthogonal sum of the tangent spaces for each factorized component of the model:
\begin{align*}
    \mathcal{T} &= \mathcal{T}_{\tilde M \mid A , \tilde X} \oplus \mathcal{T}_{ A \mid \tilde X} \oplus \mathcal{T}_{\tilde X} \\
    \mathcal{T}_{\tilde X} &= \{h_{\tilde X}(\tilde x) \in L^2(P) : E_P[h_{\tilde X}(\tilde X)] = 0 \} \\
    \mathcal{T}_{ A \mid \tilde X} &= \{h_{A\mid \tilde X}(a, \tilde x) \in L^2(P) : E_P[h_{A\mid \tilde X}( A , \tilde X)\mid \tilde X] = 0 \} \\
    \mathcal{T}_{\tilde M \mid A , \tilde X} &= \{h_{\tilde M \mid A , \tilde X}(\tilde m, a, \tilde x) \in L^2(P) : E_P[h_{\tilde M \mid A , \tilde X}(\tilde M, A, \tilde X)\mid A , \tilde X] = 0 \}.
\end{align*}
A score $h$ under this nonparametric model can therefore be factorized as
\[ h(\tilde M, A, \tilde X) = h_{\tilde M \mid A, \tilde X}(\tilde M, A, \tilde X) + h_{A \mid \tilde X}(A, \tilde X) + h_{\tilde X}(\tilde X),\]
where all summands are pairwise orthogonal.

Denoting a pathwise derivative with respect to $h$ as $\nabla_h$, we then have that, by linearity of pathwise derivatives,
\begin{align*}
\nabla_h \theta &= \nabla_{ h_{\tilde M \mid A, \tilde X}} \theta + \nabla_{h_{A \mid \tilde X}} \theta + \nabla_{h_{\tilde X}} \theta.
\end{align*}
Crucially, notice then that $\nabla_{h_{A\mid \tilde X}} \theta = 0$ for any score $h$ at $P$:
\begin{align*}
    \nabla_{h_{A\mid \tilde X}} \theta &= \frac{d}{d \epsilon} \theta(P_{\epsilon,h_{A\mid \tilde X}}) \\
    &= \frac{d}{d \epsilon}\int_{\tilde{\mathcal{X}}} \left(\int_{\tilde{\mathcal{M}}} \tilde m \, p(\tilde m\mid 1, \tilde x)\, d\tilde m \right) p(\tilde x)\, d\tilde x \\
    &=  0 .
\end{align*}
As such, we have that
\begin{align*}
\nabla_h \theta &= \nabla_{h_{\tilde M \mid A,\tilde X}} \theta +  \nabla_{h_{\tilde X}} \theta
\end{align*}
for any score $h$. 

Alternatively, if we had made Assumption \ref{kas}, then our statistical model would be semiparametric, which we can denote $\mathcal{P}_\pi$, where the distribution factorizes as
\begin{align*}
    P(\tilde M, A, \tilde X) &= P(\tilde M \mid A, \tilde X) (A\pi(\tilde X)+(1-A)(1-\pi(\tilde X))) P(\tilde X)
\end{align*}
for known validation score function $\pi$. For $\mathcal{P}_\pi$ and $P \in \mathcal{P}_\pi$, the corresponding tangent space $\mathcal{T}_\pi$ therefore factorizes as
\begin{align*}
    \mathcal{T}_\pi &= \mathcal{T}_{\tilde M \mid A , \tilde X} \oplus \mathcal{T}_{\tilde X}.
\end{align*}
Any score $h_\pi$ under this semiparametric model can therefore be factorized as
\[ h_\pi(\tilde M, A, \tilde X) = h_{\tilde M \mid A, \tilde X}  + h_{\tilde X},\]
where the summands are again orthogonal. 
Again by the linearity of the pathwise derivative operator, we have that
\[
\nabla_{h_\pi} \theta =  \nabla_{h_{\tilde M \mid A,\tilde X}} \theta +  \nabla_{h_{\tilde X}} \theta.
\]
As such, we have learned that
\begin{align*}
\boxed{\nabla_h \theta =  \nabla_{h_{\tilde M \mid A,\tilde X}} \theta +  \nabla_{h_{\tilde X}} \theta = \nabla_{h_\pi} \theta.}
\end{align*}

By the definition of pathwise differentiability, we know that
\[
\nabla_{h_\pi} \theta = E[\varphi_\pi h_\pi]
\]
holds, where $\varphi_\pi$ is the efficient influence function for $\theta(P)$ under $P \in \mathcal{P}_\pi$, by the Riesz representation theorem. However, if $\varphi_\pi$ is the EIF under $\mathcal{P}_\pi$, then it lies in the tangent space $\mathcal{T}_\pi$, which is orthogonal to  $\mathcal{T}_{A\mid \tilde X}$ and all of the scores $h_{A\mid \tilde X}$ in it. This means we can write
\begin{align*}
    \nabla_{h_\pi} \theta &= E[\varphi_\pi h_\pi] + 0 \\
    &= E[\varphi_\pi h_\pi] + E[\varphi_\pi h_{A\mid \tilde X}] \\
    &= E[\varphi_\pi (h_\pi + h_{A\mid \tilde X})] \\
    &= E[\varphi_\pi h].
\end{align*}

Putting everything together then, we have that
\[
\nabla_h \theta = \nabla_{h_\pi} \theta = E[\varphi_\pi h] .
\]
Importantly then,
\[ \boxed{\nabla_h \theta  = E[\varphi_\pi h]} \]
which, again by the definition of pathwise differentiability, tells us that the EIF for the semiparametric model $\mathcal{P}_\pi$ induced by Assumption \ref{kas}, which is $\varphi_\pi$, is the same as the nonparametric EIF, which is the random variable that would satisfy this equality. 

As a result, we conclude that it suffices to find the EIF for a given MAR-S mean functional $\theta(P)$ under a nonparametric model $\mathcal{P}$ in order to find the EIF for $\theta(P)$ under the semiparametric model $\mathcal{P}_\pi$ induced by Assumption \ref{kas}.\end{proof}

\Propmmfeif*
\begin{proof}

In order to derive the efficient influence function for functional $\theta (P) = E_P\left[\tilde{\mu}(\tilde X)\right]$ under the semiparametric statistical model $\mathcal{P}_\pi \ni P$ induced by Assumption \ref{kas}, we first first find the EIF for $E_P\left[\tilde{\mu}(\tilde X)\right]$ under a nonparametric statistical model $\mathcal{P} \ni P$, and then apply Lemma \ref{lem:mmf}.

To find the nonparametric EIF for the functional of interest, we will (1) find a candidate EIF by treating the all data as discrete and making use of the influence function operator  \citep{kennedy_semiparametric_2023}, and then (2) check this is indeed a valid influence function in the general, non-discrete case using the definition of pathwise differentiability.

Treating all data as discrete, application of the influence function operator to our functional of interest yields:
\begin{align*}
    \IF(E_P\left[\tilde{\mu}(\tilde X)\right]) &= \IF(\sum_{\tilde x \in \mathcal{\tilde X}} \tilde{\mu}(\tilde x)P(\tilde X=\tilde x)) \\
    &= \sum_{\tilde x \in \mathcal{\tilde X}} \IF(\tilde{\mu}(\tilde x)P(\tilde X=\tilde x)) \\
    &= \sum_\mathcal{\tilde X} \left[ \IF(\tilde{\mu}(\tilde x))P(\tilde X=\tilde x) +  \tilde{\mu}(\tilde x)\IF (P(\tilde X=\tilde x))\right] \\
    &= \sum_{\tilde x \in \mathcal{\tilde X}} \left[ \frac{\mathbb{I}\{\tilde X=\tilde x, A=1\}}{P(\tilde X=\tilde x, A=1)} (\tilde{M} - \tilde{\mu}(\tilde x)) P(\tilde X=\tilde x) +  \tilde{\mu}(\tilde x)(\mathbb{I}\{\tilde X=\tilde x\}-P(\tilde X=\tilde x))\right] \\
    &= \sum_{\tilde x \in \mathcal{\tilde X}} \left[ \frac{A\mathbb{I}\{\tilde X=\tilde x\}}{P(\tilde X=\tilde x, A=1)} (\tilde{M} - \tilde{\mu}(\tilde x)) P(\tilde X=\tilde x) +  \tilde{\mu}(\tilde x)\mathbb{I}\{\tilde X=\tilde x\}-\tilde{\mu}(\tilde x)P(\tilde X=\tilde x)\right] \\
    &=  \frac{A}{P(\tilde X, A=1)} (\tilde{M} - \tilde{\mu}(\tilde X)) P(\tilde X) +  \tilde{\mu}(\tilde X)-E\left[\tilde{\mu}(\tilde X)\right] \\
    &=  \frac{A}{P(A=1 \mid \tilde X)} (\tilde{M} - \tilde{\mu}(\tilde X)) +  \tilde{\mu}(\tilde X)-E\left[\tilde{\mu}(\tilde X)\right] \\
    &=  \frac{A}{\pi(\tilde X)} (\tilde{M} - \tilde{\mu}(\tilde X)) +  \tilde{\mu}(\tilde X)-\theta. 
\end{align*}
Rearranging, our candidate EIF $\varphi$ (which would be the nonparametric EIF if all data were discrete) is
\[
\varphi :=  \tilde{\mu}(\tilde X) + \frac{A}{\pi(\tilde X)} (\tilde{M} - \tilde{\mu}(\tilde X)) -\theta.
\]

We now apply the definition of pathwise differentiability to check that this is indeed the (efficient) influence function for $\theta$. In particular, the definition of pathwise differentiability tells us that 
\[
\nabla_h \theta = E[\varphi h].
\]
For the left-hand side, we have that
\[
\nabla_h \theta = \frac{d}{d\epsilon} \int_{\mathcal{\tilde X}} \left(\int_\mathcal{\tilde{M}} \tilde{m} \, p(\tilde{m}\mid 1, \tilde x)(1+\epsilon h_{\tilde{M}\mid A, \tilde X})\,d\tilde{m} \right) p(\tilde x)(1+\epsilon h_{\tilde X})\,d\tilde x
\]
and for the right-hand side, we have that
\begin{align*}
    E[\varphi h] &= E\left[ \left( \tilde{\mu}(\tilde X) + \frac{A}{\pi(\tilde X)} (\tilde{M} - \tilde{\mu}(\tilde X)) -\theta \right)h\right] \\
    &= E\left[ \left( \tilde{\mu}(\tilde X) + \frac{A}{\pi(\tilde X)} (\tilde{M} - \tilde{\mu}(\tilde X)) \right)h\right] - \theta E[ h] \\
    &= E\left[ \left( \tilde{\mu}(\tilde X) + \frac{A}{\pi(\tilde X)} (\tilde{M} - \tilde{\mu}(\tilde X)) \right)h\right] \\
    &= E\left[ \tilde{\mu}(\tilde X)h\right] + E\left[\frac{A}{\pi(\tilde X)} (\tilde{M} - \tilde{\mu}(\tilde X)) h\right] \\
    &= E\left[ \tilde{\mu}(\tilde X)E[h\mid \tilde X]\right] + E\left[E\left[(\tilde{M} - \tilde{\mu}(\tilde X))h\frac{A}{\pi(\tilde X)} \mid \tilde X \right]\right]\\
    &= E\left[ \tilde{\mu}(\tilde X)E[h\mid \tilde X]\right] + E\left[E\left[(\tilde{M} - \tilde{\mu}(\tilde X))h\frac{1}{\pi(\tilde X)} \mid \tilde X, A= 1 \right]P(A=1 \mid \tilde X)\right] \\
    &= E\left[ \tilde{\mu}(\tilde X)E[h\mid \tilde X]\right] + E\left[E\left[(\tilde{M} - \tilde{\mu}(\tilde X))h \mid \tilde X, A= 1 \right]\right].
\end{align*}
As discussed in Lemma \ref{lem:mmf}, we know that in a fully nonparametric model $\mathcal{P}$ we can factorize the score as
\[ h( \tilde{M}, A, \tilde X) = h_{\tilde{M} \mid A, \tilde X}( \tilde{M}, A, \tilde X) + h_{A \mid  \tilde X}(A ,  \tilde X) + h_{\tilde X}( \tilde X).\]
As such, we have that
\begin{align*}
    E[\varphi h] = E[\varphi h_{\tilde{M} \mid A, \tilde X}] + E[\varphi h_{A \mid  \tilde X}] + E[\varphi h_{\tilde X}]
\end{align*}
and by linearity of the pathwise derivative we also have that
\begin{align*}
\nabla_h \theta &= \nabla_{h_{\tilde{M} \mid A, \tilde X}} \theta + \nabla_{h_{A\mid  \tilde X}} \theta + \nabla_{h_{\tilde X}} \theta.
\end{align*}
Checking equality summand by summand, we have:
\begin{itemize}
    \item $h_{\tilde X}$ direction: 
    \begin{align*}
        E[\varphi h_{\tilde X}(\tilde X)] &= E\left[ \tilde{\mu}(\tilde X)h_{\tilde X}(\tilde X)\right] + E\left[E\left[(\tilde{M} - \tilde{\mu}(\tilde X)) \mid \tilde X, A= 1 \right] h_{\tilde X}(\tilde X)\right] = E\left[ \tilde{\mu}(\tilde X)h_{\tilde X}(\tilde X)\right] \\
        \nabla_{h_{\tilde X}} \theta &= \int_{\mathcal{\tilde X}} \left(\int_\mathcal{\tilde{M}} \tilde{m} p(\tilde{m}\mid1, \tilde x)\,d\tilde{m} \right)  h_{\tilde X}(\tilde x) p(\tilde x)\,d\tilde x = E[\tilde{\mu}(\tilde X)h_{\tilde X}(\tilde X)]
    \end{align*}
    and so $\nabla_{h_{\tilde X}} \theta = E[\varphi h_{\tilde X}( \tilde X)] = E[\tilde{\mu}(\tilde X)h_{\tilde X}(\tilde X)]$.
    \item $h_{A\mid \tilde X}$ direction: 
    \begin{align*}
        E[\varphi h_{A\mid \tilde X}] &= E\left[ \tilde{\mu}(\tilde X)E[h_{A\mid \tilde X}\mid \tilde X]\right] + E\left[E\left[(\tilde{M} - \tilde{\mu}(\tilde X))h_{A\mid \tilde X} \mid \tilde X, A= 1 \right]\right]  \\
        &= E\left[E\left[(\tilde{M} - \tilde{\mu}(\tilde X)) \mid \tilde X, A= 1 \right]h_{A\mid \tilde X}\right] = 0  \\
        \nabla_{h_{A\mid \tilde X}} \theta &= \frac{d}{d\epsilon} \int_{\mathcal{\tilde X}} \left(\int_\mathcal{\tilde{M}} \tilde{m} p(\tilde{m}\mid1, \tilde x)\,d\tilde{m} \right) p(\tilde x)\,d\tilde x = 0
    \end{align*}
    and so $\nabla_{h_{A\mid \tilde X}} \theta = E[\varphi h_{A\mid \tilde X}]=0$.
    \item $h_{\tilde{M}\mid A, \tilde X}$ direction: 
    \begin{align*}
        E[\varphi h_{\tilde{M}\mid A, \tilde X}] &= E\left[ \tilde{\mu}(\tilde X)E[h_{\tilde{M}\mid A, \tilde X}\mid \tilde X]\right] + E\left[E\left[(\tilde{M} - \tilde{\mu}(\tilde X))h_{\tilde{M}\mid A, \tilde X} \mid \tilde X, A= 1 \right]\right]  \\
         &= E\left[ \tilde{\mu}(\tilde X)E[h_{\tilde{M}\mid A, \tilde X}\mid \tilde X]\right] + E\left[E\left[\tilde{M}h_{\tilde{M}\mid A, \tilde X}\mid \tilde X, A= 1 \right]\right]  \\
         &= E\left[ \tilde{\mu}(\tilde X)E[E[h_{\tilde{M}\mid A, \tilde X}\mid \tilde X, A]\mid \tilde X]\right] + E\left[E\left[\tilde{M}h_{\tilde{M}\mid A, \tilde X}\mid \tilde X, A= 1 \right]\right]  \\
         &= E\left[E\left[\tilde{M}h_{\tilde{M}\mid A, \tilde X}\mid \tilde X, A= 1 \right]\right]  \\
        \nabla_{h_{\tilde{M} \mid A, \tilde X}} \theta &= \frac{d}{d\epsilon} \int_{\mathcal{\tilde X}} \left(\int_\mathcal{\tilde{M}} \tilde{m} (1+\epsilon h_{\tilde{M}\mid A, \tilde X}(\tilde{m}, 1, \tilde x))p(\tilde{m}\mid1, \tilde x)\,d\tilde{m} \right) p(\tilde x)\,d\tilde x \\
        &= \frac{d}{d\epsilon} \int_{\mathcal{\tilde X}} \left(\int_\mathcal{\tilde{M}} \tilde{m} \epsilon h_{\tilde{M}\mid A, \tilde X}(\tilde{m}, 1, \tilde x)p(\tilde{m}\mid1, \tilde x)\,d\tilde{m} \right) p(\tilde x)\,d\tilde x \\
        &= \int_{\mathcal{\tilde X}} \left(\int_\mathcal{\tilde{M}} \tilde{m}  h_{\tilde{M}\mid A, \tilde X}(\tilde{m}, 1, \tilde x)p(\tilde{m}, 1, \tilde x)\,d\tilde{m} \right) p(\tilde x)\,d\tilde x \\
        &= E[E[\tilde{M}h_{\tilde{M}\mid A, \tilde X}\mid A= 1, \tilde X]]
    \end{align*}
    and so $\nabla_{h_{\tilde{M}\mid A, \tilde X}} \theta = E[\varphi h_{\tilde{M}\mid A, \tilde X}] =  E[E[\tilde{M}h_{\tilde{M}\mid A, \tilde X}\mid A= 1, \tilde X]]$.
\end{itemize}
Because all summands are equal, the definition of pathwise differentiability holds for the candidate $\varphi$, and so indeed 
\[
\varphi :=  \tilde{\mu}(\tilde X) + \frac{A}{\pi(\tilde X)} (\tilde{M} - \tilde{\mu}(\tilde X)) -\theta
\]
is the influence function (or curve) for $\theta$, and because $\mathcal{P}$ is nonparametric, it is also the EIF.

By Lemma \ref{lem:mmf}, we also know that $\varphi$ is the EIF under the semiparametric model $\mathcal{P}_\pi$ induced by Assumption \ref{kas}.

Finally, through Assumption \ref{aoo} and the definition of the MARS mean functional, we have that
\[
\pi(\tilde X) := P(A=1 \mid \tilde X) = P(A=1 \mid X, U, V) = P(A = 1 \mid X) := \pi(X)
\]
and so
\[
\varphi =  \tilde{\mu}(\tilde X) + \frac{A}{\pi( X)} (\tilde{M} - \tilde{\mu}(\tilde X)) -\theta.
\]
\end{proof}

\Proprsq*
\begin{proof}
We know from classic semiparametric theory that the semiparametric variance lower bound for $E_P\left[\tilde \mu(\tilde X_i)\right]$ will be given by $\text{Var}(\varphi(W_i))$ in our setting, where, per Proposition \ref{prop:mmfeif}, 
\[ \varphi = \tilde \mu(\tilde X) + \frac{A}{\pi(X)}(\tilde M - \tilde \mu(\tilde X)) - \theta. \]

Classic results in semiparametric inference for missing data tell us that, with some careful algebra,
\begin{align*}
    \text{Var}(\varphi)  &= \text{Var}\left(\tilde \mu(\tilde X) + \frac{A}{\pi(X)}(\tilde M - \tilde \mu(\tilde X)) - \theta\right)\\
    &= \text{Var}\left(\tilde \mu(\tilde X) + \frac{A}{\pi(X)}(\tilde M - \tilde \mu(\tilde X))\right)\\
    &= \text{Var}\left(\tilde \mu(\tilde X)\right) + \text{Var}\left(\frac{A}{\pi(X)}(\tilde M - \tilde \mu(\tilde X))\right)\\
    &= \text{Var}(E[\tilde M^* \mid \tilde X]) + E\left[ \frac{\text{Var}(\tilde M^* \mid \tilde X)}{\pi(X)}\right] .
\end{align*}
Then, by the law of total variance, we have that
\begin{align*}
   \text{Var}(\varphi) &= \text{Var}(E[\tilde M^* \mid \tilde X]) + E\left[ \frac{\text{Var}(\tilde M^* \mid \tilde X)}{\pi(X)}\right] \\
  &= \text{Var}(\tilde M^*) - E[\text{Var}(\tilde M^* \mid \tilde X)] + E\left[ \frac{\text{Var}(\tilde M^* \mid \tilde X)}{\pi(X)}\right] \\
   &= \text{Var}(\tilde M^*) +E\left[ \frac{\text{Var}(\tilde M^* \mid \tilde X)}{\pi(X)}-\text{Var}(\tilde M^* \mid \tilde X)\right] \\
   &= \text{Var}(\tilde M^*) + E[(\pi(X)^{-1}-1)\text{Var}(\tilde M^* \mid \tilde X)].
\end{align*}

Then, under a constant validation score $\pi$,
\begin{align*}
   \text{Var}(\varphi) 
   &= \text{Var}(\tilde M^*) + E[(\pi(X)^{-1}-1)\text{Var}(\tilde M^* \mid \tilde X)] \\
   &= \text{Var}(\tilde M^*) + E[(\pi^{-1}-1)\text{Var}(\tilde M^* \mid \tilde X)] \\
   &= \text{Var}(\tilde M^*) + (\pi^{-1}-1)E[\text{Var}(\tilde M^* \mid \tilde X)] \\
   &= \text{Var}(\tilde M^*) + (\pi^{-1}-1)\operatorname{Var}(\tilde M^*)(1 - R^2(\tilde X)) \\
   &= \text{Var}(\tilde M^*)\left(1+ \left(\frac{1}{\pi}-1 \right) \left(1 - R^2(\tilde X) \right) \right),
\end{align*}
which uses the fact that $E[\operatorname{Var}(\tilde M^* \mid \tilde X)]:= \operatorname{Var}(\tilde M^*)(1 - R^2(\tilde X))$, where $R^2(\tilde X)$ the nonparametric $R^2$.
\end{proof}

\subsection{The Geopolitical Risk Index} \label{sec:gpr}

An intellectual successor to \citet{baker_measuring_2016}, \citet{caldara_measuring_2022} construct a quantitative index measuring geopolitical risk (GPR). Similar to the EPU index, the GPR index at each point in time is calculated as the (normalized) share of articles from prominent newspapers that discuss rising geopolitical risks, based on a detailed keyword-based text query.\footnote{The query is Boolean and can be interpreted as a binary keyword-based classifier. However, unlike the EPU index, there is no explicit training algorithm: \citet{caldara_measuring_2022} constructed the query using data external to the analytic sample and did not update it based on the analytic sample. For the full query and details of its construction, see Section A.3 of the online appendix in \citet{caldara_measuring_2022}.}
The GPR index estimates the probability that a newspaper article discusses rising geopolitical risks, at a specific point in time for a given set of newspapers. This involves using newspaper text to estimate a functional of missing structured data (the expected value of a binary indicator for articles discussing geopolitical risk).

We follow an identical formal setup to the one described in the previous section and produce GPR indices (for $\mathcal{T} = \{1900, 1901,\dots,2015\}$), with and without application of the MAR-S framework. We again form multiple indices based on two distinct imputation functions: the original keyword query for GPR, and a deep neural classifier based on Microsoft's DeBERTaV3 model \citep{he_debertav3_2021}.\footnote{Specifically, we use the \texttt{large} variant.} As with the EPU index, we remain faithful to the definition of geopolitical risk crafted by \cite{caldara_measuring_2022} by leveraging only the validation labels that were created for their human audit sample. The labeled data from this human audit sample is iid randomly sampled from a large universe of ProQuest articles specified by \citet{caldara_measuring_2022}, and we use the entirety of this universe of articles as unlabeled data for imputing the index.\footnote{The universe of articles in question conditions on first meeting certain keyword inclusion requirements; this influences the interpretation of downstream estimands and estimates, and motivates correction by a known period-by-period constant, under certain assumptions. See appendix B.6 of \citet{caldara_measuring_2022} for more details. The validation score function using this universe is $\pi=0.00067$.} We employ an identical data splitting procedure as for the MAR-S EPU indices. In the estimation sample, the lightweight deep neural classifier achieves an accuracy of 81.0\%, and the keyword query achieves an accuracy of 66.2\%, using the human audit validations.

\begin{figure}[ht]
\centering
\caption{GPR and MAR-S GPR Indices}
\includegraphics[width=0.9\textwidth]{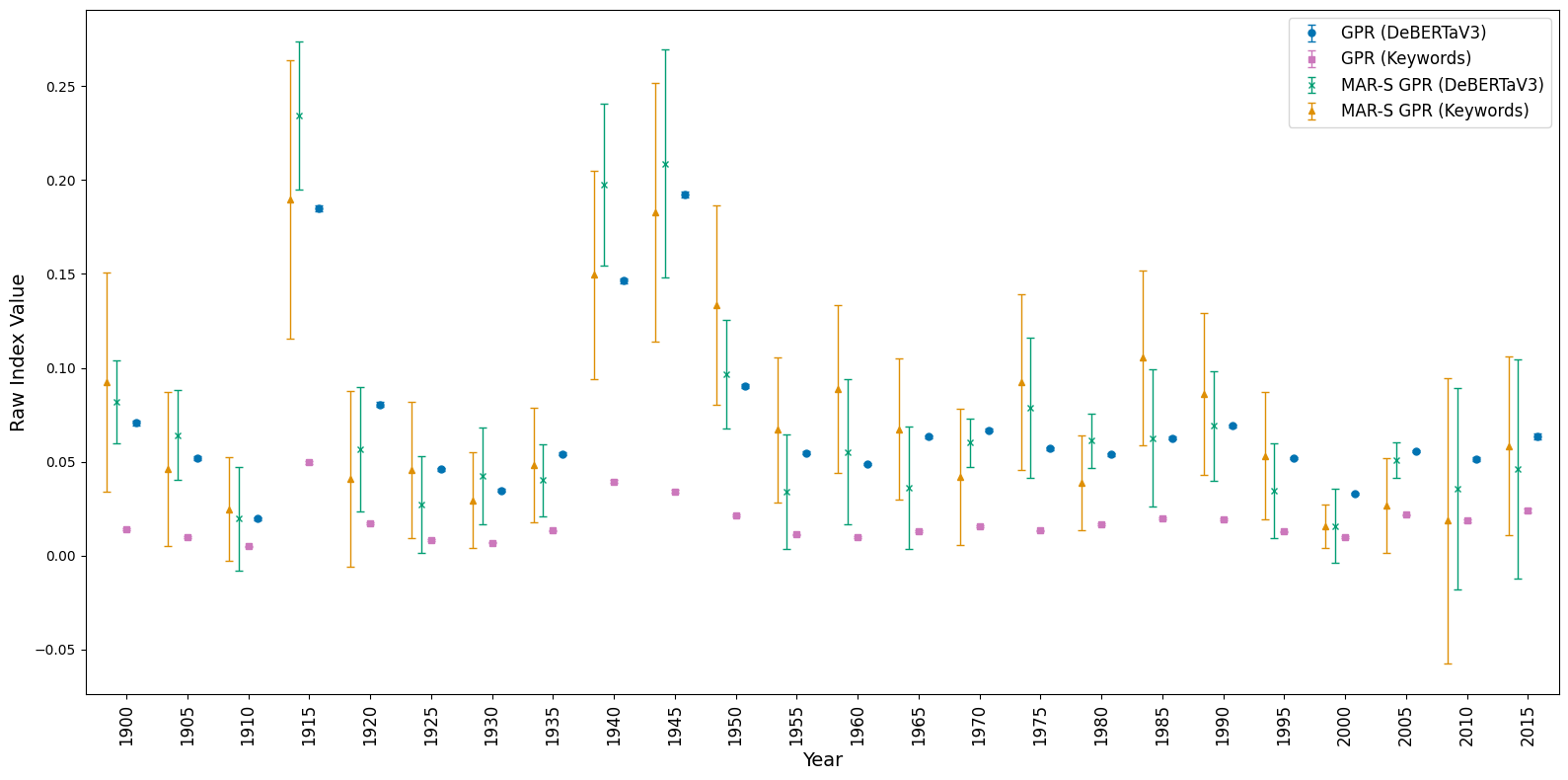}
\label{fig:gpr}

\footnotesize
\raggedright 
This figure plots unadjusted GPR indices - computed using a keyword classifier or DeBERTa classifier - along with the MAR-S adjusted indices - again computed using a keyword classifier or DeBERTa classifier. 95\% confidence intervals are plotted for each index. To maintain legibility, the indices are plotted at five year intervals. 
\end{figure}

Figure \ref{fig:gpr} plots the MAR-S GPR and unadjusted GPR indices, constructed using neural and keyword classifiers. The uncorrected GPR keyword index systematically and substantially underestimates geopolitical risk relative to the validation sample, as many articles in the audit sample discuss rising geopolitical risk but do not contain the requisite keywords. While level differences are large, variation across all indices is driven primarily by the World Wars (and September 11th, to a lesser extent). 
Confidence intervals are (erroneously) much narrower for the non-MAR-S indices than for the MAR-S indices. Moreover, the more accurate neural classifier generates narrower MAR-S confidence intervals than the keyword classifier.\footnote{Compared to estimates that only use the validation labels, the neural MAR-S GPR estimates are on average shorter (average width of 0.063 versus 0.069) and the keyword MAR-S GPR estimates are on average longer (average width of 0.082 versus 0.069).} Regression analyses in \citet{caldara_measuring_2022}  could be approached similar to the example above from \citet{baker_measuring_2016}, by using ME-LS with the debaised GPR index.

In supplementary materials, \citet{caldara_measuring_2022} show robustness of their target estimates to different keyword queries that make different tradeoffs between recall and precision, a common approach to address researcher degrees of freedom. Using neural networks significantly expands researcher degrees of freedom, relative to a world where data are created by some external source. Correcting for imputation bias addresses concerns that different choices (\textit{e.g.,} using a different prompt, model, training data set, or choice of hyperparameters) will lead to different estimates, by producing valid point estimates and confidence intervals as long as the relatively weak MAR-S assumptions are met. We view this as a significant advantage, as it is much easier to interpret - and plausibly easier to produce - a single debiased estimate than a lengthy appendix exploring different first-step estimators.

\subsection{Robustness versus Validation} \label{ssec:robustness}

\begin{figure}[!htbp]
    \centering
    \begin{subfigure}[t]{0.48\textwidth}
        \centering
        \includegraphics[width=\textwidth]{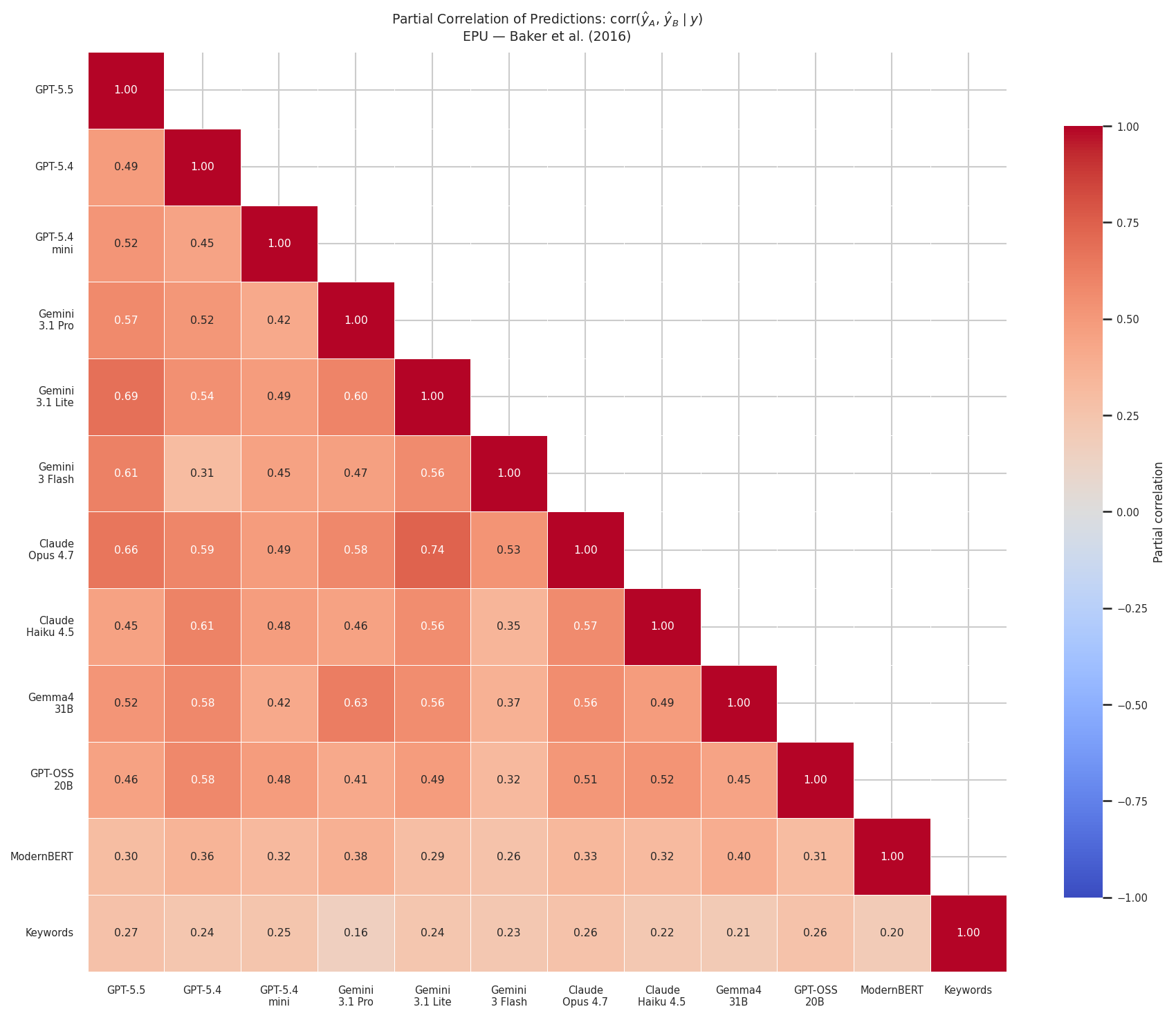}
        \caption{Correlation between model predictions of the EPU index, conditional on the validation data provided by \citet{baker_measuring_2016}.}
        \label{fig:epu_corr}
    \end{subfigure}
    \hfill
    \begin{subfigure}[t]{0.48\textwidth}
        \centering
        \includegraphics[width=\textwidth]{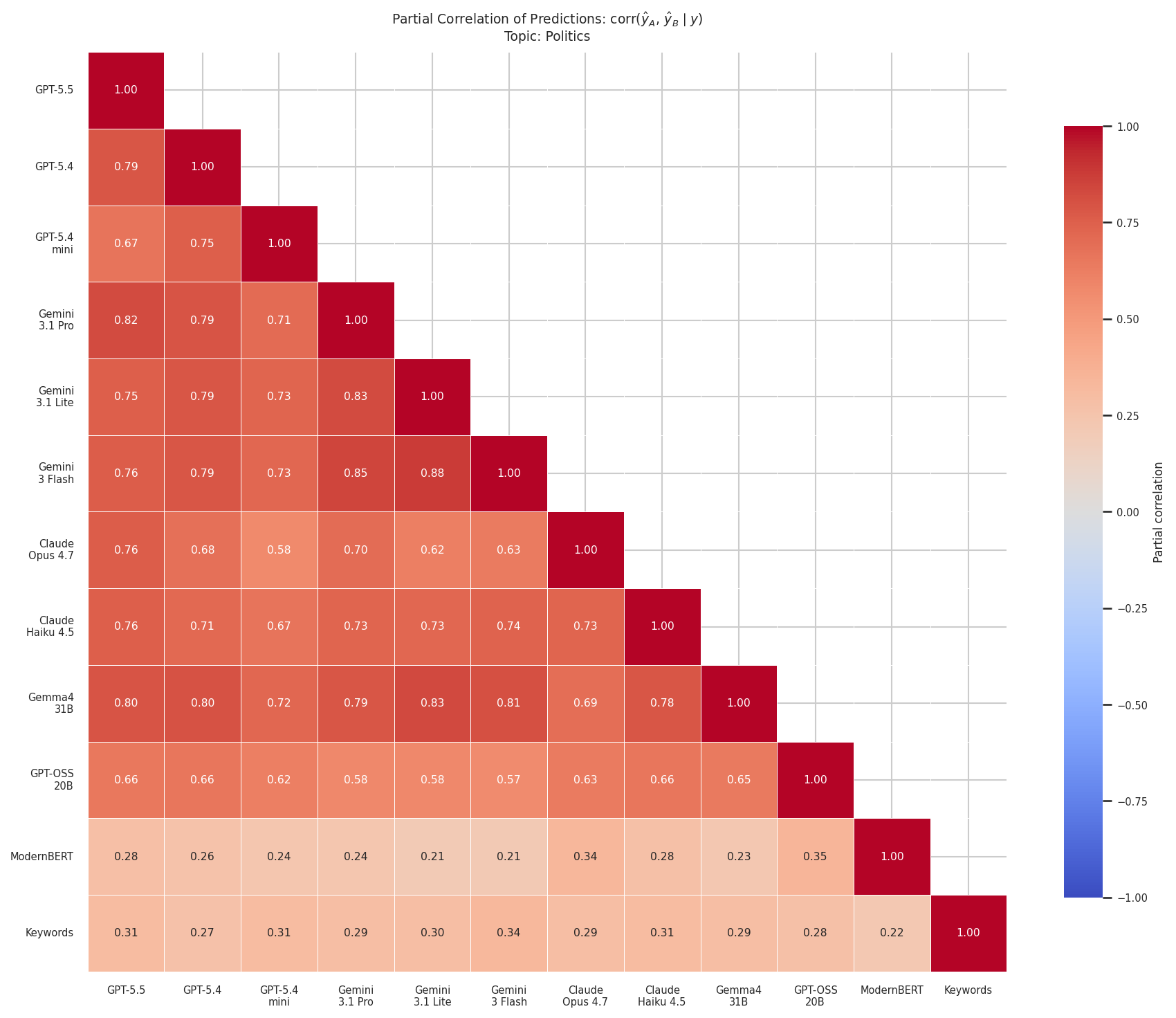}
        \caption{Correlation between model predictions of the political news index, conditional on the validation data provided by \citet{dell_american_2023}.}
        \label{fig:pol_corr}
    \end{subfigure}
    \caption{Correlation of Imputed Measures}
    \label{fig:corr_measures}
\end{figure}

Could researchers dispense with MAR-S and instead simply show robustness to using different imputation functions, or use one measure to instrument another? This requires the imputation errors to be independent. Panel (a) of Figure \ref{fig:corr_measures} reports pairwise error correlations for EPU measures imputed using different frontier LLMs, fine-tuned ModernBERT, and the original keyword classifier of \citet{baker_measuring_2016}, conditional on the \citet{baker_measuring_2016} validation labels. Panel (b) reports the analogous correlations for the politics classification task. Imputation errors are clearly correlated, particularly across frontier LLMs. %, just as the literature suggests \citep{Kim2025CorrelatedErrors}. 
Hence, correlated measures may reflect correlated errors, rather than revealing the underlying construct.

\end{document}

%% file: tab_estimators_revised.tex
\begin{table}[p]
\vspace*{-0.4in}
\centering
\caption{MAR-S estimators for common estimands}
\label{tab:mars_estimators}

\footnotesize
\renewcommand{\arraystretch}{1.35}
\setlength{\tabcolsep}{4pt}

\begin{tabular}{@{}
    >{\raggedright\arraybackslash}m{0.14\textwidth}
    >{\raggedright\arraybackslash}m{0.28\textwidth}
    >{\raggedright\arraybackslash}m{0.21\textwidth}
    >{\raggedright\arraybackslash}m{0.31\textwidth}
@{}}
\toprule
\textbf{Application}
& \textbf{Step 1: Define} \newline (moment, observed info $\tilde X$)
& \textbf{Step 2: Construct} \newline (imputation target)
& \textbf{Step 3: Derive} \newline (estimator) \\
\midrule

\multicolumn{4}{@{}l}{\textit{General}} \\

Just-identified moments
& $E[m(W;\theta)]=0$, \newline $W=(M^*,V)$, \newline $\tilde X=(X,U,V)$
& $\mathcal E(\tilde X;\vartheta)=E[m(W;\vartheta)\mid\tilde X]$, \newline approximated by $\eta(\tilde X;\vartheta)$
& $m_\eta=\eta-\frac{A}{\pi(X)}[\eta-m]$, \newline $\bar m_\eta(\vartheta)=\frac{1}{|\mathcal I|}\sum_i m_\eta(M_i,A_i,\tilde X_i;\vartheta)$, \newline $\hat\theta=\arg\min_\vartheta\bar m_\eta(\vartheta)'\bar m_\eta(\vartheta)$ \\

Over-identified moments
& $E[m(W;\theta)]=0$, \newline $d_m>d_\theta$, \newline $W$, $\tilde X$ as above
& $\mathcal E$ and $\eta$ are $d_m$-vectors, as above
& $m_\eta$, $\bar m_\eta$ as above, \newline $\hat\theta=\arg\min_\vartheta\bar m_\eta(\vartheta)'\hat Q\,\bar m_\eta(\vartheta)$, \newline $\hat Q$: GMM weighting matrix \\
\midrule

\multicolumn{4}{@{}l}{\textit{Means}} \\

Mean
& $m=M^*-\theta$, \newline $\tilde X=(X,U)$
& $\mu(\tilde X)=E[M^*\mid\tilde X]$, \newline approximated by $\hat\mu(\tilde X)$
& $m_\eta=\widetilde M-\theta$, where \newline $\widetilde M_i=\frac{A_i}{\pi(X_i)}[M_i-\hat\mu(\tilde X_i)]+\hat\mu(\tilde X_i)$, \newline $\displaystyle\hat\theta=\frac{1}{|\mathcal I|}\sum_i\widetilde M_i$ \\

Difference in conditional means
& $m_j=\mathbf 1\{K=j\}(M^*-\theta_j)$, $j\in\{k,l\}$, \newline
  $\theta_j=E[M^*\mid K=j]$, $\theta=\theta_k-\theta_l$, \newline
  $\tilde X=(X,U,K)$
& $\mu(\tilde X)=E[M^*\mid\tilde X]$
& $m_{\eta,j}=\mathbf 1\{K=j\}(\widetilde M-\theta_j)$, \newline
  $\displaystyle\hat\theta_j=\frac{\sum_i\mathbf 1\{K_i=j\}\widetilde M_i}{\sum_i\mathbf 1\{K_i=j\}}$, \quad $\hat\theta=\hat\theta_k-\hat\theta_l$ \\

Conditional mean, missing conditioning variable
& $M^*\in\{0,1\}$, $\theta=E[Y\mid M^*=1]$, \newline
  $m=M^*(Y-\theta)$, \newline
  $\tilde X=(X,U,Y)$ 
& $\mu(\tilde X)=P(M^*=1\mid\tilde X)$, \newline approximated by $\hat\mu(\tilde X)$
& $m_\eta=\widetilde M(Y-\theta)$, \newline
  $\displaystyle\hat\theta=\frac{\sum_i\widetilde M_iY_i}{\sum_i\widetilde M_i}$, \newline
  where $\sum_i\widetilde M_i/|\mathcal I|$ adjusts $P(M^*=1)$ \\
\midrule

\multicolumn{4}{@{}l}{\textit{Linear regression}} \\

Missing outcome
& $m=R(M^*-R'\theta)$, \newline $\tilde X=(X,U,R)$ 
& $\mu(\tilde X)=E[M^*\mid\tilde X]$, \newline approximated by $\hat\mu(\tilde X)$
& $m_\eta=R(\widetilde M-R'\theta)$, \newline
  $\displaystyle\hat\theta=\Big(\sum_iR_iR_i'\Big)^{-1}\sum_iR_i\widetilde M_i$, \newline
   \\

Missing regressor
& $m=R^*(Y-R^{*\prime}\theta)$, $R^*=\binom{M^*}{R}$, \newline
  $\tilde X=(X,U,R,Y)$ 
& $\mu(\tilde X)=E[M^*\mid\tilde X]$ and $\mu_2(\tilde X)=E[(M^*)^2\mid\tilde X]$, \newline
  approximated by $\hat\mu,\hat\mu_2$
& $m_\eta=\begin{pmatrix}\widetilde M\\R\end{pmatrix}Y-\begin{pmatrix}\widetilde{M^2}&\widetilde M R'\\R\widetilde M&RR'\end{pmatrix}\theta$, \newline
  $\widetilde{M^2}_i=\frac{A_i}{\pi(X_i)}[M_i^2-\hat\mu_2(\tilde X_i)]+\hat\mu_2(\tilde X_i)$, \newline
  $\displaystyle\hat\theta=\Big[\sum_i\begin{pmatrix}\widetilde{M_i^2}&\widetilde M_iR_i'\\R_i\widetilde M_i&R_iR_i'\end{pmatrix}\Big]^{-1}\sum_i\binom{\widetilde M_i}{R_i}Y_i$ \\
\midrule

\multicolumn{4}{@{}l}{\textit{Linear IV (just identified)}} \\

Missing outcome
& $m=Z(M^*-R'\theta)$, \newline
  $\tilde X=(X,U,R,Z)$ 
& $\mu(\tilde X)=E[M^*\mid\tilde X]$, \newline approximated by $\hat\mu(\tilde X)$
& $m_\eta=Z(\widetilde M-R'\theta)$, \newline
  $\displaystyle\hat\theta=\Big(\sum_iZ_iR_i'\Big)^{-1}\sum_iZ_i\widetilde M_i$ \\

Missing instrument
& $m=Z^*(Y-R'\theta)$, $Z^*=\binom{M^*}{C}$, $R=\binom{D}{C}$, \newline
  $\tilde X=(X,U,R,Y)$
& $\mu(\tilde X)=E[M^*\mid\tilde X]$, \newline approximated by $\hat\mu(\tilde X)$
& $m_\eta=\binom{\widetilde M}{C}(Y-R'\theta)$, \newline
  $\displaystyle\hat\theta=\Big[\sum_i\binom{\widetilde M_i}{C_i}R_i'\Big]^{-1}\sum_i\binom{\widetilde M_i}{C_i}Y_i$ \\
\bottomrule
\end{tabular}

\begin{minipage}{\textwidth}
\footnotesize

\textit{Notes:} All sums are over $i\in\mathcal I$; $\hat\mu,\hat\mu_2$ are fixed given the tuning sample. $\widetilde M_i$ is defined in the mean row. In the just-identified case, the GMM estimator solves $\bar m_\eta(\hat\theta)=0$. For over-identified GMM, $\hat Q$ is a positive definite weighting matrix. In the missing-instrument row, $D$ is the endogenous regressor and $C$ the exogenous controls, so $M^*$ instruments for $D$.
\end{minipage}
\thispagestyle{empty}
\end{table}